\documentclass[a4paper,11pt]{article}
\pdfoutput=1 % if your are submitting a pdflatex (i.e. if you have
\usepackage{jcappub}
\usepackage{aas_macros}
\usepackage[cp1252]{inputenc}
\usepackage[T1]{fontenc} % if needed

\usepackage{bm}

\newcommand{\tmp}{\textcolor{black}}

\newcommand\numberthis{\addtocounter{equation}{1}\tag{\theequation}}
\renewcommand{\vec}[1]{{\bm{#1}}}
\newcommand{\vnn}{\widehat{\vec{n}}}
\newcommand{\vk}{\vec{k}}
\newcommand{\meankappa}{\bar{\kappa}}
\newcommand{\bkd}{b_{\kappa,\delta}}
\newcommand{\bkg}{b_{\kappa,\Gamma}}
\newcommand{\fdf}{\tilde{\delta}_f}
\newcommand{\fdd}{\tilde{\delta}}
\newcommand{\fdj}{\tilde{\delta}_j}
\newcommand{\fdg}{\tilde{\delta}_\Gamma}
\newcommand{\fdE}{\tilde{\delta}_E}

\newcommand{\fhmn}{\widetilde{\delta h}_{mn}}

\newcommand{\HI}{H{\sc ~i}}
\newcommand{\HII}{H{\sc ~ii}}
\newcommand{\dr}{\mathrm{d}r}
\newcommand{\ds}{\mathrm{d}s}
\newcommand{\dg}{\mathrm{d}g}

\title{\boldmath On the accuracy of common moment-based radiative transfer methods for simulating reionization}

\author[a,1]{Xiaohan Wu,\note{Corresponding author.}}
\author[b]{Matthew McQuinn,}
\author[a]{Daniel Eisenstein}

\affiliation[a]{Harvard-Smithsonian Center for Astrophysics, 60 Garden Street, Cambridge 02138, MA, USA}
\affiliation[b]{Astronomy Department, University of Washington, Seattle, WA 98195, USA}

\emailAdd{xiaohan.wu@cfa.harvard.edu}
\emailAdd{mcquinn@uw.edu}
\emailAdd{deisenstein@cfa.harvard.edu}

\abstract{Modern cosmological simulations of reionization often treat the radiative transfer by solving for the monopole and dipoles of the intensity field and by making some ansatz for the quadrupole moments to close the system of equations. We investigate the accuracy of the most common closure methods, i.e. Eddington tensor choices. We argue that these algorithms are the most likely to err \tmp{after reionization} and study \tmp{qausi-analytic} test problems that mimic these situations: large-scale fluctuations in the post-reionization ionizing background and radiative transfer in a predominantly ionized medium with discrete absorbers. We show that the usual closure methods, OTVET and M1, over-ionize self-shielding absorbers when fixing the background photoionization rate, leading to $30-40\%$ higher emissivity to balance the increased recombination rate.  This over-ionization results in a simulation run with these algorithms having a factor of $\sim 2$ lower average metagalactic photoionization rate relative to truth for a given ionizing emissivity.  Furthermore, these algorithms are unlikely to reproduce fluctuations in the ionizing background on scales below the photon mean path: OTVET tends to overpredict the fluctuations there when the simulation box is smaller than twice the mean free path and underpredict otherwise, while M1 drastically underpredicts these fluctuations.  As a result, these numerical methods are likely not sufficiently accurate to interpret the Ly$\alpha$ forest opacity fluctuations observed after reionzation.  We also comment on ray tracing methods, showing that a high number of angular directions need to be followed to capture fluctuations in the post-reionization ionizing background accurately. Lastly, we argue that the strong dependence of the post-reionization ionizing background on the value of the reduced speed of light found in many simulations signals that the ionizing photon mean free path is several times larger in such simulations than the observationally measured value.}

\begin{document}
\maketitle
\flushbottom

\section{Introduction}
\label{sec:intro}

Much is still unknown about the era when the ionizing photons from the first stars and galaxies ionized the intergalactic medium (IGM), reionization. Owing to the non-linearity of this process, numerical simulations are necessary to interpret most reionization observables, including the Lyman-$\alpha$ forest, kinetic Sunyaev-Zeldovich effect, Lyman-$\alpha$ emitters, and 21-cm radiation \citep{trac11, mcquinn16}.
Interpreting these observations is thus limited by the accuracy of the simulations.  Simulating the growth and overlap of the ionized bubbles requires performing radiative transfer (RT), but solving the full six-dimensional RT equation often is prohibitive.  As a result, various approximate methods that reduce the dimensionality of the problem have been devised.

Methods that reduce the dimensionality of the problem by taking angular moments of the RT equation, only following the monopole or dipole moments and making an ansatz for the higher moments, are used in a significant fraction of all reionization simulations \citep[e.g.][]{gnedin14, chardin15, chardin17, chardin18a, chardin18b, keating18, keating19, keating20, kulkarni19, roy20, wu19a, wu19b, aubert10, ocvirk16, ocvirk18, rosdahl18, katz18, so14, norman15, norman18, petkova11, bauer15, finlator09b, finlator11, finlator12, finlator18}. Specifically, these algorithms must assume some form for one additional angular multipole beyond what they are computing to close the system equations, which for the dipole-moment equations is the quadrupole moments, mathematically represented by the ``Eddington tensor''.  The two most popular closure approximations consist of calculating the Eddington tensor as if all sources are optically thin (the OTVET algorithm \cite{gnedin01}) or an approximation that uses the local ratio of the radiative flux to energy density to interpolate between a highly anisotropic tensor that is anticipated at large ratios and an isotropic one at low ones (the M1 algorithm \cite{levermore84}). This work examines the accuracy of these closure approximations when simulating reionization, with a particular focus on \tmp{post-reionization}. The other numerical method used for simulating reionization explicitly traces rays through the simulation volume, but ray tracing codes often use a limited pixelization in angular coordinates.  Our work also has some bearing on the loss of accuracy from such pixelization.

Simulating \tmp{the post-reionization era} is necessary to interpret one of our primary observables of reionization, the Lyman-$\alpha$ forest \citep[e.g.][]{fan06, mcgreer15, bosman18, becker15}. Indeed, reionization simulations have been used to interpret the forest and place some of the strongest constraints on the timing of reionization \citep{keating20, kulkarni19, wu19b}, and simulations of reionization are often calibrated to match the $z\lesssim6$ Lyman-$\alpha$ forest observations  \citep{keating19, wu19b}. This calibration takes the form of varying parameters that adjust the source emissivities to reproduce the mean transmission observed in the forest, which is found to evolve dramatically at $z>6$ before setting onto a power-law like relation that is set by the metagalactic \HI\ photoionization rate \citep{fan06}.  Matching the evolution of forest transmission in a simulation is thought to indicate that the simulated reionization is ending near the correct time and that the simulation is capturing the post-reionization ionizing background properly.  More recently, simulations have been used to investigate the large scatter in the forest opacity at $z\sim5-6$ between different spatial regions, which is thought to owe to large fluctuations in the ultraviolet background \cite{becker18, meiksin20}.  So far, simulations have had difficulty reproducing the magnitude of this scatter \cite{chardin15, chardin17, keating18, keating19, keating20, gnedin17, meiksin20}, possibly owing to the simulation volumes being too small to capture the rare bright sources that dominate the ionizing background fluctuations \cite{meiksin20}. Such inferences from the forest require radiative transfer simulations, but the more diffusive propagation for standard Eddington tensor closure approximations may affect the conclusions.  This study addresses whether moment-based RT methods can capture properly the mean transmission in the forest as well as its fluctuations.

There are hints that the modeling errors from these approximate RT methods are less severe during the bulk of reionization. For instance, although moment based algorithms may not fully capture the shadowing behind opaque clouds, this failure likely does not significantly impact the evolution of ionized volume and mass fractions or on the morphology of reionization and, indeed,  \cite{norman15} found that a moment method reproduced a similar reionzation morphology to a full ray-tracing calculation\tmp{, albeit without a quantitative comparison}.  Furthermore,  semi-analytic models of reionization based on the excursion set formalism produce very similar reionization morphologies as ray-tracing codes do, being $\gtrsim50\%$ accurate when predicting the power spectrum of the ionization field \cite{zahn07, zahn11}, despite having following no radiative transfer whatsoever.  %\footnote{This also hints that IGM temperature fluctuations are less affected by the RT methods, since temperature fluctuations are imprinted according to the growth of the ionized bubbles in the bulk of reionization.}
A counter-argument is that Lyman-limit systems are not necessarily resolved by the simulations that were used to show agreement with the excursion set models, and the abundance of these systems is thought to cap the bubble size during reionization \cite{furlanetto05}\tmp{, potentially affecting the simulated 21-cm power spectrum at factors of $\sim2$ level \cite{sobacchi14, shukla16}}.  The radiative transfer method can affect whether the effect of self-shielding absorbers is captured, and our calculations do shed light on this issue.

There has been one previous effort to compare radiative transfer methods, the RT code comparison project of \cite{iliev06, iliev09}.  This comparison showed that most RT codes are in reasonable agreement with each other for simulating simple problems such as an \HII\ region around a single source and a shadow behind a dense absorber, which are more relevant to the bulk of reionization \cite[although OTVET and flux limited diffusion methods do not accurately capture shadowing; see][]{petkova09, norman15}.  However, none of the test problems in \cite{iliev06, iliev09} mimic the context of the post-reionization IGM, when there are many streams of radiation coming from multiple directions.  Here we develop test problems targeted towards this phase.

In this work, we design toy problems to investigate the ionization structure of absorbers of ionizing photons and ionizing background fluctuations. 
 %The former hints upon the degree to which moment-based RT methods recover the post-reionization photoionization rate, emissivity, and photon mean free path, which affects the calibration of reionization models. 
In order for the toy problems to be analytically tractable, we consider
\begin{enumerate}
    \item a single source or a single absorber with spherical/planar symmetry;
    \item problems where the fluctuations are perturbative, a characteristic that applies to post-reionization ionizing backgrounds.
\end{enumerate}
With these tests, we show that the Eddington tensor approximations are able to reproduce the correct solution to the \HII\ region expansion problem with a single source, supporting our conjecture that RT algorithms err more when simulating the end of reionization.  When considering a single isolated absorber, we are able to make inferences about how well these algorithms capture the ionization of the IGM after reionization.  We find that moment-based RT methods are likely to be less ionized within ionized regions (and hence less transmissive in the Ly$\alpha$ forest) relative to the correct solution, when fixing the emissivity.  
Furthermore, in the perturbative limit thought to hold soon after reionization, we show that these approximate RT schemes predict a substantially different spectrum for fluctuations in the photoionization rate.  %This leads to the conclusion that the Eddington tensor approximations used by existing reionization codes should not be used to study opacity fluctuations in the post-reionizaiton IGM.

This paper is organized as follows. In Section~\ref{sec:rt_review} we review the moment-based RT methods and different forms of the Eddington tensors used in the literature. Sections~\ref{sec:stromgren}, \ref{sec:spherical_absorber}, and \ref{sec:Gamma_fluc} present our toy problems. We summarize our work in Section~\ref{sec:conclusions}.

\section{A review of the moment-based RT implementations in the literature}
\label{sec:rt_review}

Let $I_\nu(\vec{x}, \vnn, t)$ denote the specific intensity at comoving position $\vec{x}$ and time $t$ moving in the direction $\vnn$. The equation of radiative transfer (RT) in an expanding universe is
\begin{equation}
\frac{1}{c}\frac{\partial I_\nu}{\partial t} + \frac{1}{a}\vnn\cdot\nabla I_\nu + \frac{1}{c}H\left( 3I_\nu - \nu\frac{\partial I_\nu}{\partial \nu} \right) = -\kappa_\nu I_\nu + j_\nu,
\label{eq:rt_full}
\end{equation}
where $\kappa_\nu$ is the absorption coefficient and $j_\nu$ is the emissivity coefficient, which we will assume to be isotropic, i.e. independent of $\vnn$. We will also assume that photons do not get significantly redshifted or diluted before they are absorbed --- equivalent to the photon mean free path being small compared to the horizon scale --- so that the terms with the Hubble function $H$ can be dropped.  This simplification is an excellent approximation for ionizing radiation at $z>3$ \citep{worseck14, meiksin18}. With this simplification, the radiative transfer equation reduces to
\begin{equation}
\frac{1}{c}\frac{\partial I_\nu}{\partial t} + \frac{1}{a}\vnn\cdot\nabla I_\nu = -\kappa_\nu I_\nu + j_\nu.
\label{eq:rt}
\end{equation}

Since we are interested in the \HI\ photoionization rate, we will focus on the frequency-integrated form of the RT equation. This is motivated by the \HI\ photoionization cross-section being sharply peaked at the Lyman limit ($\sigma_\nu\propto\nu^{-3}$). Additionally, stellar radiation, which dominates reionization, is also relatively soft with $j_\nu \sim \nu^{-2}$ and cuts off at $\sim4$~Ry, further justifying a monochromatic treatment. Moreover, \cite{meiksin18} showed that solving the full frequency-dependent RT equation only leads to sub-percent differences when computing the fluctuations in the photoionization rate at $z=2-3$ compared to the frequency-integrated approach.
We therefore will solve a frequency-averaged RT equation by integrating equation~\ref{eq:rt}, weighted by $\sigma_\nu$:
\begin{equation}
\frac{1}{c}\frac{\partial f}{\partial t} + \frac{1}{a} \vnn \cdot \nabla f = -\kappa f + j,
\label{eq:rt_f}
\end{equation}
where
\begin{equation}
f \equiv \int \frac{I_\nu(\vec{x},\vnn)}{h_P\nu} \sigma_\nu\; \mathrm{d}\nu,\quad
j \equiv \int \frac{j_\nu}{h_P\nu} \sigma_\nu\; \mathrm{d}\nu,\quad
\kappa \equiv \frac{\int \frac{I_\nu(\vec{x},\vnn)}{h_P\nu} \sigma_\nu \kappa_\nu\; \mathrm{d}\nu}{\int \frac{I_\nu(\vec{x},\vnn)}{h_P\nu} \sigma_\nu\; \mathrm{d}\nu}.
\label{eqn:momentdefs}
\end{equation}

\subsection{The Eddington tensor}

The moment-based RT equations take the zeroth and first angular moments of equation~\ref{eq:rt_f}, which yields respectively the following equations
\begin{gather}
\frac{1}{c} \frac{\partial E}{\partial t} + \frac{1}{a} \nabla\cdot\vec{F} = -\kappa E + j;
\label{eq:rt_monopole}
\\
\frac{1}{c} \frac{\partial \vec{F}}{\partial t} + \frac{1}{a} \nabla\cdot\mathbf{P} = -\kappa \vec{F},
\label{eq:rt_dipole}
\end{gather}
where 
\begin{equation}
E = \frac{1}{4\pi} \int f \; \mathrm{d}\Omega;\quad
\vec{F} = \frac{1}{4\pi} \int \vnn f \; \mathrm{d}\Omega;\quad
\mathbf{P} = \frac{1}{4\pi} \int \vnn\otimes\vnn f\; \mathrm{d}\Omega,
\end{equation}
and $\Omega$ is the solid angle.
Here $E$ is the photon energy density, $\vec{F}$ the photon flux, and $\mathbf{P}$ the radiation pressure tensor. We have assumed that the source term $j$ is isotropic. Often rather than $\mathbf{P}$, equation~\ref{eq:rt_dipole} is expressed in terms of the Eddington tensor $h_{mn}$\footnote{Note that although we dropped the frequency dependence, formally $h_{mn}$ should be $h_{mn,\nu}$.}, which is defined as
\begin{equation}
E h_{mn} = P_{mn}.
\end{equation}
The full solution to $P_{mn}$ and $h_{mn}$ is
\begin{gather}
P_{mn} = \int j(\vec{x'}) e^{-\tau(\vec{x}, \vec{x'})} \frac{(x_m-x_m')(x_n-x_n')}{|\vec{x}-\vec{x'}|^4}\; \mathrm{d}^3\vec{x'},
\label{eq:Pij_exact}
\\
h_{mn} = P_{mn} / \mathrm{Tr}(P_{mn}),
\end{gather}
where $\tau(\vec{x}, \vec{x'})$ is the optical depth between points $\vec{x}$ and $\vec{x'}$. This expression ignores the light-travel time delay, which is a good approximation when the photon mean propagation time is much smaller than the lifetime of the sources and the evolutionary timescale of the source population.\footnote{In reality, starbursts occur at $\sim10$~Myr timescales and sources at higher redshifts are even more busty. This can make equation~\ref{eq:Pij_exact} less accurate. However, for $\gtrsim10$~comoving Mpc ionized bubble sizes that contain numerous sources, the timescale that $P_{mn}$ varies corresponds to the timescale that the emissivity within a bubble changes. In the limit of many sources, the total emissivity changes in a bubble on a timescale comparable to the Hubble time.} The trace of $h_{mn}$ equals $1$, with all eigenvalues being $1/3$ for an isotropic radiation field. Invoking the same rational for ignoring light travel in the Eddington tensor, we also focus on the time-dependent solution of the RT equation. In this limit, the moment equations can be easily combined into one second order differential equation for $E$:
\begin{equation}
\frac{1}{a^2} \frac{\partial}{\partial x_m} \left( \frac{1}{\kappa} \frac{\partial E h_{mn}}{\partial x_n}\right) - \kappa E + j = 0.
\label{eq:rt_diffusion}
\end{equation}

While Eqn. \ref{eq:rt_diffusion} has a simple diffusion-like form, the complexity is hidden in the Eddington tensor $h_{mn}$, for which an exact calculation is often prohibitive. Evaluating equaticon~\ref{eq:Pij_exact} in numerical simulations requires integrating along many sightlines to obtain $\tau$, which results in an unsatisfying $\mathcal{O}(N^{5/3})$ scaling, where $N$ is the number of grid cells \citep{gnedin01}. It is therefore desirable to `close' the equation with approximate forms of the Eddington tensor that can lower the computational cost significantly.  We are aware of no astrophysics code that has closed at a higher order tensor, such as the three-index tensor that appears when taking the quadruple moment of the RT equation.  We note that \cite{finlator09} calculates the Eddington tensor by evaluating equation~\ref{eq:Pij_exact} using a long characteristics ray-tracing method that takes into account sources in 26 replicas of the periodic simulation box.  This method gives more accurate Eddington tensors than the approximate ones described below, but is computationally expensive so the Eddington tensors are not updated at every time-step.

However, an approximate Eddington tensor can likely lead to the violation of causality ($|\vec{F}| \le E$), under the diffusive approximation to the photon flux
\begin{equation}
F_m = -\frac{1}{a\kappa} \frac{\partial E h_{mn}}{\partial x_n}.
\end{equation}
In treatments of RT that assume an isotropic Eddington tensor, a flux limiter is often included to enforce $|\vec{F}| \le E$ as required by eqn.~\ref{eqn:momentdefs}, which algorithms can be increasingly stressed to satisfy in the presence of large spatial gradients of $E$ \cite{levermore81}. Using $\vec{R} \equiv \nabla E/ \left[\kappa E \right]$, a flux limiter $\lambda(R)$ is introduced where $R = |\vec{R}|$ and $\lambda(R) R \to 1^-$ as $R\to\infty$, so that $\vec{F} = -\lambda(R) \vec{R} E$ always satisfies causality. In the anisotropic diffusion treatment, one can similarly introduce a flux limiter with $F_m = -\left[ \lambda(R)/\kappa \right]\; \partial (E h_{mn})/\partial x_n$ \cite{petkova09}.

Below, we review popular approximations for the Eddington tensor that have been used in the literature.

\subsubsection{OTVET closure}
\label{sec:otvet_review}

The optically thin Eddington tensor (OTVET) has been described or used in \cite{gnedin01, gnedin14, petkova09, petkova11}. The main idea is to calculate the Eddington tensor assuming no attenuation so that $\tau=0$ in equation~\ref{eq:Pij_exact} \citep{gnedin01}. More formally, the OTVET approximation replaces equation~\ref{eq:Pij_exact} for the momentum flux tensor with
\begin{equation}
P_{mn}(\vec{x}) = \int j(\vec{x'}) G_{mn}(\vec{x}-\vec{x'})\; \mathrm{d}^3\vec{x'},
\label{eq:Pij_otvet}
\end{equation}
where the Green's function kernel is
\begin{equation}
G_{mn}(\vec{x}) = \frac{x_m x_n}{|\vec{x}|^4}.
\end{equation}

As equation~\ref{eq:Pij_otvet} is a convolution, it can be evaluated in $N\log N$ time with the Fast Fourier transform. However, Fourier transforming $G_{mn}$ leads to a divergent zeroth mode (see Appendix~\ref{sec:FT_deltahij}), implying that the Eddington tensor is isotropic everywhere in a periodic volume. This isotropy occurs for the same reason the sky is infinitely bright in an infinite static universe (Olber's paradox). In practice, implementations of OTVET only include image sources out to half a box size away at a given location in the otherwise-periodic simulation to achieve a non-trivial Eddington tensor.\footnote{For instance, \citep{gnedin14} calculates the Eddington tensor by setting up a source grid and a $G_{mn}$ grid in $x$-space, performing the discrete Fourier transform to $k$-space, and transforming their multiplication back to $x$-space. This method ensures that the Eddington tensor is exact out to half a box size away in the case of a single source (see Appendix~\ref{sec:FT_deltahij} for details).} The maximum source distance that is used to calculate the Eddington tensor is thus capped by the simulation box size.

The OTVET algorithm of \cite{gnedin14} can lead to OTVET misestimating the degree of anisotropy of the true Eddington tensor. To understand this error, we estimate how far away from a source the Eddington tensor becomes roughly isotropic.  Assume that the total emissivity is dominated by sources with number density $n_*$ and luminosity $L_*$, and that the photon mean free path is $\lambda_{\rm mfp}$. The radius at which the more or less isotropic ionizing background starts dominating over the flux from the source can be estimated by calculating the proximity region of a source
\begin{gather}
\frac{L_*}{4\pi r^2} = \frac{1}{4\pi}n_* L_* \lambda_{\rm mfp}, \\
\implies r \sim 1/\sqrt{n_* \lambda_{\rm mfp}}.
\end{gather}
Since OTVET assumes an unattenuated background when calculating the Eddington tensor and follows sources out to half a box length, the OTVET Eddington tensor becomes approximately isotropic for $r$ that corresponds to taking $\lambda_{\rm mfp}\rightarrow L_{\rm box}/2$ in the above equation, where $L_{\rm box}$ is the szie of the simulation box, leading to the Eddington tensor becoming isotropic at an incorrect scale if $\lambda_{\rm mfp} \neq L_{\rm box}/2$.    %For the exact Eddington tensor, $\lambda_{\rm mfp}$ can the typical size of ionized bubbles during reionization ($\sim10$ comoving Mpc \cite{furlanetto05}), or the observed mean free path after reionization (e.g. $\sim80$ comoving Mpc at $z=5$ \cite{worseck14}). For OTVET, $\lambda_{\rm mfp}$ in the above equation is instead roughly the simulation box size of $L$. Note that the above estimate ignores source clustering. This is a good approximation when the mean free path (or box size, if using OTVET) is large, so that sources contributing to the Eddington tensor are abundant and distributed roughly isotropically within one mean free path.

Let us estimate the critical scale $r$ above which OTVET transitions to an isotopic Eddington tensor. During reionization, for $n_*=0.3, 0.01, 10^{-4}$~Mpc$^{-3}$, which are the comoving number densities of $10^9, 10^{10}, 10^{11}\ M_\odot$ halos at $z=8-9$ respectively, then for a box size of $L_{\rm box}=20$~comoving Mpc, OTVET predicts that at $r \gtrsim 0.6, 3.2, 32$~comoving Mpc away from a source respectively, the Eddington tensor transitions to isotropy.  These sizes are comparable to a single \HII\ region around a source, but are smaller than the typical bubble sizes simulation find throughout reionization.
After reionization, if we take $\lambda_{\rm mfp} = 80$~comoving Mpc \cite{worseck14} and $n = 1, 0.08, 0.004$~Mpc$^{-3}$, which correspond to the comoving number density of $10^9, 10^{10}, 10^{11}\ M_\odot$ halos at $z=5$ respectively, then the true sizes of proximity region are $0.1, 0.4, 1.8$~comoving Mpc respectively. These values are much smaller than the photon mean free path, leading to the exact Eddington tensor being isotropic in the vast majority of the post-reionization IGM. However, for a simulation using OTVET, if the box size is much smaller (larger) than $\lambda_{\rm mfp}$, e.g. $L_{\rm box}=20(1000)$~comoving Mpc, the extent over which the Eddington tensor is isotropic will be enlarged (shrunken) by a factor of $2.8$ ($2.5$), since OTVET implies an effective $\lambda_{\rm mfp}=10(500)$~Mpc for such a box size. OTVET thus overestimates the degree of anisotropy of the exact Eddington tensor after reionization in a small box simulation, while underestimates it in a large box simulation. As we will show in Section~\ref{sec:Gamma_fluc}, this leads to significant errors when estimating the fluctuations in the post-reionization ionizing background at small scales.

\subsubsection{M1 closure}
\label{sec:M1_review}

While the M1 closure is motivated by capturing how an isotropic black body transforms when boosted into different inertial frames \citep{levermore84}, a situation not applicable to reionization, it has been widely adopted in radiative transfer calculations of reionization owing to its simplicity as the Eddington tensor in M1 is a local function of $E$ and $\vec{F}$.  M1 is implemented in a number of cosmological simulation codes \cite{aubert08, rosdahl13, rosdahl15, aubert15, kannan19}, and it has been used in the reionization simulations of \cite{bauer15, ocvirk16, ocvirk18, rosdahl18, keating18, keating19, keating20, kulkarni19, wu19a, wu19b}.  % probably not enough citation here but so many papers...

M1 starts with a decomposition for the Eddington tensor \citep{levermore84}
\begin{equation}
h_{mn} = \frac{1-\chi}{2}\delta^K_{mn} + \frac{3\chi-1}{2} \hat{\mathrm{n}}_m \hat{\mathrm{n}}_n;\quad
\hat{\mathrm{n}}_m = \frac{F_m}{|\vec{F}|},
\label{eq:hij_chi}
\end{equation}
where $\delta^K_{mn}$ is the Kronecker delta. This form assumes that the specific intensity is symmetric around the direction of $\vec{F}$, with the Eddington tensor having an eigenvalue $\chi$. M1 chooses a particular form for $\chi$ given by
\begin{equation}
\quad
\chi = \frac{3+4g^2}{5+2\sqrt{4-3g^2}};\quad
g = \frac{|\vec{F}|}{E}.
\end{equation}
This relation between $\chi$ and $g$ is obtained by assuming that the specific intensity is isotropic in some inertial frame, and transforming back to the lab frame. This relation also ensures that $|\vec{F}| \le E$  \citep[i.e. the M1 scheme is flux-limited][]{levermore84}.

\begin{figure}[tbp]
\centering 
\includegraphics[width=0.7\textwidth]{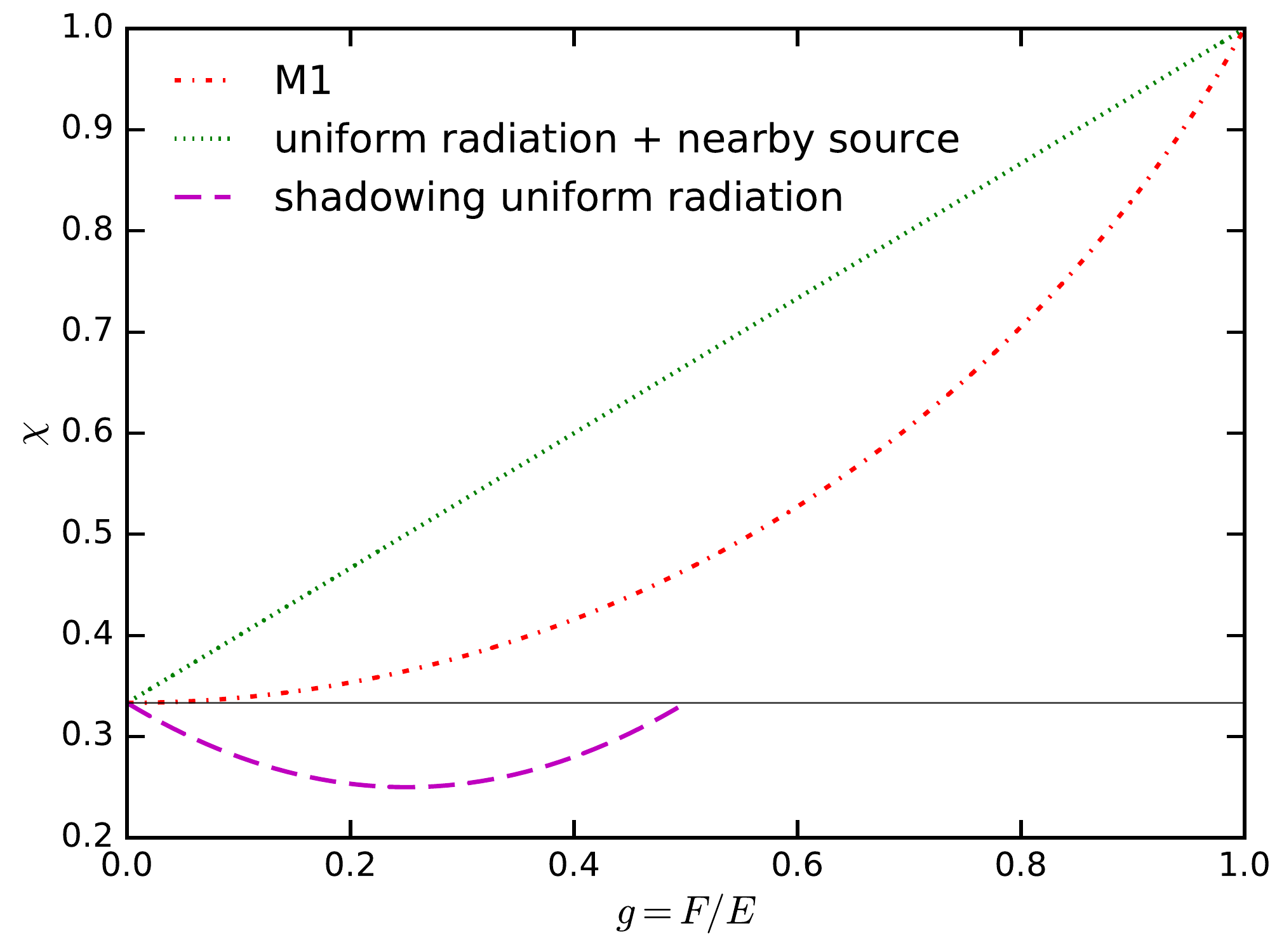}
\hfill
\caption{\label{fig:chi_g} $\chi(g)$ relation that determines the Eddington tensor in equation~\ref{eq:hij_chi}, in different contexts related to reionization. The green dotted line represents $\chi(g)$ near a point source in a uniform radiation background. Magenta dashed line shows $\chi(g)$ outside a spherical absorber with sharp boundary and infinite opacity which shadows uniform radiation coming from infinity. The M1 $\chi(g)$ relation is illustrated by the red dot-dashed line. The thin black line show $\chi=1/3$, the value that corresponds to an isotropic Eddington tensor.}
\end{figure}

% While the M1 form of $\chi(g)$ has not been motivated specifically for reionization,\footnote{M1 can be seen as an interpolation between the isotropic diffusion limit where $g = |\vec{F}|/E \ll 1$ and $h_{mn}=\delta^K_{mn}/3$, and the free-streaming limit where $g=1$, $h_{mn} = \hat{\mathrm{n}}_m \hat{\mathrm{n}}_n$, limits that may have some applicability to reionization.} equation~\ref{eq:hij_chi} is more generic and applies to some contexts in the overlap phase or after reionization. 
To understand how accurately the M1 Eddington tensor describes the radiation field in the overlap phase or after reionization, let us consider the form of the Eddington tensor in two scenarios: a point source in a uniform radiation background, and a spherical absorber with sharp boundary and infinite opacity shadowing uniform radiation coming from infinity. In both cases, the radiation field is symmetric around the direction of $\vec{F}$ and so the Eddington tensor can be written in the form of equation~\ref{eq:hij_chi}. In the first case, $g\to1$ near the source and $g\to0$ far outside the proximity region, and the Eddington tensor satisfies $\chi(g) = (1+2g)/3$. In the latter case, outside the absorber $g\to0$ at large radii, and increases to $1/2$ near the boundary of the absorber. However, the exact solution's $\chi$ is not monotonic, and $\chi=1/3$ both at large radii and right outside the boundary of the absorber, instead taking the form $\chi(g) = (4g^2 - 2g + 1)/3$. The green dotted and magenta dashed lines in Figure~\ref{fig:chi_g} show these $\chi(g)$ relations, corresponding to point source in uniform radiation field and the shadowing of a spherically symmetric absorber exposed to a uniform radiation background, respectively. The M1 $\chi(g)$ relation is illustrated by the red dot-dashed line. Owing to its monotonicity, the M1 form of $\chi(g)$ better represents the former case. Although M1 can be seen as an interpolation between the isotropic diffusion limit where $g = |\vec{F}|/E \ll 1$ and $h_{mn}=\delta^K_{mn}/3$, and the free-streaming limit where $g=1$, $h_{mn} = \hat{\mathrm{n}}_m \hat{\mathrm{n}}_n$, it does not capture the radiation field around absorbers. In fact, in Appendix~\ref{sec:absorber_eqns} and Section~\ref{sec:spherical_absorber} we will show that M1 over-ionizes absorbers because it cannot capture such non-monotonic $\chi(g)$.

For OTVET, we have shown that outside the ``proximity region'' the Eddington tensor becomes isotropic, for box sizes corresponding to cosmological volumes. For M1, in large bubbles ($\gtrsim10$~comoving Mpc) or when the photon mean free path is large, $\left(\mathrm{d}\chi/\mathrm{d}g\right)\big|_{g=0}=0$ indicates that the Eddington tensor tends to isotropic at linear order in density, ignoring source clustering. We will show in Section~\ref{sec:Gamma_fluc} that this leads to biases in the simulated ionizing background fluctuations.

\subsection{Comparison to ray-tracing methods}
\label{sec:raytracing_review}

In addition to moment-based approaches, more accurate ray-tracing methods are also used for simulating reionization \cite{abel02, mellema06, mcquinn07, trac07, wise11, pawlik08, pawlik11, petkova11code}. The long-characteristics method integrates the RT equation from each source cell to each gas cell, while the short-characteristics method performs the integration only along lines that connect nearby cells, with the boundary conditions at the cell faces obtained by interpolation.
\tmp{Both methods have been used extensively for studying the bulk of reionization \cite[e.g.][]{iliev06sim, mellama06sim, jensen13, giri19, trac08, battaglia13}.}

However, the $\mathcal{O}(N^2)$ scaling of long-characteristic ray-tracing method is computationally expensive, encountering difficulty simulating through the era of overlap and past the end of reionization. Adaptive ray-tracing mitigates this by splitting and merging rays, and at the end of reionization rays are limited by some algorithms to span a finite number of solid angles \cite{trac07, wise11, pawlik08}. Such a ray limiting scheme makes the algorithm effectively behave like the short-characteristics method\tmp{, since in essence both methods sample a finite number of directions.}  We explore the effect of ray binning further in Section~\ref{sec:Gamma_fluc} and Appendix~\ref{sec:arctan_higher_order}.

% In moment-based RT codes, ray-tracing methods have also been used to compute the Eddington tensor. For instance, \cite{finlator09} calculates the Eddington tensor by evaluating equation~\ref{eq:Pij_exact} using a long characteristics method. After each time-step, any cell in which $E$ changes by more than a certain threshold is flagged for update of the Eddington tensor. A ray-tracing code then calculates the total optical depth from the cell to each source cell. The simulation box is replicated 26 times around the actual computational volume to account for periodic boundary condition. This method yields more accurate Eddington tensors than M1 and OTVET, albeit more time-consuming. Especially, this method alleviates the errors that OTVET makes in the shapes of the ionized regions in the presence of multiple sources \citep[Fig.~5 of][]{finlator09}. This implementation has been used in the simulations of \cite{finlator09b, finlator11, finlator12, finlator18}.

\section{Moment-based methods versus exact solution: expansion of ionized bubbles}
\label{sec:stromgren}

We first consider the expansion of an \HII\ region in a uniform medium with a single point source at the origin. This test is most relevant for the growth of isolated ionized bubbles early on in reionization. %Albeit not directly related to the post-reionization regime or the overlap phase that is our primary focus, this problem hints upon simulating the post-reionization IGM temperature fluctuations, which are imprinted by the growth of the ionized bubbles and can likely be detected from observations of the Lyman-$\alpha$ forest \cite{keating18, wu19b}. Moreover, whether the photoionization rate inside ionized bubbles can be accurately captured when assuming an approximate Eddington tensor might affect the simulated Lyman-$\alpha$ transmission in the bubbles. This impacts constraining the IGM neutral fraction using the proximity zones of quasars or Lyman-$\alpha$ emitters and the Lyman-$\alpha$ damping wing \cite{bolton11, keating15, davies18dampingwing, wang20, mason20}, although no moment-based code has been used for such studies.
\tmp{Albeit not directly related to the post-reionization regime that we are most interested in, we include this test problem for the completeness of examining the accuracy of moment-based RT methods.  We will also briefly comment on the performance of moment-based RT methods when simulating the overlap of ionized bubbles and when more complex small-scale physics is involved in simulating the bulk of reionization, but a detailed examination is beyond the scope of this paper.}

How the growth of the ionized bubbles might be affected by an approximate Eddington tensor can be understood by considering photon conservation. In the absence of recombinations, which is a good approximation during the bulk of reionization since the ionized bubbles keep on growing until recombinations balance ionizations near the end of reionization \cite{furlanetto05}, photon conservation implies that every ionizing photon should result in one ionization of a neutral atom. Since the moment-based RT algorithms conserve photons, at each snapshot the size of the ionized bubble can simply be calculated by equating the number of photons to the number of hydrogen atoms inside the bubble. This implies that the propagation of I-front is not impacted by the Eddington tensor approximation. Since the volume-filling fraction of ionized bubbles is captured by all Eddington tensor approximations, this conservative property may further suggest that moment-based RT methods capture many of the \tmp{\it gross} properties about the bulk of reionization as studies with excursion set models (which have no radiative transfer) suggest that many of these properties are driven by the clustering of sources and the volumetric ionization \cite{zahn11}. %thus suitable for studying the post-reionization temperature fluctuations \cite{keating18, wu19b}, imprints of reionization on the CMB \cite{roy20}, and 21-cm signal during reionization \cite{kaurov16}.

Another property of interest is the photoionization rate profile inside an ionized bubble.  We will consider two limiting cases, one where the Eddington tensor is purely radial and the other where the Eddington tensor is isotropic. The purely radial Eddington tensor produces the exact solution to the growth of an ionized bubble around a single point source whereas the isotropic one would not.  At outlined in Section~\ref{sec:otvet_review}, the OTVET Eddington tensor will transition from radial to isotropic at some radius in the \HII\ region, with the radius depending on the size of the simulation box.   % The isotropic Eddington tensor is the limit of OTVET when there are abundant sources in a simulation volume. A more realistic approximation of the OTVET Eddington tensor during reionization can be obtained by assuming a point source in a uniform ionizing background, where the Eddington tensor transitions from being radial near the source to being isotropic outside the ``proximity region'' (see Section~\ref{sec:otvet_review} for details). However, the isotropic Eddington tensor provides an estimate of the maximum error that moment-based RT methods can likely make.

Let us consider the maximum error that approximate Eddington tensors can make on the photoionization rate profile inside an ionized bubble by assuming an isotropic Eddington tensor.  Inside the ionized bubble where the opacity is close to 0, equation~\ref{eq:rt_monopole} implies that $\vec{F} = \hat{\vec{r}}/r^2$, where $\hat{\vec{r}}$ is the unit vector along the radial direction.  Meanwhile, the relation $\vec{F} = -(\nabla E) / [3\kappa]$ as indicated by equation~\ref{eq:rt_dipole} suggests that $E$ has to be constant so that $\vec{F}$ does not diverge, since $\kappa \to 0$.  This violates the condition $|\vec{F}| \le E$ at small radii. Therefore the flux-limiters that the moment-based RT codes use must ensure that $|\vec{F}|=E=1/r^2$ inside ionized bubbles where opacity is negligible.  This discussion is specific to OTVET and flux-limited diffusion methods using the isotropic Eddington tensor, since M1 is naturally flux-limited and one can verify that $|\vec{F}|=E$ is an allowed solution by M1.
Since the gas experience a sharp transition from highly ionized to neutral at the I-front, the differences at the ionized bubble's edge from different approximations are not observationally relevant. These conclusions are consistent with \cite{norman15}, who tested the flux-limited diffusion method using the \HII\ region expansion problem. We thus expect moment-based RT methods to be accurate enough for quasar proximity zone studies.

While our analytic study is limited to a single source, \cite{gnedin01, finlator09} showed that OTVET distorts the ionized bubbles when multiple sources are present.  In the case of bubble overlap, M1 also has trouble simulating two colliding beams traveling in opposite directions \cite{rosdahl13}.  Thus, in more complex geometries the moment algorithms' solutions will differ more from the exact compared to the uniform \HII\ region problem.
\tmp{In the photon-counting limit, \cite{zahn07, zahn11} showed that how an algorithm advects radiation in \HII\ regions during reionization has little effect on the large-scale observables and thus the algorithmic choice may not matter to sufficiently capture the bulk of reionization \cite[but see][]{molaro19}.  Our investigations in this section therefore only have bearings on the gross properties of the radiation field during reionization.  In addition to potential problems in simulating the overlap of ionized bubbles, when more complex small-scale physics is involved, e.g. gas clumping and I-front trapping, differences in the algorithms will likely also enlarge the discrepancies in the predicted observables.
For instance, the lagging behind of I-fronts in the presence of self-shielding dense absorbers may introduce more jaggedness into the shape of the bubble edges \cite{daloisio20}, which could potentially alter the large-scale 21-cm power spectrum.  However, the inability of OTVET and flux limited diffusion methods to cast shadows behind dense absorbers indicates that these algorithms produce much smoother reionization morphologies.
The increased recombination rate owing to gas clumping also alters the size distribution of the ionized bubbles at factors of $\sim2-3$ level, leading to factors of $\sim2$ drop in the large-scale power of the 21-cm power spectrum \cite[e.g.][]{sobacchi14, shukla16, mao20}.  This effect is especially important during the second half of reionization, and the spatial structure of the overlap phase could be considerably more complex if the absorption systems are abundant \cite{alvarez12}.
A thorough investigation into how well different RT algorithms, especially the moment-based ones, capture these effects requires numerical simulations.  We therefore leave the examination of the accuracy of moment-based RT methods on simulating the bulk of reionization to future work.}

Despite the above uncertainties in simulating the bulk of reionization, in the post-reionization regime where a self-shielding absorber faces radiation from all directions, a simplified test problem may assist in understanding the performance of moment-based algorithms on capturing the physics of the self-shielding regions.
We thus aim to understand the ionization of dense absorbers in a uniform UV background with moment-based RT methods in the next section.

%\matt{I commented out summary because section was so short that felt redundant, but it's up to you.  I moved multiple sources into above paragraph}
%In summary, we find that M1 and OTVET recover the exact solution to the single source HII region expansion problem. We expect the propagation of the I-front and the radial profile of the photoionization rate inside the ionized bubble of a single source to be independent of the Eddington tensor assumed, and that an incorrect Eddington tensor likely only slightly affects the ionization at the edge of the bubble. Moment-based RT methods are thus accurate enough for studying the Lyman-$\alpha$ transmission in the proximity zones of quasars and Lyman-$\alpha$ emitters during the bulk of reionization. 
%These effects likely only make modest errors when simulating the bulk of reionization, given that these algorithms conserve photons \cite{norman15}. We therefore expect the global ionized fraction and the morphology of reionization to be only moderately affected by these Eddington tensor approximations. This motivates that the IGM temperature and the relic temperature fluctuations from inhomogeneous reionization are simulated relatively correctly by the moment-based RT methods. Comparisons between simulations and observations of the Lyman-$\alpha$ forest regarding the IGM temperature is thus relatively robust.

\section{Moment-based methods versus exact solution: a spherical absorber with uniform radiation from infinity}
\label{sec:spherical_absorber}

In this section, we study the predictions of different Eddington tensors on the ionization structure of absorbers of ionizing photons.  \tmp{These absorbers are systems that have substantial Lyman-continuum optical depths and so self shield, often termed Lyman-limit systems.}  In ionized regions, dense absorbers set the photon mean free path and total number of recombinations, playing an important role in regulating the amplitude of the post-reionization ionizing background \citep[e.g.][]{mcquinn16} and in the growth of ionized bubbles near the end of reionization \cite{mhr00, miraldaescude03, furlanetto05}.
%Of particular importance are the Lyman-limit systems where self-shielding of ionizing radiation starts to take place, which are absorbers with \HI\ column densities of $10^{17.2}<N_{\rm HI}<10^{19}$~cm$^{-2}$.  A significant fraction of the total recombination in the IGM happens in the Lyman-limit systems, although lower column density absorbers could contribute to half or even more of the total recombination, depending on the slope of the \HI\ column density distribution \cite{miraldaescude03}. We study how an absorber with a certain density profile is ionized by an isotropic ionizing background, assuming that the radiation is governed by the moment RT equations with Eddington tensor approximations.
Since reionization simulations often adjust their source emissivity to match the Lyman-$\alpha$ forest transmission (which is shaped by the emissivity times the mean free path), whether different Eddington tensor approximations correctly captures the ionization of absorbers also affects whether simulations calibrate to the correct source emissivity. %All these quantities can be constrained by observations of the Lyman-$\alpha$ forest, and the calibration of the simulation parameters would in turn affect the interpretation of the resulting reionization models.

To understand the ionization of absorbers with these radiative transfer algorithms, we study a toy problem where a spherical absorber with monomial density profile is exposed to an otherwise uniform ionizing background.  While simple, we think this toy problem captures the essential features of isolated absorbers in ionized regions.  Radiation is roughly uniform owing to the large photon mean free path of tens of comoving megaparsec, since there are numerous galaxies within a mean free path \cite{furlanetto05, sobacchi14, worseck14}; the mean free path at the late stages of reionization (when bubble sizes are large) and after reionization is roughly set by the abundance of Lyman-limit systems that self-shield themselves from the radiation background.  Moreover, dense absorbers are mostly associated with low-mass galaxies with negligible star formation rate \cite{rahmati14}, especially at higher redshifts when the mean density is higher.  Therefore, it is likely the case that no local source substantially alters the radiation field around absorbers.  Approximating the absorbers as spherical is motivated by Lyman-limit systems being associated with halo-like overdensities  \cite{mcquinn11, munoz16}. %Moreover, although absorbers can have extended structures, after reionization the overdense absorbers are approximately in hydrostatic equilibrium, and thus have a well-defined characteristic size \cite{schaye01, rahmati18}. %Since Lyman-limit systems are associated with collapsed halos, their density profile can be taken to be monomial.
Additionally, \cite{zheng02} showed that a singular isothermal sphere density profile ($n_{\rm H} \propto r^{-2}$) can reproduce the rough properties of observed column density distribution after reionization, with lower column density absorbers corresponding to larger impact parameters. A final simplification to our test problem is dropping the time dependence, which is likely an excellent approximation owing to the short timescale to reach photoionization equilibrium  ($\sim10^5$~yrs at $z\sim5$). %Our toy problem provides some assessment of the errors that the Eddington tensor approximations make about properties of the IGM at the end of reionization.
%\tmp{We note that although a large fraction of the above mentioned motivations is based on the current consensus of the still unclear physical nature of Lyman-limit systems, our test problem is not specifically designed for studying the ionization of a Lyman-limit system.  Our focus is simply on the dense absorbers that self-shield from the radiation field and thus shape the ionizing background after reionization, for which reason we avoid calling these absorbers as Lyman-limit systems.  Our results based on this test problem therefore also has bearings beyond the scope of Lyman-limit systems.}

Assuming a spherical absorber with monomial density profile in photoionization equilibrium with radiation coming uniformly from infinity, we calculate the radial profiles of $E$ (proportional to the photoionization rate), $F$, and neutral fraction ($x_{\rm HI}$) given by the exact solution, OTVET, and M1. Here we take $F = -|\vec{F}|$ since $\vec{F}$ points radially inward. Following \cite{zheng02}, we initialize the $x_{\rm HI}$ profile by assuming that the absorber is optically thin and in photoionization equilibrium with the ionizing background. We next update the opacity ($\kappa$) profile and calculate a new $E$ profile by solving the time-independent RT equation with the updated $\kappa$ profile, where we use $\kappa = \sigma n_{\rm H} x_{\rm HI}$ and $\sigma=6.3\times10^{-18}$~cm$^{-2}$ is the photoionization cross section of our monochromatic $13.6$~eV radiation. The $x_{\rm HI}$ profile is then updated again assuming the absorber in photoionization equilibrium with the new $E$ profile, and used to update the $\kappa$ profile. These steps are iterated until the fractional change in the $x_{\rm HI}$ profile is less than $10^{-3}$ at every grid point.

In each iteration, solving for $E$ and $F$ for the different radiative transfer methods reduces to solving ordinary differential equations. For the exact solution, we integrate along each direction to get the optical depth $\tau$ at each radius, and then integrate $e^{-\tau}$ over all solid angles. For the solutions using OTVET and M1, we have derived a set of differential equations for $\log E$ and $g = -F/E$ in Appendix~\ref{sec:absorber_eqns}, using a change of variable $\ds/\dr = -\kappa$. We therefore utilize equations~\ref{eq:spherical_dlogEds} and \ref{eq:spherical_dgds} to obtain the solution to $E$ and $F$. For OTVET, the Eddington tensor is isotropic everywhere, since the absorber is illuminated from all directions. To solve these equations, we use a root finding method for OTVET and an explicit integration for M1. The structure of the differential equations in M1 requires a certain boundary condition at finite $s$ be fulfilled, which determines the point where we start integrating. While the equations we solve (\ref{eq:spherical_dlogEds} and \ref{eq:spherical_dgds}) do not involve any flux limiter, they naturally give $g\le1$ at all radii.

Owing to self-shielding, the IGM experiences a sharp transition between being highly ionized in the diffuse gas and becoming neutral in dense absorbers of ionizing photons \cite{zheng02, mcquinn11, rahmati13}. Since this transition occurs inside the radius where the optical depth is of the order of unity, the total recombination is dominated by the gas at outer radii. Therefore the $E$, $F$, and $x_{\rm HI}$ profiles are well characterized by a single parameter, the self-shielding radius, where the optical depth is of order 1. In this case the solutions are expected to be self-similar with respect to reasonable changes in the amplitude of the density profile or the photoionization rate, which we will demonstrate below. We calculate solutions with hydrogen number density profile $n_{\rm H}=0.01\left({r}/{R}\right)^m$ and $\Gamma(r=\infty)=5\times10^{-13}\ {\rm s}^{-1}$ \cite{becker13, wyithe11, calverley11, daloisio18, wu19b}, where $r$ is the distance from the center of the absorber, $R$ defines the ``size'' of the absorber, and $\Gamma$ is the photoionization rate. Here $n_{\rm H} = 0.01$~cm$^{-3}$ is roughly 200 times the mean density of the universe at $z=5-6$ and is the self-shielding density found in \cite{rahmati13}. We adopt $R = 2-20$~kpc, which corresponds to the virial radii of $10^8 - 10^{11}\ M_\odot$ halos at $z\sim5$. We use the case-B recombination rate $\alpha_B = 2.6\times10^{-13}\ {\rm s^{-1}}$ at $10^4$~K, and for simplicity ignore temperature variations within the absorber. In reality the denser interior of the absorber is expected to be just somewhat colder, and the case-A recombination rate is more appropriate for describing the gas outside the surface where $\tau=1$. However, we expect our major conclusions will not change if more realistic parameters are adopted. Finally, we include singly ionized helium, so the electron number density is a factor of $1.08$ higher than the \HII\ number density.

\begin{figure}[tbp]
\centering 
\includegraphics[width=\textwidth]{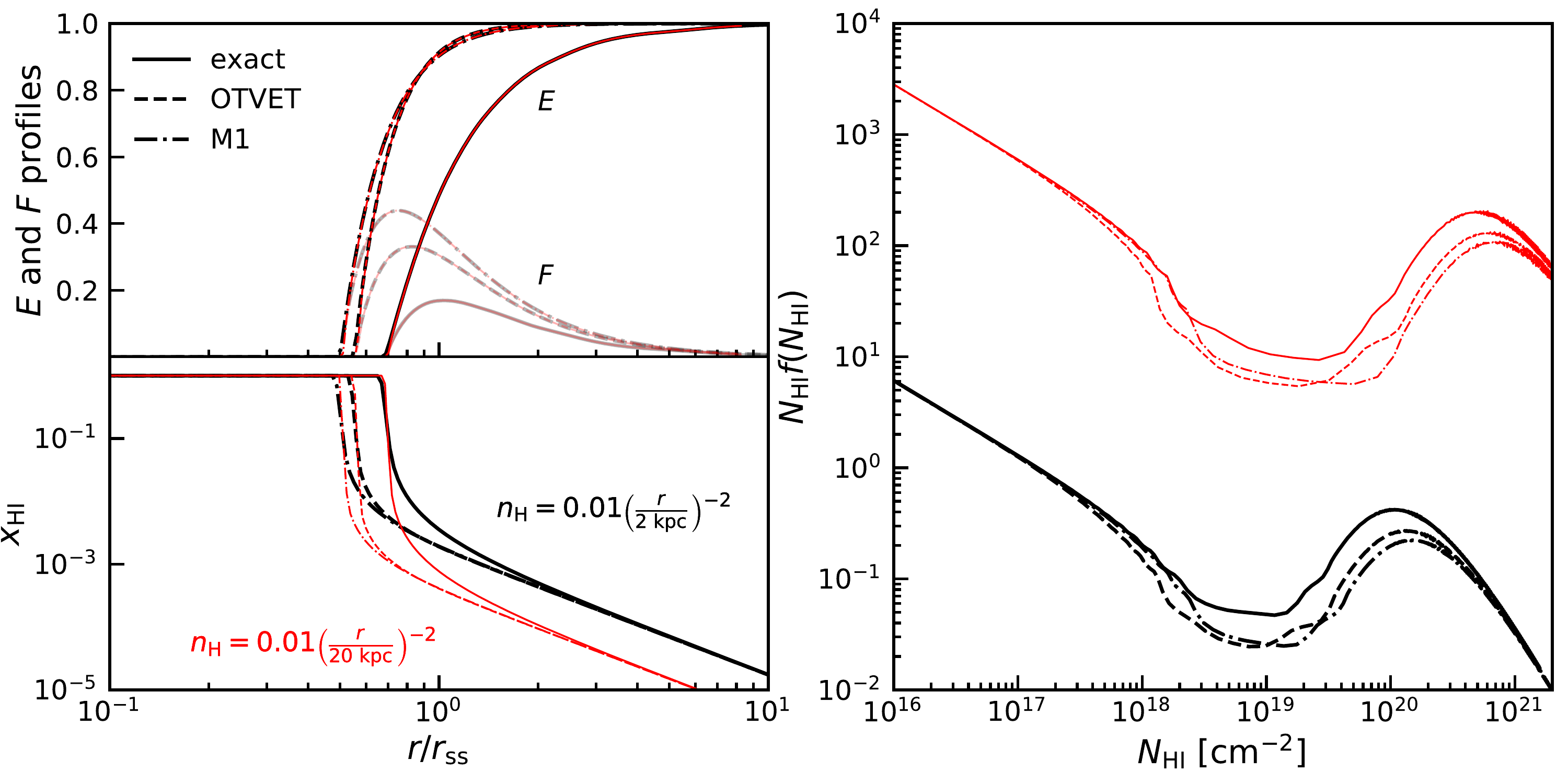}
\hfill
\caption{\label{fig:absorber_profiles} Left: radial profiles of $E$, $|F|$ (top panel), and the \HI\ fraction ($x_{\rm HI}$; bottom panel) of two spherical absorbers with isothermal density profiles, assuming absorbers in photoionization equilibrium with a radiation field coming uniformly from infinity. The two absorbers have density profiles $n_{\rm H} = 0.01 (2\ {\rm kpc}/r)^{-2}$~cm$^{-3}$ and $n_{\rm H} = 0.01 (20\ {\rm kpc}/r)^{-2}$~cm$^{-3}$, and are differentiated by black and red lines respectively. The profiles are given in units of the self-shielding radius of the exact solution ($r_{\rm ss}$), defined as the radius where the optical depth is 1. The $E$ and $|F|$ profiles almost overlap completely for the two absorbers. Solid, dashed, and dot-dashed lines represent the exact solution, the OTVET solution, and the M1 solution, respectively. Darker and lighter colors in the top panel show the $E$ and $|F|$ profiles, respectively.
Right: the \HI\ column density distributions (in arbitrary units) calculated from the $x_{\rm HI}$ profiles.
For a fixed incident radiation field, OTVET and M1 over-ionize absorbers, which we show impacts the relation between the ionizing background, mean free path, and emissivity of the sources in simulations using these algorithms. % which requires $30-40\%$ higher emissivity to balance the higher total recombination than the exact solution.
}
\end{figure}

We first consider the isothermal density profile with $n_{\rm H} \propto r^{-2}$, which roughly reproduces the observed column density distribution after reionization \cite{zheng02} and matches the slope of the probability distribution function of high density gas in simulations \cite{mhr00, mcquinn11}. The left panel of Figure~\ref{fig:absorber_profiles} shows the radial profiles of the radiation (top panel) and $x_{\rm HI}$ (bottom panel) given by the exact solution (solid lines), OTVET (dashed lines), and M1 (dot-dashed lines). The profiles are normalized to the self-shielding radius ($r_{\rm ss}$) of the exact solution, defined as the radius where the optical depth is 1. The black and red lines represent solutions with $R=2$~kpc and 20~kpc respectively. Darker and lighter colors in the top left panel illustrate the monopole moment of the radiation ($E$) and the dipole moment ($|F|$), respectively. The $E$ and $|F|$ profiles of the two absorbers almost overlap despite the 3 orders of magnitude difference in the mass inside $R$, demonstrating the self-similarity of the solutions. The peaks of the $|F|$ profiles predicted by OTVET and M1 are a factor of $>2$ higher than that given by the exact solution, indicating more photon flux penetrating into the absorber. Therefore, OTVET and M1 over-ionize self-shielding absorbers for a fixed incident radiation field.  %, when a simulation is calibrated to match the background photoionization rate or the Lyman-$\alpha$ forest transmission.

Over-ionization of absorbers leads to a higher total recombination rate and, therefore, to a higher post-reionization emissivity, as the emissivity should be in balance with recombinations.  This balance holds because the emissivity evolves on timescales much longer than the time for photons to travel one mean free path, and the photon mean free path at the redshifts of interest in this work ($z\sim5$) is much smaller than the horizon, making the terms on the right-hand side of equation~\ref{eq:rt} dominate over those on the left-hand side \cite{meiksin18}.  The right panel of Figure~\ref{fig:absorber_profiles} demonstrates the increased total recombination rate in the OTVET and M1 solutions further, which shows the \HI\ column density distributions in arbitrary units computed from the $x_{\rm HI}$ profiles of the two absorbers. OTVET and M1 predict lower abundance of high column density absorbers, thus raising the corresponding emissivity. We calculate the differences in the total recombination rate by integrating $\int (1-x_{\rm HI})^2 n^2_{\rm H} r^2\; \dr$, and find that OTVET and M1 lead to a factor of $\sim1.3$ and $\sim1.4$ more recombination than the exact solution respectively. These numbers only differ by $2\%$ for the two absorbers shown in Figure~\ref{fig:absorber_profiles} owing to self-similarity of the solutions. In other words, in order for a simulation using OTVET or M1 to be able to reproduce the observed photoionization rate or Lyman-$\alpha$ forest transmission, the emissivity needs to be a factor of $1.3-1.4$ higher than the true value\footnote{\tmp{Note that the gas is still highly ionized at the self-shielding radius where the optical depth is $\sim1$, which corresponds to the formal definition of Lyman-limit systems.  The gas at around this radius is also where most of the recombinations take place.  Thus our test problem does not suffer from the conceptual problem of mistaking damped Lyman-$\alpha$ systems (where the gas turns fully neural) with Lyman-limits.}}. Conversely, if the emissivity of the simulation is set to match the observations \cite[e.g.][]{becker13}, we find that the predicted $\Gamma(r\to\infty)$ by OTVET and M1 is a factor of $\sim0.4$ lower than the exact solution. Since the Lyman-$\alpha$ forest transmission traces fluctuations in the ionizing background, a $\sim60\%$ lower photoionization rate should result in a factor of $\sim2.5$ increase in the optical depth, thus reducing the forest transmission. Although currently there still lack observational constraints on the emissivity, it can be constrained with future Lyman-$\alpha$ forest observations or with star formation density observations and estimates for the escape fraction of ionizing photons. Simulations using OTVET and M1 thus are unlikely to reproduce the true relation between their sources' emissivity and the average photoionization rate (which largely sets the Lyman-$\alpha$ forest transmission).

Because of this inconsistency when using M1 and OTVET, the effective photon mean free path that would be inferred by taking the ratio of the photoionization rate to the emissivity in simulations with these algorithms should be biased low. When a simulation with OTVET or M1 is calibrated to match a fixed photoionization rate, the effective mean free path of the simulation is lower by $\sim30\%$ relative to the true value. When the simulation is calibrated with a fixed emissivity, the effective mean free path is smaller by a factor of $\sim2.5$.

The above finding may seem inconsistent with the right panel of Figure~\ref{fig:absorber_profiles}, where the lower abundance of high column density systems indicated by OTVET and M1 implies longer photon mean free paths in simulations using these algorithms.  However, calculating the mean free path by integrating over the \HI\ column density distribution assumes that radiation still behaves as rays with the approximate Eddington tensors, which is likely violated when using M1 and OTVET.  Shooting rays across the simulation box and calculating the optical depth along the rays is also the method used in some previous studies using the M1 algorithm \citep[e.g.][]{chardin15, keating19, keating20} to measure the mean free path.  We thus compute a mean free path by calculating $(2 \pi) \int r (1 - e^{-\tau(r)})\; \dr$, which is the cross section of the absorber assuming radiation behaves like rays.  We find that with this approach of calculating the mean free path, the OTVET and M1 methods overestimates the mean free path by $10\%$ compared to the exact solution, since the cross section of the absorber is reduced with these Eddington tensors.  This seemingly controversial finding with that of the effective mean free path defined before is likely caused by the radiation being more diffusive when using OTVET and M1, so that despite the lower \HI\ column density distribution function indicating longer mean free paths, the higher recombination rate implies shorter ones.

Over-ionizing absorbers at a fixed photoionization rate also leads to underpredicting the \HI\ content after reionization, traced by the high column density gas that self-shields itself from the ionizing background. The high-redshift \HI\ mass density ($\Omega_{\rm HI}$) has been constrained by observations of damped Lyman-$\alpha$ systems (defined as systems with \HI\ column densities $>2\times10^{20}$~cm$^{-2}$) \cite[e.g.][]{prochaska09, crighton15, bird17}. Uncertainties in $\Omega_{\rm HI}$ can propagate into uncertainties in the \HI\ 21-cm intensity fluctuations, affecting predictions about future \HI\ intensity mapping observations \cite{padmanabhan16}. We find that when fixing the background photoionization rate, simulations using M1 and OTVET results in lower $\Omega_{\rm HI}$ by $\sim20-30\%$. Although no reionization simulation has been compared against the observed $\Omega_{\rm HI}$ or used to predict the post-reionization 21-cm signal, we point out that there is potential bias in the simulated $\Omega_{\rm HI}$ introduced by M1 and OTVET.

Since a $30-40\%$ higher emissivity is required to balance the total recombination rate when fixing the photoionization rate, simulations using M1 and OTVET likely underpredict the duration of reionization by a similar amount as they need to spuriously increase the emissivity to latch on to the forest transmission. Often simulations tune the emissivity by adjusting one parameter, such as the escape fraction of ionizing photons from galaxies or from the birth clouds of star particles \citep[e.g.][]{gnedin14, wu19a, wu19b}, so that the entire history is affected. The effect may be more complex in simulations that adjust multiple parameters to tune the emissivity  \citep[e.g.][]{keating18, keating19, keating20, kulkarni19}.

Finally, to bracket the range of potential biases, let us consider the absorber to have a shallower density profile with $n_{\rm H} \propto r^{-1.7}$ rather than $n_{\rm H} \propto r^{-2}$ profile considered so far. The shallower profile gives a steeper power-law slope of $-1.83$ for the \HI\ column density distribution, roughly consistent with the findings of simulations examining optically thin columns at $z\sim 4-6$, while an isothermal density profiles give a column density distribution of slope $-1.67$ \citep{mcquinn11, rahmati13}. %Moreover, at $z\sim5-6$, the total recombination in the IGM is dominated by gas with lower column densities \citep{munoz16, rahmati18}, making a shallower density profile a better approximation.
For a $r^{-1.7}$ density profile, we find that when fixing the background photoionization rate, OTVET and M1 yield a factor of $\sim1.2$ higher total recombination than the exact solution. When fixing the total emissivity, OTVET and M1 produce a ionizing background that is a factor of $\sim0.5$ lower in amplitude than the exact solution. These differences are smaller than those found for the $r^{-2}$ density profile because lower column density regions (for which radiative transfer is less important) are weighted more heavily in the total recombination rate. %Our estimates using the isothermal density profile thus provides a rough upper limit for the effect of M1 and OTVET on the ionization of absorbers.

To summarize, we find that the moment-based RT methods with M1 and OTVET will err at the tens of percent level in reproducing the relation between the photoionization rate, emissivity, and photon mean free path in the post-reionization IGM. When M1 and OTVET simulations are calibrated to match the Lyman-$\alpha$ forest transmission or the ionizing background, the emissivity is overestimated by $20-40\%$ to balance the total recombination rate because absorbers are over-ionized, where the range owes to making different motivated assumptions about the density profile of ionizing absorbers. This over-ionization also results in the effective photon mean free path (defined as the ratio of the photoionization rate to the emissivity) of the simulations being lower by a similar amount than the true value. If simulations adopt a fixed emissivity, the resulting photoionization rate and effective mean free path are underpredicted by $50-60\%$ when using M1 and OTVET.
Among other effects, these biases result in an undeprediction of the simulated duration of reionization when calibrating to the photoionization rate inferred from the Lyman-$\alpha$ forest, or they result in a more opaque Lyman-$\alpha$ forest when calibrating to observations of the sources' emissivities.

\section{Moment-based methods versus exact solution: fluctuations in the post-reionization ionizing background in a static IGM}
\label{sec:Gamma_fluc}

Finally, we turn our attention to how well fluctuations in the ionizing background after reionization are captured in moment-based RT methods.  This exploration also has bearing upon the performance of ray-tracing methods (especially those using short characteristics). Ionizing background fluctuations have been compelling at explaining the excess scatter in the $z\sim5-6$ Lyman-$\alpha$ forest opacity \cite{davies16, davies18, becker18, daloisio18, nasir20}. Simulations of reionization have thus attempted to capture ionizing background fluctuations, in addition to the relic temperature fluctuations owing to patchy reionization \cite{chardin15, chardin17, kulkarni19}. Both of these effects may be testable with future Lyman-$\alpha$ forest observations and can thus put constraints on reionization models \cite{onorbe19}.

The post-reionization ionizing background fluctuations are thought to quickly become small after reionization \citep{croft99, croft04, meiksin04, mcdonald05, slosar09}\citep[but see][]{maselli05, mesinger09}, allowing them to be calculated by solving the RT equation with linear perturbation theory. Moreover, at the post-reionization redshifts of interest in this work ($z\sim5$), the photon mean free path is much smaller than the Hubble radius, making the time-independent solution to the RT equation a good approximation to the full time-dependent solution \citep{meiksin18}. 
\tmp{We thus focus on a problem setup where small ionizing background fluctuations on large-scales are characterized by the time-independent solution to the RT equation\footnote{\tmp{A comparison between the exact solution to the ionizing background fluctuations using linear perturbation theory (equation~\ref{eq:deltaE_exact}) and the long characteristics ray-tracing method of \cite{davies16} showed that the two methods agree well on the 3D power spectrum of the photoionization rate fluctuations on scales well below the photon mean free path (private communication with Fred Davies).  This agreement holds even in the relatively strongly fluctuating (order unity) regime at $z > 5.5$, further justifying our approach of solving the time-independent RT equation with linear perturbation theory.}}.}
In this section we show that common Eddington tensor approximations lead to significant errors on estimating the ionizing background fluctuations at scales smaller than the photon mean free path, and that recovering the true ionizing background fluctuations require resolving the radiation field with far more than the four angular directions followed by Eddington tensor closure algorithms.

\subsection{Exact solution}

We first derive the full solution to the time-independent RT equation using linear perturbation theory, which we term the exact solution. The approach in this section follows \cite{pontzen14, gontcho14, meiksin18}. We will write the overdensity in the quantity $X$ as $\delta_X \equiv X/\bar{X}-1$, writing the volume average as $\bar{X}$, which is valid in the limit where the fluctuations about the mean are small. 

In the time-independent case, equation~\ref{eq:rt_f} becomes 
\begin{equation}
\frac{1}{a} \vnn \cdot \nabla f = j - \kappa f.
\label{eq:rt_f_tindependent}
\end{equation}
Ignoring all terms of order $\delta$ yields the zeroth-order solution
\begin{equation}
\meankappa\bar{f} = \bar{j},
\end{equation}
where $\meankappa, ~\bar{f}, ~\bar{j}$ are the spatially averaged quantities. This equation reflects a balance between emissivity and recombination.

Expanding equation~\ref{eq:rt_f_tindependent} to first order and simplifying using the zeroth order solution, we get
\begin{gather*}
%\frac{1}{a} \bar{f} \vnn \cdot \nabla \delta_f = \bar{j}(1+\delta_j) - \meankappa\bar{f}(1+\delta_\kappa)(1+\delta_f) \\
\frac{1}{a\meankappa} \vnn \cdot \nabla \delta_f = \delta_j - \delta_\kappa - \delta_f. \numberthis \label{eq:rt_delta}
\end{gather*}
Since the opacity fluctuations must either trace the density fluctuations ($\delta$) or the photoionization rate fluctuations ($\delta_\Gamma$; as we assume there are no other long-range fields of relevance), we can further expand $\delta_\kappa = \bkd\delta + \bkg\delta_\Gamma$, ignoring any shot noise term that would be small to the extent the sources of opacity are abundant.  These bias coefficients encapsulate how the non-perturbative small-scale fields trace the large-scale overdensities (and so our approach can be thought of as an effective perturbation theory to linear order in overdensities and lowest order in derivatives).  We could similarly expand $\delta_j$ in terms of these quantities, but we choose to keep our equation in terms of $\delta_j$.  For the calculations here, we will assume the sources are not modulated by the photoioinization rate -- which should be a good approximation ignoring recombination radiation --, and model $\delta_j$ as a linear in $\fdd$ plus a shot noise term, as in the halo model \cite{cooray02}.  

Therefore,
\begin{equation}
\frac{1}{a\meankappa} \vnn \cdot \nabla \delta_f = \delta_j - \bkd\delta - \bkg\delta_\Gamma - \delta_f.
\label{eq:transformthis_f}
\end{equation}
The Fourier transform of the above equation gives, after some rearranging,
\begin{gather*}
%-\frac{i \vnn \cdot \vk}{a\meankappa} \fdf = \fdj - \bkd\fdd - \bkg\fdg - \fdf \\
\fdf = \frac{\fdj - \bkd\fdd - \bkg\fdg}{1 - i \vnn \cdot \vk/a\meankappa}, \numberthis
\end{gather*}
where tilde's denote the Fourier transform. Integrating both sides over all solid angles, we get
\begin{align*}
\fdE = \frac{\bar{f} \int \mathrm{d}^2\vnn\; \fdf}{\int \mathrm{d}^2\vnn\; \bar{f}} = \frac{1}{4\pi} \int \mathrm{d}^2\vnn\; \fdf %= \frac{1}{4\pi}\int\frac{\mathrm{d}\cos\theta \mathrm{d}\phi}{1 - i k\cos\theta/a\meankappa} (\fdj - \bkd\fdd - \bkg\fdg) \\
= S(k) (\fdj - \bkd\fdd - \bkg\fdg), \numberthis
\end{align*}
where $S(k) = (a\meankappa/k)\ \arctan(k/\left[a\meankappa\right])$ and we assumed that $\delta_j$ is isotropic. Since the photoionization rate in the monochromatic limit is the angle-averaged intensity, i.e. $\Gamma \propto E$, we find
\begin{equation}
\fdE = \frac{\fdj - \bkd\fdd}{\bkg + S^{-1}(k)}.
\label{eq:deltaE_exact}
\end{equation}

In the uniform mean free path case, the opacity bias factors are zero, resulting in $\fdE = \fdj S(k)$. Our solution in this case is the full solution to time-independent radiative transfer equation (equation~\ref{eq:rt_f_tindependent}), as in this case this equation is linear in $f$.  We note that much treatment in the literature of the post-reionization ionizing background is in this uniform mean free path limit \citep{meiksin04, mcquinn14}. \tmp{This solution also captures the proximity zone, where the source term $j$ dominates over the absorption term $\kappa f$.} This suggests that this section's results have scope beyond the perturbative limit.\footnote{Going to one higher order in derivatives in the effective linear-in-$\delta$ theory would result in additional terms with $\bkd\fdd \rightarrow \bkd\fdd + b_{\kappa, k^2\delta} k^2 \fdd +b_{\kappa, k^4\delta} k^4 \fdd + ...$ in eqn.~\ref{eq:deltaE_exact} (and a similar term if we expanded $\delta_j$). Such terms could be important at higher wavenumbers, and we choose not to follow them here.  Numerical calculations suggest the bias coefficients associated with these higher derivatives in the opacity are small \citep{mcquinn14}.}

\subsection{Eddington tensor approximations}

We now calculate the spectrum of ionizing background fluctuations for moment-based RT methods with the Eddington tensor approximations. We work with the moment equations~\ref{eq:rt_monopole} and \ref{eq:rt_dipole} and again drop the time-dependent derivatives. Note that $\Gamma \propto E$.  The zeroth order solution for for the moment equations are
\begin{gather}
\meankappa \bar{E} = \bar{j};\quad \bar{\vec{F}} = 0.
\label{eq:soln_0th_E_F}
\end{gather}
which are just the zeroth and first moments of equation~\ref{eq:rt_f_tindependent} for isotropically emitting sources.
%Thus $\vec{F}$ is a first order quantity.

We define $\vec{\delta}_\vec{F} \equiv \vec{F}/\bar{E}$, and again $\delta_E = E / \bar{E}-1$. Expanding equations~\ref{eq:rt_monopole} and \ref{eq:rt_dipole} to first order gives
\begin{gather*}
% \frac{1}{a} \bar{E} \;\nabla \cdot \delta_\vec{F} = -\meankappa \bar{E} (1+\delta_\kappa)(1+\delta_E) + \bar{j}(1+\delta_j); \\
\frac{1}{a\meankappa} \frac{\partial}{\partial x_m} \delta_{F_m} = -\delta_\kappa - \delta_E + \delta_j \numberthis \\
% \frac{1}{a} \bar{E} \frac{\partial}{\partial x_n} (1+\delta_E) (\bar{h}_{mn} + \delta h_{mn}) = -\meankappa\bar{E}(1+\delta_\kappa)\delta_{F_m} \\
\frac{1}{a\meankappa} \frac{\partial}{\partial x_n} \delta_E \bar{h}_{mn} +  \frac{1}{a\meankappa} \frac{\partial}{\partial x_n} \delta h_{mn} = -\delta_{F_m} \numberthis,
\end{gather*}
where repeated indices are summed. Here $\bar{h}_{mn} = \delta^K_{mn}/3$, where $\delta^K_{mn}$ is the Kronecker delta, and
\begin{equation}
\delta h_{mn} = \frac{\delta P_{mn}}{{\rm Tr}(\bar{P}_{mn})} - \frac{\bar{P}_{mn}}{{\rm Tr}(\bar{P}_{mn})^2}{\rm Tr}(\delta P_{mn}).
\end{equation}
Eliminating $\vec{\delta}_\vec{F}$ from the above equations, we get
\begin{equation}
-\frac{1}{3a^2\meankappa^2} \nabla^2 \delta_E - \frac{1}{a^2\meankappa^2} \frac{\partial^2}{\partial x_m \partial x_n} \delta h_{mn} = \delta_j - \delta_\kappa - \delta_E \\
\end{equation}
The Fourier transform of the above equation gives
\begin{equation}
\fdE = \frac{\fdj - \bkd\fdd - \frac{k_m k_n}{a^2\meankappa^2} \fhmn}{1+\bkg+\frac{k^2}{3a^2\meankappa^2}},
\label{eq:deltaE_hij}
\end{equation}
where we have used $\Gamma \propto E$ and again our bias expansion for $\delta_\kappa$.  Equation~\ref{eq:deltaE_hij} is the general expression for ionizing background fluctuations, regardless of the form of the Eddington tensor.\footnote{Our time-independent solutions are valid on scales much smaller than the horizon.  For low wavenumbers, evolutionary effects become important and solving the full time-dependent RT equation is required to avoid a formal divergence in the time-independent solution \cite{meiksin18}.} In the uniform mean free path case, we show in Appendix~\ref{sec:FT_deltahij} that the exact Eddington tensor recovers the relation $\fdE = \fdj S(k)$.

Because the moment equations are derived from angular moments of the same linear equation the exact solution applies (equation~\ref{eq:rt_delta}), one might think the linear bias coefficients should be the same as for the exact case in the limit that both are treating the same source and opacity fields.  However, formally our zeroth moment equation assumed $\overline{\kappa E} = \meankappa \bar{E}$, but this average depends on how the radiation field overlaps with the \HI\ -- which depends on algorithm and has the effect in the linear solution of rescaling bias coefficients \cite[see][]{mcquinn18}.  Indeed, when one does perturbation theory for the ionizing background using the exact equations we are putting in an effective average for $\meankappa$.\footnote{The standard expression for the effective $\meankappa$ is $\int \mathrm{d}N_{\rm HI}\; f(N_{\rm HI}) (1- \exp[-\sigma N_{\rm HI}])$ motivated by Poissonian absorbers with column density distribution $f(N_{\rm HI})$.  Our results in \S\ref{sec:spherical_absorber} suggest that different algorithms likely result in $\sim10\%$ differences in the effective $\meankappa$.}  We ignore this complication here, but note that \S\ref{sec:spherical_absorber} suggests that the $\meankappa \bar{E}$ could differ by tens of percent between different radiative transfer algorithms.  Rescaling the biases by similar amounts has a minimal effect on our results.

A caveat of the above derivation is that we do not include the possible effects of a flux limiter, which is an additional element that can be relevant in implementations of OTVET. For a spherical absorber sitting in an ionizing background, $|\vec{F}| \le E$ is naturally satisfied without the need to invoke any flux limiter (Section~\ref{sec:spherical_absorber}), but a flux limiter is likely required to ensure $|\vec{F}| \le E$ in the proximity zone of a source (Section~\ref{sec:stromgren}). We therefore expect our formalism to fail at high enough wavenumbers, especially for the isotropic Eddington tensor. However, to keep the RT equation at linear order in $\delta$'s, the flux limiter should only be expanded to zeroth order. The effects of a flux limiter on $\vec{\delta}_{\vec{F}}$ can thus only enter at quadratic and higher orders in the overdensity. Therefore, our linear order solutions are unaffected. The possible contribution of a flux limiter to the power spectrum of $E$ is suppressed relative to the linear order solution at perturbative wavenumbers, and deviations from the exact solution of $\fdE$ owing to an approximate Eddington tensor is unlikely to be fixed by the inclusion of a flux limiter.

We examine predictions by OTVET and M1 below, focusing on the 3D and 1D power spectra of $E$ (equivalent to $\Gamma$). The 3D power spectrum is defined by $\langle \fdE^*(\vec{k}') \fdE(\vec{k}) \rangle = P^{\rm 3D}_E(k) (2\pi)^3 \delta^D(\vec{k} - \vec{k}')$, where the angle brackets denote an ensemble average and $\delta^D(\vec{k} - \vec{k}')$ represents the Dirac delta function, and the 1D power spectrum is obtained by $P^{\rm 1D}_E(k) = \int_k^\infty \mathrm{d}k'/[2\pi]\; k' P^{\rm 3D}_E(k')$.  The 1D power spectrum characterizes the spectrum of fluctuations along a skewer through the Universe, being most applicable to the Ly$\alpha$ forest. We generate power spectra for the source term ($\fdj$) using the halo model at $z=5.5$ \cite{cooray02}, assuming luminosity proportional to halo mass and a minimum halo mass ($M_{\rm min}$) for producing ionizing photons. $z\gtrsim5.5$ is the redshift at which the Lyman-$\alpha$ forest transmission shows large spatial scatter on $50/h$~Mpc scale.
We adopt $\bkg=-2/3$, which is related to the slope of the \HI\ column density distribution, and $\bkd=1$, since Lyman limit systems are abundant and so likely good tracers of the matter distribution \cite{gontcho14}.\footnote{Focusing on equations~\ref{eq:deltaE_exact} and \ref{eq:deltaE_isotropic}, $\fdE$ is primarily determined by $\fdj$. This is because the amplitude of fluctuations in the sources is much larger than that in the sinks, owing to their larger bias. The $\bkg$ term in the denominator changes somewhat the amplitude of $\Gamma$ fluctuations, but has a minimal effect on its shape.}

\subsubsection{M1 and isotropic}

For M1, since $\mathrm{d}\chi/\dg=0$ when $g=0$, the Eddington tensor is thus isotropic at linear order in the density. The solution to $\fdE$ corresponds to that assuming an isotropic Eddington tensor:
\begin{equation}
\fdE = \frac{\fdj - \bkd\fdd}{1 + \bkg + k^2/[3a^2\meankappa^2]}.
\label{eq:deltaE_isotropic}
\end{equation}
This expression reproduces the exact solution at large scales $k \lesssim a\meankappa = 1/\lambda_{\rm mfp}$, since $\arctan(x) \approx x-x^3/3$ for small $x$. However, as we show below, this isotropic solution significantly underestimates the small-scale fluctuations in $\Gamma$.

\begin{figure}[tbp]
\centering 
\includegraphics[width=\textwidth]{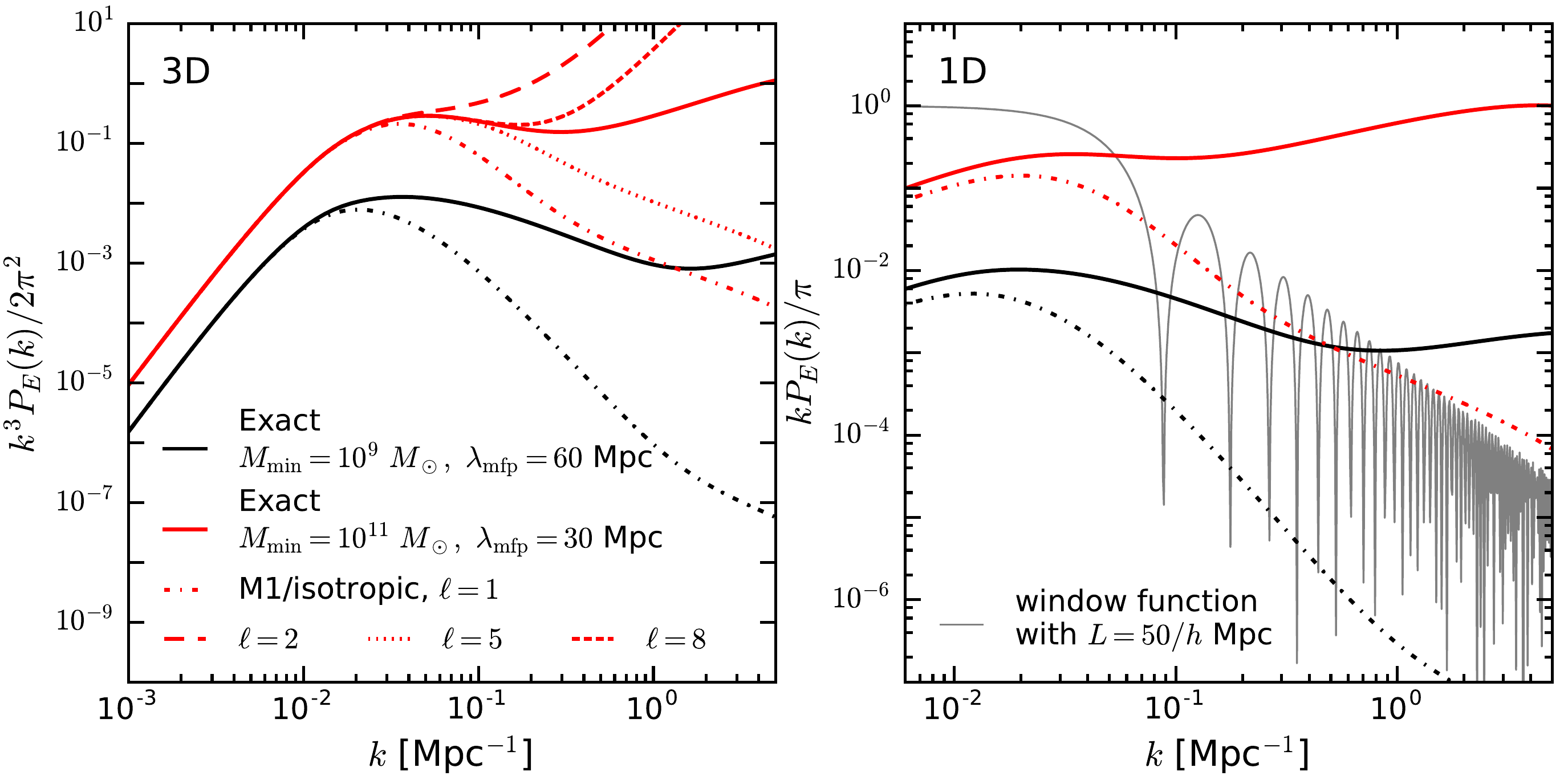}
\hfill
\caption{\label{fig:flucmfp_M1} 3D (left) and 1D (right) dimensionless power spectra of the post-reionization photoionization rate, assuming $\bkd=1$, and $\bkg=-2/3$. Solid lines show the exact solution with different source power spectra and photon mean free paths $\lambda_{\rm mfp}$. The source power spectra are calculated using the halo model at $z=5.5$, assuming luminosity proportional to halo mass and a minimum halo mass ($M_{\rm min}$) for producing ionizing photons. Black and red lines correspond to $M_{\rm min}=10^9 M_\odot, \; \lambda_{\rm mfp}=60$~comoving Mpc and $M_{\rm min}=10^{11}\ M_\odot, \lambda_{\rm mfp}=30$~comoving Mpc respectively. Dot-dashed lines represent the solutions given by M1, which reduces to an isotropic Eddington tensor at linear order in density. Long dashed, dotted, and short dashed lines illustrate the solutions when closing the moment equations at orders $\ell=2,5,8$ respectively, while the isotropic Eddington tensor correspond to closing the moment equations at $\ell=1$. Moment-based or ray-tracing methods only reproduce the exact solution at 3D wavenumbers $k\lesssim\ell/\lambda_{\rm mfp}$ and fair even worse in 1D, leading to a bias in the simulated ionizing background fluctuations which affects interpreting the Lyman-$\alpha$ forest observations. The grey line in the right panel shows the squared window function $W^2(k)=\mathrm{sinc}^2(kL/2)$ with $L=50/h$~Mpc, where $50/h$~Mpc is the typical scale that the variations in the Lyman-$\alpha$ forest transmission are measured. The 1D case represents the variance along a line segment with size $\sim 2\pi/k$ and is applicable to Lyman-$\alpha$ forest observations.}
\end{figure}

Figure~\ref{fig:flucmfp_M1} compares the 3D (left panel) and 1D (right panel) power spectra of $E$ (equivalent to the photoionization rate) as indicated by the exact solution (solid lines) and the M1/isotropic Eddington tensor solution (dot-dashed lines). Different colors represent solutions with different source power spectra and photon mean free paths. Black and red lines use $M_{\rm min}=10^9 M_\odot,\; \lambda_{\rm mfp}=60$~comoving Mpc and $M_{\rm min}=10^{11}\ M_\odot, \lambda_{\rm mfp}=30$~comoving Mpc respectively.  Here 60~Mpc is the observed mean free path in \cite{worseck14}, while 30~Mpc takes into account that the observed values may be biased high by a factor of $\approx2$ owing to the quasar proximity effect \cite{daloisio18}. The comparison of dashed (M1/isotropic) and solid (exact solution) curves in Figure~\ref{fig:flucmfp_M1} show that the isotropic Eddington tensor approximation substantially underestimates the fluctuations in the ionizing background at scales smaller than the mean free path.

Our finding suggests that simulations with M1, which are commonly used to study the scatter in the Lyman-$\alpha$ forest transmission \cite{chardin15, chardin17, keating18, keating19, keating20, kulkarni19}, substantially underpredict sub-mean free path fluctuations in the ionizing background\footnote{Our finding is in qualitative agreement with numerical simulations.  A comparison between the long characteristics method of \cite{davies16} and a simulation with M1 using the same emissivity and opacity fields showed that the overall level of ionizing background fluctuations in the simulation with M1 is noticeably weaker (private communication with Fred Davies).}. We therefore estimate the amount of underestimation in the variation of the photoionization rate by M1 on $L=50/h$ comoving Mpc scale, which is the typical scale that the variations in the Lyman-$\alpha$ forest transmission are measured \cite{becker15, bosman18, eilers18}. The variance of the photoionization rate is given by $\sigma_E^2 \propto \int P^{\rm 1D}_E(k) W^2(k)\; \mathrm{d}k$, where $P^{\rm 1D}_E(k)$ is the 1D power spectrum of $E$, and $W(k) = \mathrm{sinc}(kL/2)$. The grey line in the right panel of Figure~\ref{fig:flucmfp_M1} illustrates $W^2(k)$. For a photon mean free path of 30 comoving Mpc, we find that M1 underestimates $\sigma_E$ on $50/h$~Mpc scales by $\approx20\%$. If the mean free path is 60~Mpc, the underestimation is boosted to $\approx30\%$. Using a $L=20/h$~Mpc window increases the underestimation by a modest factor of $5-8\%$.  These differences are smaller than expected from the dramatic differences in the power spectrum seen in the figure because much of the variance is driven by $k \lesssim \lambda_{\rm mfp}^{-1}$ where the algorithms agree.  Note that we have implicitly assumed that a simulation using M1 is able to reproduce the true mean free path, while Section~\ref{sec:spherical_absorber} has illustrated that simulations with M1 likely underestimate the effective mean free path by $\sim30\%$ when calibrated to match the post-reionization ionizing background. However, simulations with the reduced speed of light approximation have shown that the volume-averaged \HI\ fraction is roughly inversely proportional to the adopted speed of light \cite{bauer15, deparis19, ocvirk19, wu19a}, indicating that the mean free path of the simulations is likely overestimated by a factor of a few from the observed values possibly because current simulations do not capture the necessary scales (see Section~\ref{sec:reduced_c} for detailed discussions). For a factor of 2 overestimation of the mean free path, we find that M1 underestimates $\sigma_E$ by $60-70\%$ compared to the exact solution with the true mean free path.

In addition to underpredicting the variance of the photoionization rate on $50/h$~Mpc scales, underestimating fluctuations in the photoionization rate on sub-mean free path scales likely affects the occurrence of high Lyman-$\alpha$ transmission as well. This may impact the statistics of the transmission spikes in simulations with M1 \cite{chardin18b}, although density fluctuations are more important for interpreting Lyman-$\alpha$ forest transmissions.

We note that the M1/isotropic solution corresponds to closing the moment equations at order $\ell=1$. In \S\ref{sec:Gamma_fluc_higherorders} we will explore the solutions to $E$ when closing the moment equations at higher orders (red long dashed, short dashed, and dotted lines in the left panel of Figure~\ref{fig:flucmfp_M1}).

\subsubsection{OTVET}

Since OTVET includes image sources within one box size from a given location in a simulation volume, the degree of isotropy of the resulting Eddington tensor is expected to depend on the box size. As we show in Appendix~\ref{sec:FT_deltahij}, the OTVET solution to $\fdE$ is
\begin{equation}
\fdE = \frac{\left(1 + \frac{2\pi^2 k}{3a^2\meankappa^2 L_{\rm O}}\right)\fdj - \bkd\fdd}{1 + \bkg + \frac{k^2}{3a^2\meankappa^2}},
\label{eq:deltaE_otvet}
\end{equation}
where $L_{\rm O} = \int_{\cal V} r^{-2}\; \mathrm{d}^3x$, ${\cal V}$ is the simulated volume, and $r=\sqrt{x^2+y^2+z^2}$ with a wrap around the box (such that $x \rightarrow L_{\rm box} - x$ if $x$ is larger than half the box size $L_{\rm box}$) to take into account periodic boundary conditions.  % integral is numerically evaluated over $[0,L/2]^3$ and multiplied by 8, with $L$ being the box size. 
Note that this equation only applies to $k \gg 2\pi/L$.
For large enough box sizes, the Eddington tensor becomes isotropic owing to the Olber's paradox, and the solution tends to equation~\ref{eq:deltaE_isotropic}. For box sizes smaller than twice the mean free path, the OTVET Eddington tensor is more anisotropic than the exact Eddington tensor, since fewer sources contribute to the Eddington tensor than in the exact solution. In this case, we expect OTVET to overestimate the amount of ionizing background fluctuations at scales smaller than the mean free path.

\begin{figure}[tbp]
\centering 
\includegraphics[width=\textwidth]{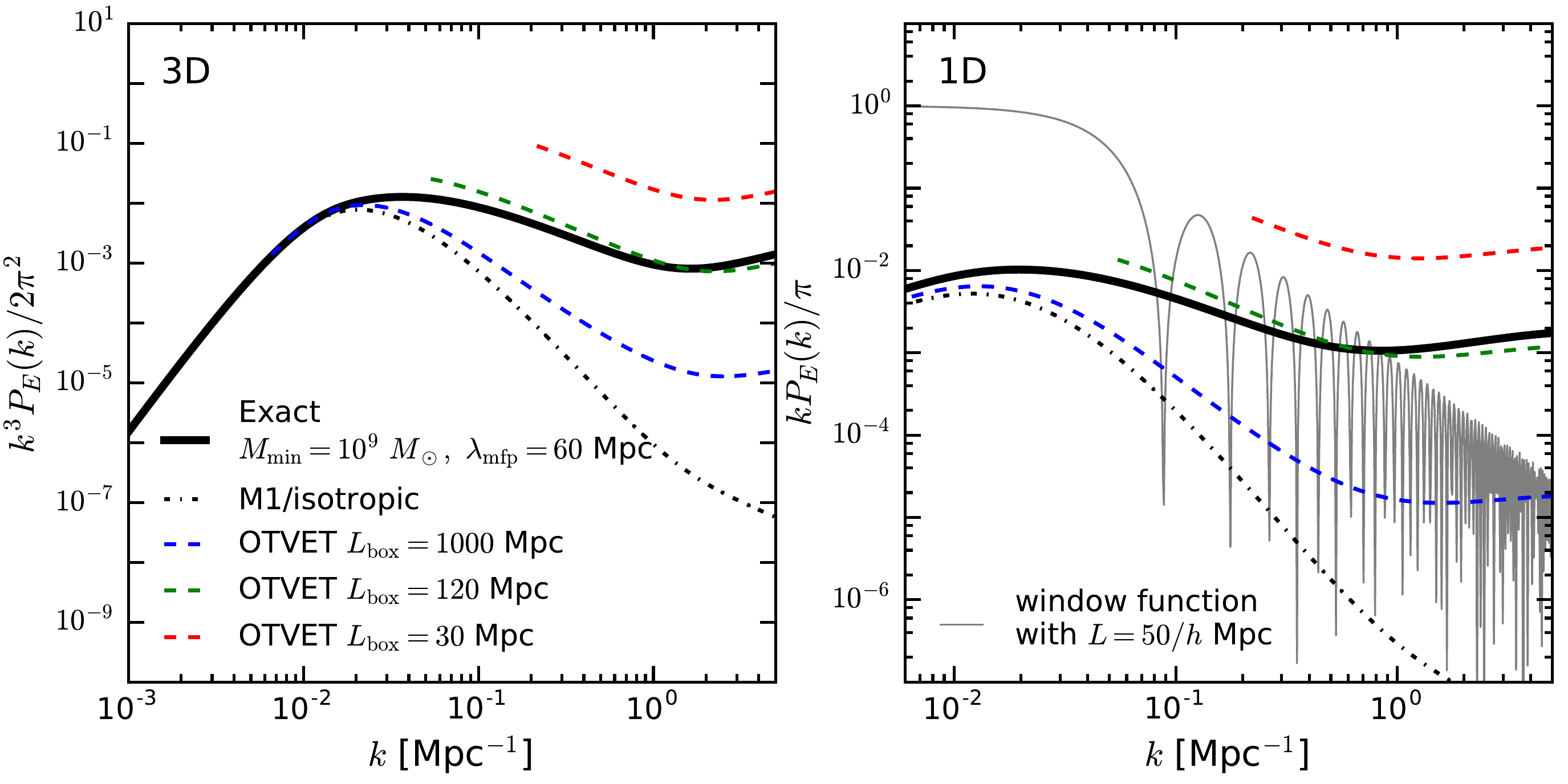}
\hfill
\caption{\label{fig:flucmfp_otvet} 3D (left) and 1D (right) power spectra of the post-reionization photoionization rate, assuming $z=5.5$, $\lambda_{\rm mfp} = 1/a\meankappa = 60$ comoving Mpc, $\bkd=1$, and $\bkg=-2/3$. The black solid and dot-dashed lines show the exact and M1/isotropic solutions respectively, using source power spectra with minimum halo mass $M_{\rm min}=10^9\ M_\odot$. Blue, green, and red dashed lines represent the OTVET solutions with box sizes $L_{\rm box}=1000,120,30$~Mpc respectively. The OTVET solutions are only shown for $k>2\pi/L_{\rm box}$. The grey line in the right panel shows the squared window function $W^2(k)=\mathrm{sinc}^2(kL/2)$ with $L=50/h$~Mpc, where $50/h$~Mpc is the typical scale that the variations in the Lyman-$\alpha$ forest transmission are measured. Simulations with OTVET overestimate (underestimate) small-scale ionizing background fluctuations when box sizes are smaller (larger) than twice the mean free path.}
\end{figure}

Figure~\ref{fig:flucmfp_otvet} shows the $z=5.5$ 3D and 1D power spectra of $E$ given by the exact solution (black solid lines), the M1/isotropic solution (black dot-dashed lines), and OTVET (dashed lines), using the source power spectrum with minimum halo mass of $10^9\ M_\odot$ and a photon mean free path of 60 comoving Mpc. We have again used $\bkd=1, \bkg=-2/3$. For OTVET, the blue, green, and red colors represent solutions with simulation box sizes $L_{\rm box} = 1000, 120, 30$ comoving Mpc respectively. For large enough box sizes, the OTVET solution tends to the M1/isotropic solution, but the underestimation in the ionizing background fluctuations at large wavenumbers is less severe because the optically thin assumption produces more anisotropy in the Eddington tensor on small scales. For box sizes smaller than twice the true mean free path, OTVET overestimates sub box-scale fluctuations in the ionizing background, since only sources within one box size contribute to the Eddington tensor, making the Eddington tensor more anisotropic. When the box size is twice the mean free path, OTVET roughly reproduces the ionizing background fluctuations.

To compute the variance of the photoionization rate on $50/h$ comoving Mpc scale, we calculate $\sigma_E^2 \propto \int P^{\rm 1D}_E(k) W^2(k)\; \mathrm{d}k$ for the exact solution and OTVET, using $P^{\rm 1D}_E(k<2\pi/L_{\rm box})=0$. We find that when the box size is much larger than the photon mean free path, e.g. $L_{\rm box}=1000$ Mpc and $\lambda_{\rm mfp}=60$ Mpc, OTVET underestimates the variations of the photoionization rate by $20-30\%$, similar to M1 as expected since both have an isotropic Eddington tensor in this limit. If the box size is smaller than the mean free path, e.g. $L_{\rm box}=30$ Mpc and $\lambda_{\rm mfp}=60$ Mpc, OTVET overestimates the variance by a factor of $3-4$. So far, small box simulations with OTVET ($L_{\rm box}=40/h$~Mpc in \cite{gnedin17}) have been mainly used to interpret the large scatter in the Lyman-$\alpha$ forest transmission at $z=5-6$. Our findings indicate that these simulations likely overpredict fluctuations in the ionizing background by factors of $2-4$ and thus should predict more scatter in the spatial transmission in the Lyman-$\alpha$ forest. Simulations with box sizes similar to twice the mean free path likely fair much better at reproducing the correct amount of photoionization rate fluctuations, e.g. the $80/h$~Mpc box simulations in \cite{gnedin14}. Those simulations have been used to study the Lyman-$\alpha$ transmission spikes \cite{garaldi19}.

\subsubsection{Closing the moment equations at higher orders}
\label{sec:Gamma_fluc_higherorders}

While the M1/isotropic solution corresponds to closing the moment equations at order $\ell=1$, we consider the solution to $E$ when closing the moment equations at higher orders (see Appendix~\ref{sec:arctan_higher_order} for full derivations),
\tmp{which corresponds to sampling the radiation intensity with more angular directions.  In the spirit of the sampling theorem where an evenly sampled function in time domain can be completely determined by the summation of the Fourier modes up to the Nyquist frequency, the sampling of different directions can be characterized by the spherical harmonics.  Taking the $\ell$-th angular moment of the RT equation, or more precisely expanding the intensity with the $\ell$-th Legendre polynomial (see Appendix~\ref{sec:arctan_higher_order}), thus corresponds to sampling $(2\ell+1)$ angular directions (the spherical harmonic function $Y^m_\ell$ has $m\in[-\ell, \ell]$).  Taking the angular moment of the RT equation up to the $\ell$-th order is therefore roughly consistent with sampling a total of $\approx \int (2\ell+1)\; \mathrm{d}\ell \approx \ell^2$ directions.  For instance, \cite{trac07} limits the maximum number of rays in a cell to 64, which roughly corresponds to closing the moment RT equations to order $\ell=8$.  Hence the accuracy of ray-tracing methods in simulating the end of reionization can be probed by studying solutions to the higher order moment equations, with the $(\ell+1)$-th order moment set to 0 to close the set of moment equations.}

The red long dashed, dotted, and short dashed lines in the left panel of Figure~\ref{fig:flucmfp_M1} show the 3D power spectra of $E$ when closing the equations at orders $\ell=2,5,8$ respectively, assuming that the source power spectrum has a minimum halo mass of $10^{11}\ M_\odot$. Moment-based RT methods are unlikely to close the equations at orders higher than these values owing to memory constraints (as we are unaware of attempts to go beyond $\ell=1$), and \tmp{as argued above} there is a correspondence between moment codes that truncate at order $\ell$ and short characteristic ray-tracing codes that pixelate the sphere with $\approx \ell^2$ directions. In addition, $\ell=8$ also mimics the (pseudo) long-characteristics ray-tracing method of \cite{trac07}, where they merge rays so that the number of rays in a cell is capped at 64. Since information on angular scales $\lesssim\pi/\ell$ is not captured when closing the moment equations at order $\ell$, moment-based methods should converge to the exact solution at wavenumbers $k\lesssim\ell/\lambda_{\rm mfp}$. Comparison between the higher moment solutions and the exact solution in Figure~\ref{fig:flucmfp_M1} is roughly consistent with this estimate. Our results imply that capturing the post-reionization ionizing background fluctuations requires following the radiation field on very small angular scales, which are currently not achieved by moment-based methods or most ray-tracing codes. This likely introduces a bias in the simulated ionizing background fluctuations which affects interpreting the Lyman-$\alpha$ forest observations.

\subsection{Summary}

Overall, we find that moment-based methods produce a qualitatively different spectrum of ionizing background fluctuations on scales smaller than the photon mean free path, in addition to somewhat over-ionizing the dense absorbers of ionizing photons as found in \S\ref{sec:spherical_absorber}. The M1 Eddington tensor tends to isotropic after reionization, leading to significantly underestimated ionizing background fluctuations on small scales. This results in M1 underestimating the variance of the photoionization rate on $50/h$~Mpc scales by $20-30\%$, if the simulations capture the observed photon mean free path. For OTVET, the degree of anisotropy of the Eddington tensor depends on the box size. Large enough boxes give more isotropic Eddington tensors, thus leading to underestimation of the ionizing background fluctuations similar to M1. Small boxes produce more anisotropy in the Eddington tensors, resulting in overestimation of the ionizing background fluctuations. For box sizes smaller than twice the observed mean free path, the overestimation of the variance of the photoionization rate on $50/h$~Mpc scales could reach a factor of $\sim2-4$ level. Moment-based methods should be used with this caution when studying the large scatter in the Lyman-$\alpha$ forest transmission. We additionally showed that ray-tracing methods may not be completely immune to these difficulties if they limit the number of angular directions that are followed. 

\tmp{One caveat we are unable to address is the transition from the overlap phase where local sources likely dominate the photoionization rate to long after reionization where the ionizing background is roughly uniform.  For the exact solution, we do not expect this to affect our discussions because near the end of reionization the photon mean free path is limited by the Lyman-limit systems instead of the rapidly growing bubble size \cite{furlanetto05}, making a rapid evolution in the ionizing background and its fluctuations unlikely during the transition phase \cite{furlanetto09}.  Even in a regime where local sources dominate the photoionization rate, the exact solution captures the proximity zone effect.
For the moment based methods, it is less clear how well our analytic solutions describe the situations in real simulations.  In addition to having trouble simulating the process of bubble overlap, it is also unclear whether the diffusion-like algorithms result in a very different time scale for the ionizing background to asymptote to the analytic time-independent solution equation~\ref{eq:deltaE_hij} compared to the exact case.  Our calculations are thus limited to the scope where the time-independent perturbative solution to the RT equation should apply.  We defer the answer to these issues to future investigation with numerical simulations.}
%although the clustering of the sources shapes the high-value tail of the ionizing background fluctuations on small-scales \cite{mesinger09}, we are mostly concerned with the ionizing background in the diffuse IGM which sets the Lyman-$\alpha$ forest transmission.  The proximity effect is therefore less of a concern for our purposes, and it is mainly the small photoionization rate fluctuations in the diffuse IGM that we need to address.

\section{Discussions}

\subsection{The photon mean free path in reionization simulations}
\label{sec:reduced_c}

\tmp{In Section~\ref{sec:spherical_absorber} and \ref{sec:Gamma_fluc} we pointed out two potential issues that reionization simulations with OTVET and M1 may encounter, namely that the simulations may err at reproducing the photon emissivity and mean free path at $\sim30\%$ level when calibrated to match a fixed photoionization rate, and that they likely significantly underestimate or overestimate small-scale ionizing background fluctuations.  Here we argue that many published simulations using the M1 algorithm likely also suffer from overshooting the empirical measurements of the mean free path \citep[e.g.][]{worseck14} by factors of a few, given the findings of multiple simulations using the reduced speed of light approximation \cite{bauer15, deparis19, ocvirk19, wu19a}.  Recent simulations with OTVET algorithm use these measurements to cap the simulations' mean free paths \cite{gnedin14, gnedin16} and so cannot overshoot these measurements.}

Specifically, these simulations found that the volume-averaged \HI\ fraction after reionization is roughly inversely proportional to the value of the speed of light adopted \cite{ocvirk19, wu19a}, or since the \HI\ fraction is inversely proportional to the background photoionization rate, the simulated ionizing background amplitude is proportional to the adopted speed of light. The ionizing background should be independent of the speed of light if the time to travel one mean free path and be absorbed is much smaller than the Hubble time.  However, the longer the time to travel one mean free path is, the more reduced the speed of light in the simulation is. Gnedin~\cite{gnedin16} showed in simulations that the anticipated size of this effect is about a factor of two at $z=5.5$ for $\tilde{c}=0.1c$ compared to $\tilde{c}=1c$, much smaller than the factor of ten scaling in \cite{ocvirk19}.  The dependence found in \cite{gnedin16} is quantitatively reproduced by solving the radiative transfer equation, which ignoring redshifting (but including dilution) gives $I_\nu(t, {\hat n}) = \int_0^t \tilde{c} j_\nu(t', {\hat n}) \exp(-\int_{t'}^t [3H(t'') + \tilde{c} \kappa(t'', {\hat n})]\; \mathrm{d}t'')\; \mathrm{d}t'$ \cite[e.g.][]{meiksin18} where $\tilde{c}$ is the assumed speed of light. One can observe that the speed of light only enters via the term in the exponential with $H$ (and an analogous result holds for moment equations).  Thus, a rough estimate of the effect of the reduced speed of light on $I_\nu$ can be obtained from $1 + 3 H(t)/[\tilde{c}\kappa(t)]$.  For $z=5$ and the observed value of $[a \kappa]^{-1} = 80$~Mpc \cite{worseck14}, this implies that $I_\nu$ differs by a factor of $\sim1.6$ for simulations using $\tilde{c}=1c$ and $0.1c$, where $c$ is the actual speed of light.  Thus, the stronger scaling found in \cite{ocvirk19} is likely because the mean free path is overestimated by a factor of at least several in the simulations of \citep{ocvirk19}.

\begin{figure}[tbp]
\centering 
\includegraphics[width=0.7\textwidth]{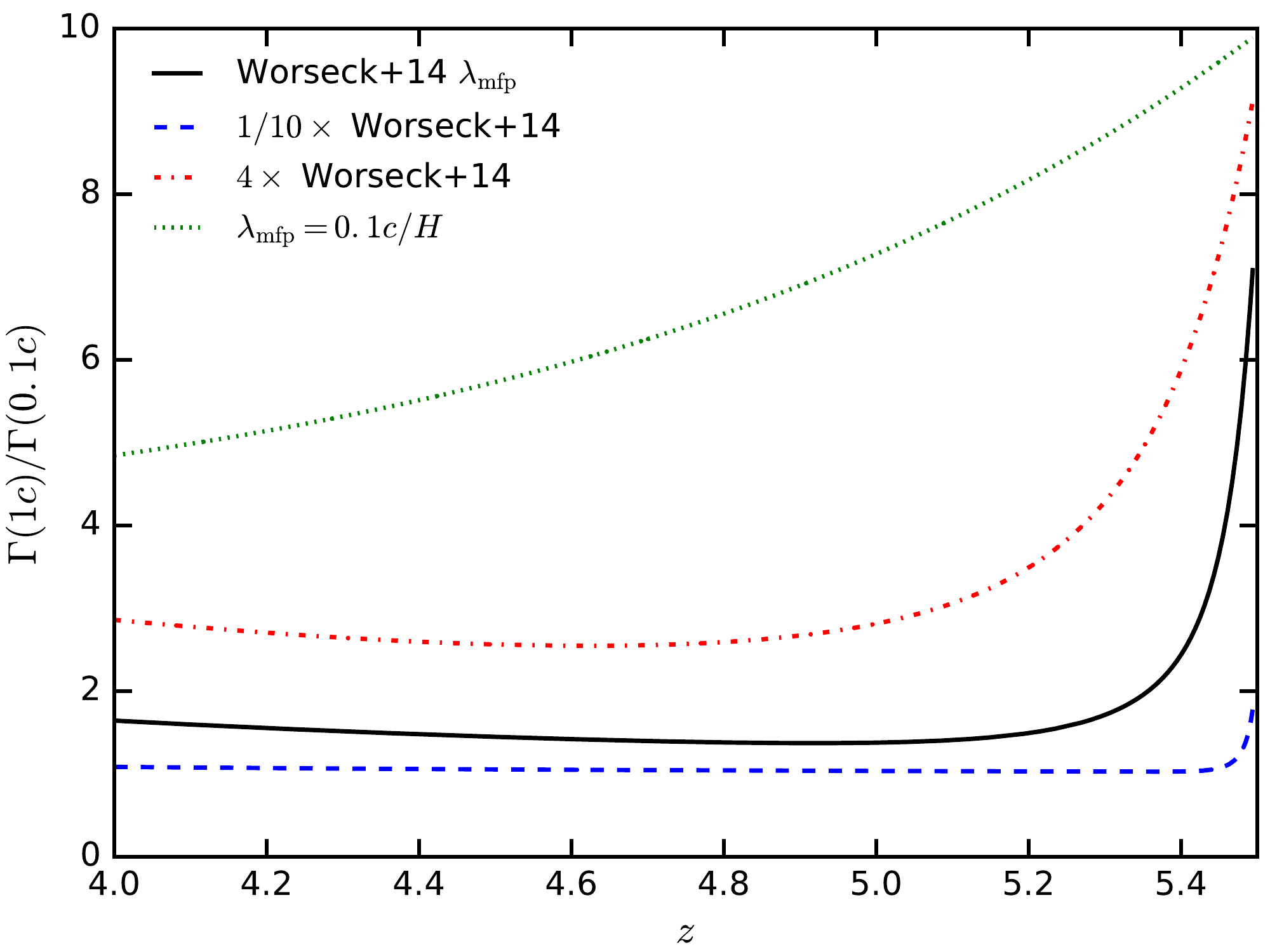}
\hfill
\caption{Estimates of the ratios of the photoionization rates in simulations with the true speed of light and a tenth of the speed of light, assuming reionization ends at $z=5.5$ and constant emissivity afterwards.  The black solid, blue dashed, red dot-dashed, and green dotted lines show the ratios as a function of redshift using photon mean free path values as measured in \cite{worseck14}, a tenth of \cite{worseck14}, four times that of \cite{worseck14}, and a tenth of the Hubble radius respectively.  Our calculations indicate that simulations which found strong scaling relations of the post-reionization volume-averaged \HI\ fraction with the value of the reduced speed of light likely overestimate the photon mean free path by at least a factor of a few, although our predictions may seem extreme.}
\label{fig:Gamma_clight}
\end{figure}

To estimate to what extent the simulations overpredict the photon mean free path, we evaluate the integral $I_\nu(t, {\hat n}) = \int_0^t \tilde{c} j_\nu(t', {\hat n}) \exp(-\int_{t'}^t  [3H(t'') + \tilde{c} \kappa(t'', {\hat n})]\; \mathrm{d}t'')\; \mathrm{d}t'$ using different values of the mean free path.  We assume that overlap occurs at $z=5.5$ so that the emissivity $j_\nu$ is zero before then, and that $j_\nu$ is constant afterwards.  Figure~\ref{fig:Gamma_clight} illustrates our predictions for the ratios of the photoionization rate ($\Gamma$) in simulations with the true speed of light and with $0.1c$.  The black solid, blue dashed, red dot-dashed, and green dotted lines show the ratios as a function of redshift using photon mean free path values as measured in \cite{worseck14}, a tenth of \cite{worseck14}, four times that of \cite{worseck14}, and a tenth of the Hubble radius ($c/H$) respectively.  The ratios of $\Gamma$ tend to the ratio of the adopted speed of light right after overlap, but become less extreme at lower $z$.  Our calculations show that if the simulations produce the same mean free path as that measured in \cite{worseck14}, the differences in $\Gamma$ in simulations with $\tilde{c} \ge 0.1c$ should be less than a factor of 2, in rough agreement with our simple estimate using $1+3H/[\tilde{c}\kappa]$ above.  Reducing the mean free path by a factor of 10 eliminates the differences in the post-reionization $\Gamma$, consistent with the analysis of \cite{gnedin16}.  However, the strong linear scaling relation between the post-reionization volume-averaged \HI\ fraction and the adopted speed of light as found in \cite{ocvirk19} is not recovered in our calculations, even when enlarging the mean free path in \cite{worseck14} by a factor of 4 or assuming the mean free path is a tenth of the Hubble radius.  This may surprisingly indicate that the photon mean free path is hardly limited in reionization simulations having trouble with the reduced speed of light.  While this prediction is likely too extreme, the rough agreement between our calculations and the findings of \cite{gnedin16} suggests that a factor of a few overestimate of the mean free path in those simulations is still a possible explanation to the reduced speed of light problem.

\tmp{In summary, we find that in addition to having trouble capturing the correct post-reionization photoionization rate, emissivity, and photon mean free path simultaneously and reproducing the ionizing background fluctuations, simulations using M1 likely also overpredict the photon mean free path by at least factors of a few.}
One possible reason is that they likely overshoot the emissivity. \cite{mcquinn11} found a steep power-law scaling of the post-reionization photoionization rate with emissivity, suggesting that a small change in emissivity could lead to much a larger change in the mean free path. Matching the Lyman-$\alpha$ forest transmission therefore requires fine-tuning of the emissivity \cite{chardin15, keating18, keating19, keating20, kulkarni19}\tmp{, which might alleviate the reduced speed of light problem to some degree}. In addition, the IGM is expected to clump on scales of $10^4-10^8\ M_\odot$ \cite{daloisio20}, which most of the cosmological simulations are unable to resolve. Given the $\sim300$~Myr relaxation time of the gas after heating \cite{daloisio20}, resolving these small-scale structures is likely still necessary to adequately limit the photon mean free path\tmp{, and thus possibly remove the reduced speed of light dependence of the simulated photoionization rate}. We defer an exploration of the resolution requirement to resolve the Lyman-limit systems to future work.

\section{Conclusions}
\label{sec:conclusions}

This paper discussed the accuracy of common moment-based radiative transfer algorithms on simulating reionization.  Specifically, it investigated the use of an approximate Eddington tensor as an ansatz for the quadrupole moment to close the system of monopole and dipole equations.   \tmp{We argued that during reionization, the large-scale growth of ionized bubbles is likely minorly affected by the choice of the Eddington tensor because of photon conservation. (However, structures on smaller scales than the bubbles, especially those driven by dense self-shielding clumps, may not be adequately captured by the RT algorithm.)} % so predictions of the simulations on observables imprinted by the morphology of reionization (e.g. relic temperature fluctuations in the IGM after reionization and the kinetic Sunyaev-Zel'dovich effect) are relatively accurate.
We considered a during-reionization example of the radiation field from a single source ionizing a uniform medium, finding that the usage of a flux limiter that caps the amplitude of the photon flux ensures that the exact solution to the radiation field is reproduced, even though OTVET may produce an Eddington tensor that becomes isotropic inside ionized bubbles. %As a result, the choice of the Eddington tensor also has a negligible effect on the photoionization rate profile in an ionized bubble, implying that moment-based RT methods are accurate enough for quasar proximity zone studies.
Thus, we suspect \tmp{(with the above caveat regarding dense absorption systems)} that moment-based RT methods are thus able to capture the \tmp{gross} properties of the \HII\ regions during reionization. We argued that their differences with the exact solution are likely to be larger at the end of reionization and just after.

We studied test problems targeted at the ionization structure of dense absorbers in ionized regions and fluctuations in the post-reionization ionizing background.  We found that for a spherical absorber in photoionization equilibrium with radiation coming uniformly from infinity, the usual closure methods, OTVET and M1, over-ionize the absorber when fixing the background photoionization rate.  For a simulation run with these algorithms, this over-ionization leads to $20-40\%$ higher emissivity required to balance the total recombination for a given background photoionization rate, or a factor of $\sim2$ lower metagalactic photoionization rate given the ionizing emissivity.  The effective mean free path of the simulations, defined as the ratio of the metagalactic photoionization rate to the emissivity, is thus underpredicted by similar amounts.  However, if one measured the mean free path by shooting rays across the simulation box in OTVET and M1, this curiously results in a $\sim10\%$ overestimation of the mean free path.  These biases indicate that simulations using OTVET and M1 likely underpredict the duration of reionization and $\Omega_{\rm HI}$ after reionization when calibrating to the Lyman-$\alpha$ forest transmission or the inferred photoionization rate, or they produce a more opaque Lyman-$\alpha$ forest when calibrating to given emissivities.

Considering linear-order fluctuations in the post-reionization ionizing background, we found that moment-based RT algorithms produce very different power spectra of the ionizing background fluctuations from the exact solution.  The M1 Eddington tensor leads to significantly suppressed power on scales smaller than the photon mean free path, leading to $20-30\%$ underestimation of the variance of the photoionization rate on $50/h$~Mpc scales.  OTVET results in a similar underprediction for large simulation boxes, but overpredicts the small-scale fluctuations in the ionizing background when the simulation box size is smaller than twice the photon mean free path, causing a factor of $2-4$ overestimation of the variance of the photoionization rate on $50/h$~Mpc scales when the box size is a half of the mean free path.
These algorithms thus should be used with caution for modeling the large spatial scatter in the Lyman-$\alpha$ forest transmission (which most likely owes to large-scale ionizing background fluctuations), and the transmission spikes which have contributions from ionizing background fluctuations on all scales.

We also investigated a curious feature found in simulations using the M1 algorithm, which the above differences do not seem sufficient to explain: several studies have found that the post-reionization volume-averaged \HI\ fraction scales essentially inversely with the adopted speed of light.  We showed that this should not occur if the mean free path is consistent with observations, concluding that in these simulations mean free paths are likely larger than the measured value by a factor of a few. %Longer mean free paths amplify the effect of the reduced speed of light shrinking the Hubble radius, which alone is unlikely to result in such strong scaling of the background photoionization rate with the value of the speed of light as found in the literature.  
Most cosmological simulations of reionization lack the resolution to resolve the $10^4-10^8\ M_\odot$ mass scales which the IGM clumps on, and future work might focus on an exploration of the resolution requirement to resolve all Lyman-limit systems.

Given the above caveats of moment-based RT algorithms with approximate Eddington tensors, more accurate ray-tracing methods might be a favored choice for simulating reionization.  However, we found that ray-tracing methods that limit the number of angular directions that they follow likely also have trouble reproducing the small-scale fluctuations in the post-reionization ionizing background, which requires resolving a large number of angular directions. 

Cosmological radiative transfer is still at a nascent state with no consensus on what algorithm is best.  This study's considerations may help motivate the design of the next generation algorithm. % therefore needs to incorporate new methods to record enough angular information of the radiation field, in addition to capturing the sub-grid recombinations and I-front trapping owing to unresolved IGM clumping.

\acknowledgments
We acknowledge useful conversations with Anson D'Aloisio, Avery Meiksin, Vid Ir\v si\v c, and Fred Davies.  We especially thank Nick Gnedin for explaining how the ART code implements OTVET.  \tmp{We also thank the anonymous referee for the valuable comments and feedback.}  MM is supported by NASA award~19-ATP19-0191.

\appendix
\section{Solving the static moment equations with an opacity profile}
\label{sec:absorber_eqns}

We solve for $E$ and $F$ profiles assuming single absorber with planar/spherical geometry and analytical $\kappa$ profiles, and radiation coming uniformly from infinity. This approximates the situation of Lyman-limit systems at the end of reionization (we define Lyman-limit systems here as systems that have substantial Lyman-continuum optical depths and so self shield). The exact solution simply corresponds to calculating the optical depth and integrating over all directions. For M1 and OTVET, we solve equations~\ref{eq:rt_monopole} and \ref{eq:rt_dipole}, or \ref{eq:rt_diffusion}. Here $F$ is a signed quantity with $|F| = |\vec{F}|$, and is positive when $\vec{F}$ points to the $+z$ direction (plane parallel case) or radially outward (spherical case).

\subsection{Plane parallel absorber}

Consider a plane parallel absorber with a monomial $\kappa$ profile $\kappa = \kappa_0 (z_0/z)^m$, and radiation comes uniformly from infinity to the $x-y$ plane. In the limit of large $m$, this problem corresponds to a plane parallel absorber with a sharp edge at $z=z_0$.

\begin{itemize}

\item Exact solution:

\begin{gather}
\tau(\mu, z) = \frac{\kappa_0 z_0^m}{(m-1)z^{m-1}\mu}; \\
E(z) = \int_0^1 \mathrm{d}\mu\; \exp(-\tau(\mu, r)); \\
|F(z)| = \int_0^1 \mathrm{d}\mu\; \exp(-\tau(\mu, r)) \mu,
\end{gather}
where $\mu = \cos(\theta)$. One can verify that $g=|F|/E$ is $0.5$ at $z\to\infty$, and $1$ at $z\to0$.

\item OTVET:

The Eddington tensor is isotropic everywhere. We do a change of variables $\ds/\dr = -\kappa$, so $s = \kappa z / (m-1)$ with a monomial $\kappa$. Equation~\ref{eq:rt_diffusion} becomes $\mathrm{d}^2E/\ds = 3E$, and the solution is
\begin{gather}
E(s) = \exp(-\sqrt{3}s); \\
|F(s)| = \frac{1}{3\kappa}\frac{\mathrm{d}E}{\mathrm{d}z} = \frac{1}{\sqrt{3}} \exp(-\sqrt{3}s).
\end{gather}
Therefore $g = |F|/E = 1/\sqrt{3}$ everywhere. This solution holds for arbitrary $\kappa$ profile.

\item M1 and general $\chi(g)$:

We first derive the solution assuming a general form of $\chi(g)$. We use the same change of variables $\ds/\dr = -\kappa$. Because of the symmetry of the system, we have $h_{zz}=\chi$. We can construct differential equations for $g$ and $E$. Equations~\ref{eq:rt_monopole} and \ref{eq:rt_dipole} become
\begin{gather}
\frac{\mathrm{d}\ln E}{\ds} = \frac{\frac{\mathrm{d}\chi}{\dg} - g}{\chi - g \frac{\mathrm{d}\chi}{\dg}}; \\
\frac{\dg}{\ds} = \frac{g^2 - \chi}{\chi - g \frac{\mathrm{d}\chi}{\dg}}.
\end{gather}
Note that these equations are independent of the form of $\kappa$. Also note that
\begin{equation}
\frac{\ds}{\dg} = \frac{\chi - g \frac{\mathrm{d}\chi}{\dg}}{g^2 - \chi}.
\end{equation}
This equation can be integrated to get $s(g)$. Since $0 \le g \le 1$, in order for $s\to\infty$ ($z\to0$) to be possible, at $z\to0$ we must get $g\to\sqrt{\chi}$, which is the singularity of the denominator. For M1 and the exact solution, this gives $g\to1$ and $\chi\to1$ at $z\to0$. For OTVET, this corresponds to $g=1/\sqrt{3}$.

Specifically for M1, $\ds/\dg$ has a zero-point at $g = 2\sqrt{3}/5 = 0.69$, and $\ds/\dg$ changes sign when crossing this zero-point. It can then be realized that in order for $s$ to be able to go to infinity, $g \ge 0.69$ is required everywhere $z>0$. However, the boundary condition at $z\to\infty$ ($s\to0$) does not have a physically motivated choice.

\end{itemize}

\subsection{Spherical absorber}

\subsubsection{monomial opacity profile}

Consider a spherical absorber at the origin with $\kappa = \kappa_0 (R/r)^m$ and radiation coming uniformly from infinity. In the limit of large $m$, this corresponds to a spherical absorber with sharp edge at $r=R$.

\begin{itemize}

\item Exact solution\footnote{The $\tau$ integral can be calculated using a recursion formula: \url{https://en.wikipedia.org/wiki/List_of_integrals_of_rational_functions}.}:

\begin{gather}
\tau(\mu, r) = \int_0^\infty \kappa(\sqrt{r^2+s^2+2rs\mu})\; \ds = \frac{\kappa_0 R^m}{r^{m-1}} \int_0^\infty \frac{\mathrm{d}x}{(1 + x^2 + 2x\mu)^{m/2}}; \\
E(r) = \frac{1}{2} \int_{-1}^1 \mathrm{d}\mu\; \exp(-\tau(\mu, r)) ;\\
|F(r)| = \frac{1}{2} \int_{-1}^1 \mathrm{d}\mu\; \exp(-\tau(\mu, r)) \mu,
\end{gather}
where $\mu = \cos(\theta)$.

\item OTVET:

The Eddington tensor is isotropic everywhere. We do a change of variables $u=\sqrt{3}\kappa r/(m-1)$ and use $E=u^\nu f(u)$, where $\nu$ is a number to be determined. Plugging these into equation~\ref{eq:rt_diffusion}, we find that when $\nu = (m+1)/[2(m-1)]$, $f(u)$ is the solution to the modified Bessel's equation. Considering the boundary condition at $r\to\infty$ ($u\to0$), we find
\begin{gather}
E(u) = E_0 u^\nu K_\nu(u); \\
|F(u)| = \frac{1}{3\kappa}\frac{\mathrm{d}E}{\dr} = - \frac{E_0}{\sqrt{3}} \left( \nu u^{\nu-1} K_\nu(u) + u^\nu \frac{\mathrm{d}K_\nu(u)}{\mathrm{d}u} \right),
\end{gather}
where the normalization $E_0 = 2^{\nu-1} \Gamma(\nu)$ comes from the asympototic behavior of $K_\nu(u\to0)$.

\item M1 and general $\chi(g)$:

We change the variable $r$ to $s$, with $\ds/\dr = -\kappa$. We keep the term
\begin{equation}
\frac{\mathrm{d} \ln r}{\mathrm{d} \ln s} = -\frac{s}{\kappa r},
\end{equation} 
instead of plugging in the value $-(m-1)$ for a monomial $\kappa$. We would like to obtain differential equations for $g=-F/E$ and $E$, for a general form of $\chi(g)$. Since $h_{rr}=\chi, h_{\theta\theta}=h_{\phi\phi}=(1-\chi)/2$, equations~\ref{eq:rt_monopole} and \ref{eq:rt_dipole} become
\begin{gather*}
\frac{\dg}{\ds} = -\frac{2g}{s}\frac{\mathrm{d}\ln r}{\mathrm{d}\ln s} - 1 - \frac{\mathrm{d}\ln E}{\ds}g; \\
\frac{\mathrm{d}\ln E}{\ds} = -\frac{3\chi-1}{s \chi}\frac{\mathrm{d}\ln r}{\mathrm{d}\ln s} - \frac{1}{\chi}\frac{\mathrm{d}\chi}{\dg}\frac{\dg}{\ds} - \frac{g}{\chi}. \numberthis \label{eq:spherical_dlogEds}
\end{gather*}
Plugging the second equation into the first one, we get
\begin{equation}
\frac{\dg}{\ds} = \frac{\frac{g^2}{\chi} - 1 + \frac{\chi-1}{\chi}\frac{g}{s}\frac{\mathrm{d}\ln r}{\mathrm{d}\ln s}}{1 - \frac{g}{\chi}\frac{\mathrm{d}\chi}{\dg}}.
\label{eq:spherical_dgds}
\end{equation}

To solve equations~\ref{eq:spherical_dlogEds} and \ref{eq:spherical_dgds}, we first integrate equation~\ref{eq:spherical_dgds} to obtain a solution for $g(s)$, and then plug this into equation~\ref{eq:spherical_dlogEds} to get $E$. However, the denominator of equation~\ref{eq:spherical_dgds} can possibly be 0. We thus examine
\begin{equation}
\frac{\ds}{\dg} = \frac{1 - \frac{g}{\chi}\frac{\mathrm{d}\chi}{\dg}}{\frac{g^2}{\chi} - 1 + \frac{\chi-1}{\chi}\frac{g}{s}\frac{\mathrm{d}\ln r}{\mathrm{d}\ln s}}.
\label{eq:spherical_dsdg}
\end{equation}
At $s=0$, the physically motivated value for $g$ is 0. The function $s(g)$ must therefore be monotonically increasing. Thus if the numerator of equation~\ref{eq:spherical_dsdg} reaches 0 at some $g$ before $s\to\infty$, the denominator must also be 0. This puts a constraint on the value of $\kappa r = -(1/s)\ \mathrm{d}\ln r/\mathrm{d}\ln s$ at this certain $g$. Also note that when $s$ is large, the denominator of equation~\ref{eq:spherical_dsdg} is dominated by $g^2/\chi-1$. In order for $s$ to be able to go to infinity, we must have $g^2/\chi\to1$ at $s\to\infty$ ($r\to0$).

One can verify that the exact solution and the OTVET solution satisfy equations~\ref{eq:spherical_dlogEds} and \ref{eq:spherical_dgds}. For the exact solution, $g\to1$ and $\chi\to1$ at $s\to\infty$. For the OTVET solution, because $\chi=1/3$, $g\to1/\sqrt{3}$ at $s\to\infty$.

Specifically for M1, $1 - (g/\chi)\ \mathrm{d}\chi/\dg$ is 0 when $g=2\sqrt{3}/5=0.69$. In order for a solution to exist and in order for $s$ to be able to go to infinity when integrating $\ds/\dg$, $\kappa r = -s\; \mathrm{d}\ln s/\mathrm{d}\ln r = 4/\sqrt{3} = 2.31$ must be satisfied at $g=0.69$. We therefore need to start the integration of equation~\ref{eq:spherical_dgds} from this singularity to $s\to0$ and $s\to\infty$. The specific value of $\dg/\ds$ at the singularity can be obtained by L'Hopital's rule. Note that $g(s\to0)=0$ and $g(s\to\infty)=1$ are satisfied.

\end{itemize}

\begin{figure}[tbp]
\centering 
\includegraphics[width=\textwidth]{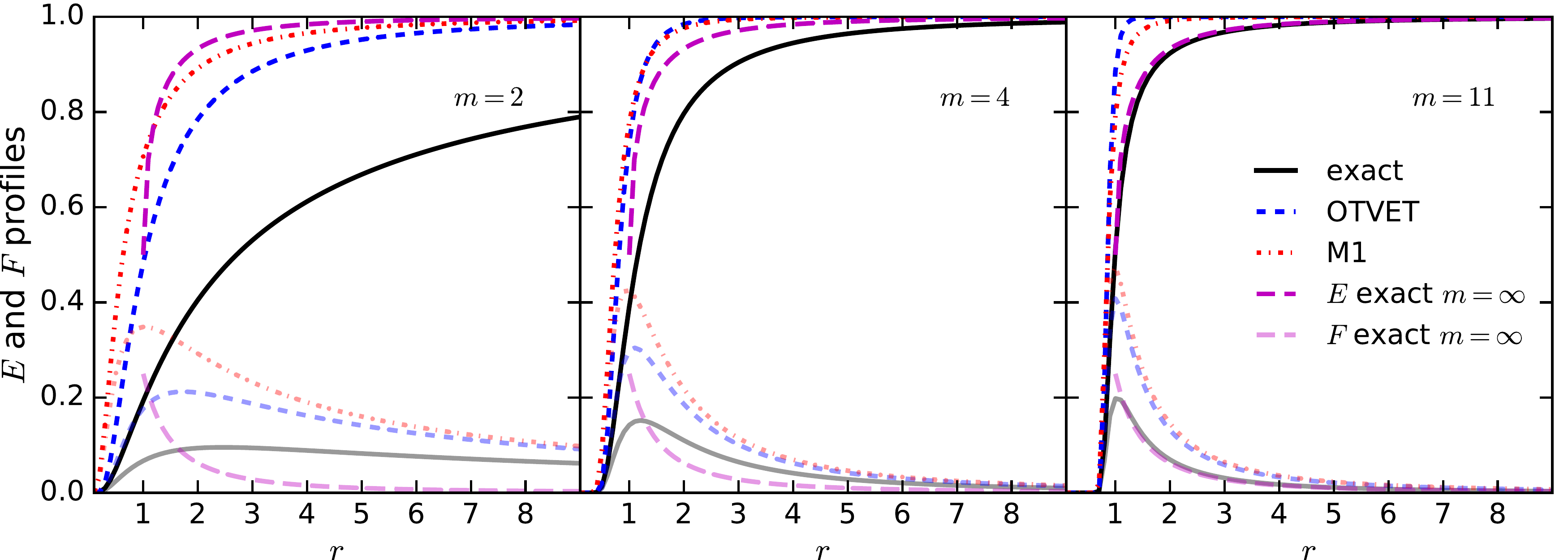}
\hfill
\caption{\label{fig:absorber_profiles_monomial} The radial profiles of $E$ and $|F|$, for a spherical absorber with monomial opacity profile $\kappa = \kappa_0 (R/r)^m$ and uniform incoming radiation from infinity. We adopt $\kappa_0=1, R=1$ with arbitrary units. The black solid, blue dashed, and red dot-dashed lines show the exact solution, OTVET, and M1 respectively. From left to right, $m=2,4,$ and $11$. Magenta long dashed lines represent the exact solution for the limiting case of a sharp absorber with $m=\infty$. In all cases, the fluxes going into the absorber predicted by OTVET and M1 are about $>2\times$ the flux of the exact solution.}
\end{figure}

Figure~\ref{fig:absorber_profiles_monomial} shows the $E$ and $|F|$ profiles for the exact solution (black solid lines), OTVET (blue dashed lines), and M1 (red dot-dashed lines), using $\kappa_0=1, R=1$ with arbitrary units. From left to right, $m=2,4,$ and $11$. Magenta long dashed lines represent the exact solution for the limiting case of a sharp absorber with $m=\infty$. In all cases, OTVET and M1 result in $>2\times$ as much flux into the absorber than the exact solution, while the $E$ profiles are flatter. In the limit of $m\to\infty$, the OTVET and M1 fluxes are roughly $\sim0.5(R/r)^2$, which is twice the flux of the exact solution.

\begin{figure}[tbp]
\centering 
\includegraphics[width=\textwidth]{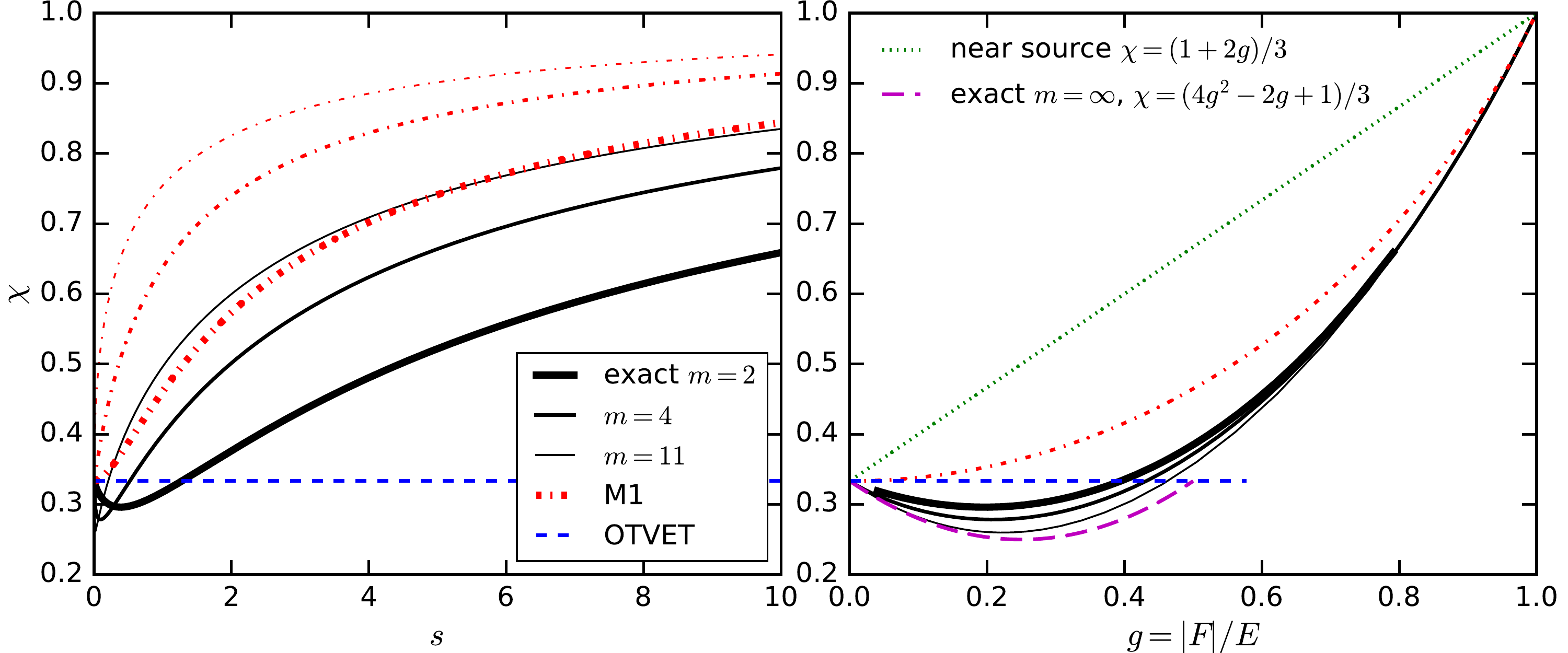}
\hfill
\caption{\label{fig:chi_monomial} $\chi$ (the radial component of the Eddington tensor $h_{rr}$) as a function of $s=\int_\infty^r \kappa\; \mathrm{d}r$ (left) and as a function of $g=|\vec{F}|/E$, for a spherical absorber with monomial opacity profile $\kappa = \kappa_0 (R/r)^m$ and uniform incoming radiation from infinity. The black solid, blue dashed, and red dot-dashed lines show the exact solution, OTVET, and M1 respectively. Lines from thick and thin represent $m=2,4,$ and $11$, respectively. The magenta long dashed line in the right panel show the exact solution for the limiting case of a sharp absorber with $m=\infty$. The green dotted line represent $\chi(g)$ near a point source in a uniform radiation background.}
\end{figure}

Figure~\ref{fig:chi_monomial} shows $\chi(s)$ (left panel) and $\chi(g)$ (right panel) for the exact solution, OTVET, and M1, for $m=2,4,11$. Notably, the exact solution does not show a monotonic $\chi(s)$ or $\chi(g)$ as M1 does. Especially, the limiting case of sharp absorber with $m=\infty$ (magenta dot-dashed line in the right panel) indicate that seeing half of the sky is the same as seeing the full sky, yielding isotropic Eddington tensors. In this case $\chi(g) = (4g^2-2g+1)/3$. In the opposite case where an observer is close to an ionizing source in an otherwise uniform ionizing background, $\chi(g)=(1+2g)/3$, shown as the green dotted line in the right panel. The monotonic behavior of the M1 $\chi(g)$ seems to fit this scenario better than the case of near an absorber. These two cases thus show opposite trend of $\chi(g)$.

\subsubsection{Opacity profile that accounts for ionization}

The section above assumes a monomial $\kappa$ profile, which is not consistent with an absorber exposed to an ionizing background. If the density profile of the absorber goes as $r^{-m}$, at large $r$ we expect $n_{\rm HI} \propto n^2_{\rm H}$, so $\kappa \propto r^{-2m}$. At small $r$ the absorber has self-shielded from the outside ionizing background, so $n_{\rm HI} \propto n_{\rm H}$ and $\kappa \propto r^{-m}$. In this section we thus include ionization of the absorber and calculate the $E$, $F$, and $x_{\rm HI}$ profiles.

We assume the absorber has a monomial density profile which goes as $r^{-m}$. Radiation comes uniformly from infinity and illuminate the absorber, which in the end reaches photoionization equilibrium. We use an iterative approach to calculate the $x_{\rm HI}$ profile and $E$ and $F$ profiles. We first assume the absorber is optically thin and is illuminated by a uniform ionizing background. An $x_{\rm HI}$ profile is calculated assuming photoionization equilibrium and used to update the $\kappa$ profile. We then use the updated $\kappa$ profile to calculate the $E$ profile, which updates the $x_{\rm HI}$ profile and thus the $\kappa$ profile again. We iterate until the fractional change in $x_{\rm HI}$ between iterations is less than $10^{-3}$ at each grid point. We use the case B recombination rate at $10^4$~K to calculate $x_{\rm HI}$, and include singly ionized helium so that $n_{\rm e} = 1.08 n_{\rm HII}$. We assume no temperature gradient inside the absorber, since $10^4$~K is roughly the equilibrium temperature of dense gas where atomic cooling balances photoheating.

For the exact solution, $E$ in each iteration is calculated by integrating the optical depth at each polar angle from infinity. For OTVET, we solve for $g$ using equation~\ref{eq:spherical_dgds} and an root finding algorithm. We first set up an evenly distributed grid of $\log(s)$ values. At each grid point of $\log(s)$, we calculate the left-hand side of equation~\ref{eq:spherical_dgds} by centered finite difference and the right-hand side using the values at the grid point. Boundary conditions $g=1/\sqrt{3}$ at the largest $s$ and $g(s)=0$ are implemented when calculating the central finite difference. We call scipy.optimize.root to find the solution of $g$ in each iteration. It is then fed into equation~\ref{eq:spherical_dlogEds} to calculate $E$.

For M1, we work out the value of $\dg/\ds$ at the singularity to be
\begin{equation}
\frac{\dg}{\ds} = \frac{2}{25}\left( -2 + \sqrt{2 \left( 2+3\frac{\mathrm{d}\kappa r}{\ds} \right)} \right).
\end{equation}
This special point is then used as starting point for integrating equations~\ref{eq:spherical_dgds} and \ref{eq:spherical_dlogEds}.  Specifically, given a $\kappa$ profile, we first find the relation $s(r) = \int_\infty^r \kappa(r')\; \mathrm{d}r'$ and the $s$ value corresponding to $\kappa r = 4/\sqrt{3}$.  The value of $\mathrm{d}\kappa r/\ds$ at the singularity is then calculated by interpolating over a grid of $\mathrm{d}(\kappa r)/\ds$ values computed by central finite difference, using the position of the singularity.

% \subsubsection{Single source at origin}
% \label{sec:stromgren_appendix}

% Assuming spherical geometry with a single source at the origin, we derive solutions for $\kappa \propto (r/R)^m$. In the limit of $m\to\infty$, this problem corresponds to a sharp absorbing wall at $r=R$ and therefore mimics the situation at the I-front of an expanding HII region.

% \begin{itemize}

% \item Exact solution:

% \begin{gather}
% \tau(r) = \int_0^r \kappa(r)\dr = \frac{\kappa r}{m+1} \\
% E = F = \frac{\exp(-\tau)}{r^2}
% \end{gather}

% \item M1 and OTVET

% To re-derive equations~\ref{eq:spherical_dlogEds} and \ref{eq:spherical_dgds}, note that $g=F/E$ and we use $\ds/\dr=\kappa$. The equations for $\mathrm{d} \log E / \ds$ and $\dg/\ds$ are then exactly the same as equations~\ref{eq:spherical_dlogEds} and \ref{eq:spherical_dgds}.
% When $g=1, \chi=1$, $E=F=s^{-2/(m+1)}\exp(-s)=\exp(-\tau)/r^2$. We recover the exact solution.

% For M1, if BC at $s=0$ is $g=1$, then we get $g=1$ at all $s$. Then M1 gives the exact solution.

% For OTVET, if again isotropic, then the solution is still $E=s^\nu K_\nu(s)$, but with $\nu = (m-1)/2(m+1)$, where $s = \sqrt{3} \kappa r / (m+1)$. Then it seems $g=0$ at $s=0$, which is unphysical?

% \end{itemize}

\section{Fourier transform of the Eddington tensor}
\label{sec:FT_deltahij}

\subsection{The exact Eddington tensor with a spatially uniform opacity}

Here we verify that in the case of a spatially uniform $\kappa$, plugging the exact Eddington tensor into equation~\ref{eq:deltaE_hij} recovers the exact solution of the radiative transfer equation (eqn.~\ref{eq:deltaE_exact}), as consistency requires. Results obtained in this section can also be used to calculate $\fhmn$ for OTVET, as shown in the next section. To evaluate the Fourier transform of the exact Eddington tensor using equation~\ref{eq:Pij_exact}, we need to calculate the Fourier transform of the kernel
\begin{equation}
G_{mn} = e^{-a\kappa r} \frac{x_m x_n}{r^4}.
\label{eq:Gij}
\end{equation}
It is convenient to work in a rotated basis where $\vec{k}$ aligns with the $z$ axis and then rotate back to an arbitrary catesian basis.
%Then to calculate $\widetilde{G}_{mn}$ in arbitrary coordinates, we multiply the expression for $\widetilde{G}_{mn}$ by the rotation matrix $A$: $A^T \widetilde{G}_{mn} A$.
In the rotated basis, the $[1,1], [2,2]$ components of $\widetilde{G}_{mn}$ equal
\begin{equation}
\pi \int_0^\infty r^2\; \dr\; \int_{-1}^1 e^{ikr\mu - a\kappa r} 
\frac{1-\mu^2}{r^2}\; \mathrm{d}\mu = 2\pi \frac{-ka\kappa + (k^2+a^2\kappa^2)\arctan(k/[a\kappa])}{k^3},
\end{equation}
where $\pi$ comes from integrating $\sin^2\phi$ or $\cos^2\phi$, and $\mu = \cos\theta$. The $[3,3]$ component of $\widetilde{G}_{mn}$ is
\begin{equation}
2\pi \int_0^\infty r^2\; \dr\; \int_{-1}^1 e^{ikr\mu - a\kappa r} 
\frac{\mu^2}{r^2}\; \mathrm{d}\mu = 4\pi \frac{ka\kappa - a^2\kappa^2 \arctan(k/[a\kappa])}{k^3}.
\end{equation}
These sum to a trace of
\begin{equation}
{\rm Tr}(\widetilde{G}_{mn}) = 4\pi \frac{\arctan(k/[a\kappa])}{k},
\end{equation}
which is the Fourier transform of $\exp(-a\kappa r)/r^2$ and invariant to our chosen coordinates.

%In the $\kappa=0$ (OTVET) limit, the $[1,1], [2,2]$ components of $\widetilde{G}_{mn}$ becomes $\pi^2/k$, and the $[3,3]$ component is 0. The trace becomes $2\pi^2/k$.

%The rotation matrix $A$ can be written as the multiplication of two other rotation matrices, i.e. $A = A_2 A_1$:
%\begin{gather}
%A_1 =
%\begin{pmatrix}
%-\frac{k_2}{\sqrt{k_1^2+k_2^2}} & \frac{k_1}{\sqrt{k_1^2+k_2^2}} & 0 \\
%-\frac{k_1}{\sqrt{k_1^2+k_2^2}} & -\frac{k_2}{\sqrt{k_1^2+k_2^2}} & 0 \\
%0 & 0 & 1
%\end{pmatrix}\quad
%A_2 =
%\begin{pmatrix}
%1 & 0 & 0 \\
%0 & \frac{k_3}{\sqrt{k_1^2+k_2^2+k_3^2}} & \frac{\sqrt{k_1^2+k_2^2}}{\sqrt{k_1^2+k_2^2+k_3^2}} \\
%0 & -\frac{\sqrt{k_1^2+k_2^2}}{\sqrt{k_1^2+k_2^2+k_3^2}} & \frac{k_3}{\sqrt{k_1^2+k_2^2+k_3^2}}
%\end{pmatrix}
%\end{gather}

%Then we calculate $(A_2 A_1)^T \widetilde{G}_{mn} A_2 A_1$. 
Rotating $G_{mn}$ back to a coordinate system where the $z$ axis can point in an arbitrary direction (such that $\vec{k}\cdot\vec{r} = kr (\sin\theta \cos\phi, \sin\theta \cos\phi, \cos\theta)$) yields the general expression for $\widetilde{G}_{mn}$, namely
\begin{equation}
\widetilde{G}_{mn} = \frac{2\pi}{k^5} \left( -\left(\delta^K_{mn} k^2 - 3 k_m k_n \right) k a\kappa + \left[ \left(\delta^K_{mn} k^2 - k_m k_n \right) k^2 + \left(\delta^K_{mn} k^2 - 3 k_m k_n \right) a^2\kappa^2 \right] \arctan \left(\frac{k}{a\kappa} \right) \right).
\label{eq:fGij_full}
\end{equation}

These equations can give an expression for $\widetilde{P}_{mn}$. In $x$-space, we have
\begin{equation}
P_{mn}(\vec{x})  \equiv \overbrace{\frac{1}{3}{\rm Tr}(\bar{P}_{mn}) \delta^K_{mn}}^{\bar{P}_{mn}}+  \overbrace{{\delta}_j \star {G}_{mn}}^{\delta P_{mn}}(\vec{x}),
\end{equation}
since the radiation field in the unperturbed background is isotropic and the second convolution owes to the anisotropy. Since $h_{mn} = P_{mn}/{\rm Tr}(P_{mn})$, we expand $P_{mn}$ to first order to obtain an expression for $
\fhmn$:
\begin{gather*}
h_{mn} = \frac{\bar{P}_{mn} + \delta P_{mn}}{{\rm Tr}(\bar{P}_{mn}) + {\rm Tr}(\delta P_{mn})} =  \bar{h}_{mn} + \delta h_{mn}, \\
{\rm where~~~~} \bar{h}_{mn} = \frac{\bar{P}_{mn}}{{\rm Tr}(\bar{P}_{mn})} = \frac{\delta^K_{mn}}{3};\quad
\delta h_{mn} = \frac{\delta P_{mn}}{{\rm Tr}(\bar{P}_{mn})} - \frac{\bar{P}_{mn}}{{\rm Tr}(\bar{P}_{mn})^2}{\rm Tr}(\delta P_{mn}) \numberthis.
\end{gather*}
Thus in Fourier space, the fluctuating part of the Eddington tensor is
\begin{gather*}
\fhmn = \frac{\widetilde{\delta P}_{mn}}{{\rm Tr}(\bar{P}_{mn})} - \frac{\delta^K_{mn}}{3}\frac{{\rm Tr}(\widetilde{\delta P}_{mn})}{{\rm Tr}(\bar{P}_{mn})} = \frac{\fdj \widetilde{G}_{mn}}{\int {\rm Tr}(G_{mn})\; \mathrm{d}^3 \vec{r}} - \frac{\delta^K_{mn}}{3}\frac{\fdj {\rm Tr}(\widetilde{G}_{mn})}{\int {\rm Tr}(G_{mn})\; \mathrm{d}^3 \vec{r}}. \numberthis
\label{eq:fhij_full}
\end{gather*}

Plugging the above expression into equation~\ref{eq:deltaE_hij}, we have
\begin{equation}
\frac{k_m k_n}{a^2\kappa^2}\fhmn = \fdj \left( 1 - \left( 1 + \frac{k^2}{3a^2\kappa^2} \right) \frac{a\kappa}{k}\arctan(k/[a\kappa]) \right),
\end{equation}
where we have used $\int {\rm Tr}(G_{mn})\; \mathrm{d}^3 \vec{r} = 4\pi / [a\kappa]$.
Taking the $\kappa$ bias factors to be 0 as appropriate in the case where this derivation applies of a uniform mean free path, we thus recover equation~\ref{eq:deltaE_exact}.

\subsection{OTVET}

The above derivation can be used to obtain an expression for OTVET directly, since OTVET takes the limit $\kappa \rightarrow 0$ in equation~\ref{eq:Gij}. The Kernel function thus becomes
\begin{equation}
\widetilde{G}^{\rm OTVET}_{mn} = \frac{\pi^2}{k} \left( \delta^K_{mn} - \frac{k_m k_n}{k^2} \right).
\end{equation}

However, the integration of ${\rm Tr}(G_{mn})$ over all space is divergent. Therefore $\fhmn=0$ in equation~\ref{eq:fhij_full}, and the Eddington tensor becomes isotropic. In the simulations of \cite{gnedin14}, this Olber's paradox is solved by only including the image sources out to half a box size away from a given location in the simulation volume, when calculating $h_{mn}$. Specifically, to calculate equation~\ref{eq:Pij_otvet} in a periodic box, a grid of $G_{mn}$ values is set up in $x$-space with a wrap-around the box to take into account the periodic boundary condition. One then performs discrete Fourier transform (DFT) to obtain $\widetilde{G}_{mn}$ in $k$-space. $P_{mn}$ is calculated by the inverse DFT of the product of $\widetilde{G}_{mn}$ and $\tilde{j}$, which is the DFT of the source grid. This approach ensures that $P_{mn}$, and thus $h_{mn}$, is exact out to half a box size away when there is only one source.
% \footnote{Another way of computing $P_{mn}$ is to directly set up a grid of $\widetilde{G}_{mn}$ in $k$ space using an analytic formula for $\widetilde{G}_{mn}$. This leads to a divergent zeroth mode in the optically thin limit, because an observer in a simulation box sees an infinite number of the periodic images of the sources. The Eddington tensor thus become isotropic owing to the Olber's paradox. One can suppress the zeroth mode, for instance by numerically evaluating $\int G_{mn}(x) \mathrm{d}\vec{x}$ only within the simulation box. This choice is a bit arbitrary and not self-consistent, and leads to $h_{mn}$ having a non-zero eigenvalue that allows radiation to propagate in the tangential direction in the case of a single source \citep[Fig.~5 of][]{gnedin01}. This approach is thus not used anymore. On the contrary, using the DFT of a $G_{mn}$ grid ensures that there is only one non-zero eigenvalue with eigenvectors pointing in the radial direction.}

The above approach of calculating $h_{mn}$ also implies that $\int {\rm Tr}(G_{mn})\; \mathrm{d}^3 \vec{r}$ is only carried out within the box size limit. To calculate $L_{\rm O} = \int {\rm Tr}(G_{mn})\; \mathrm{d}^3 \vec{r} = \int [1/r^2]\; \mathrm{d}^3 \vec{r}$, we numerically evaluate $8 \times \int_{[0,L/2]^3} 1/(x^2+y^2+z^2)\; \mathrm{d}x\mathrm{d}y\mathrm{d}z$, where $L$ is the box size. If we ignore changes in $\widetilde{G}_{mn}$ owing to convolution with the window function, which arises from DFT of the simulation box, then we get
\begin{equation}
\frac{k_m k_n}{a^2\kappa^2}\fhmn = -\fdj \frac{2\pi^2 k}{3a^2\kappa^2 L_{\rm O}}.
\end{equation}
This formula should hold for high wavenumbers, i.e. $k\gg2\pi/L$. This gives the OTVET solution for $\fdE$, equation~\ref{eq:deltaE_otvet}.

% Clearly, larger box sizes result in more isotropic Eddington tensor and the transfer function tends to the isotropic solution of $T^{\rm ISO} = 1/(1+k^2/3a^2\kappa^2)$. The above approach of calculating $h_{mn}$ also implies that the photon mean free path of a simulation is capped by the box size $L$. Moreover, how isotropic the Eddington tensor is depends on $L$. For instance, for a physical value of $\lambda_{\rm mfp}=80$~comoving Mpc, using $L=20$~Mpc enlarges the proximity region by a factor of 2, while adopting $L=1000$~Mpc shrinks it by a factor of $3.5$.

\section{Taking higher angular moments of the RT equation}
\label{sec:arctan_higher_order}

Section~\ref{sec:Gamma_fluc} showed that closing the moment equations at the second moment and assuming an isotropic Eddington tensor leads to significant underestimation of ionizing background fluctuations at $k>a\meankappa$. At $k \ll a\meankappa$, the isotropic solution recovers the exact solution, because $\arctan(x) \approx x - x^3/3$ for small $x$. This motivates us to close the moment equations at higher order.

Consider the Fourier transform of equation~\ref{eq:transformthis_f}:
\begin{equation}
-\frac{i}{a\meankappa} \vnn \cdot \vec{k} \fdf = \fdj - \bkd\fdd - \bkg\fdg - \fdf.
\end{equation}
The goal is to expand $\fdf$ with Legendre polynomials. We therefore define $I_\ell = 1/2\; \int_{-1}^1 \fdf P_\ell(\mu)\; \mathrm{d}\mu$, where $\mu = \vnn \cdot \hat{\vec{k}}$. Multiplying the above equation by $P_\ell(\mu)/2$ and integrating over $\mu$, we get:
\begin{gather*}
-\frac{ik}{a\meankappa} \frac{1}{2} \int \fdf \mu P_\ell(\mu)\; \mathrm{d}\mu = (\fdj - \bkd\fdd - \bkg\fdg) \delta^K_{\ell,0} - I_\ell; \\
-\frac{ik}{a\meankappa} \frac{\ell}{2\ell+1} I_{\ell-1} + I_\ell -\frac{ik}{a\meankappa} \frac{\ell+1}{2\ell+1} I_{\ell+1} = \delta^K_{\ell,0} (\fdj - \bkd\fdd - \bkg\fdg), \numberthis
\end{gather*}
where in the second line we have used the recurrence relation $\mu P_\ell(\mu) = \ell/(2\ell+1)\; P_{\ell-1}(\mu) + (\ell+1)/(2\ell+1)\; P_{\ell+1}(\mu)$. Fluctuations in the photoionization rate (and $E$) is thus given by $I_0$, and note that $\fdg = I_0$.

\begin{figure}[tbp]
\centering 
\includegraphics[width=.7\textwidth]{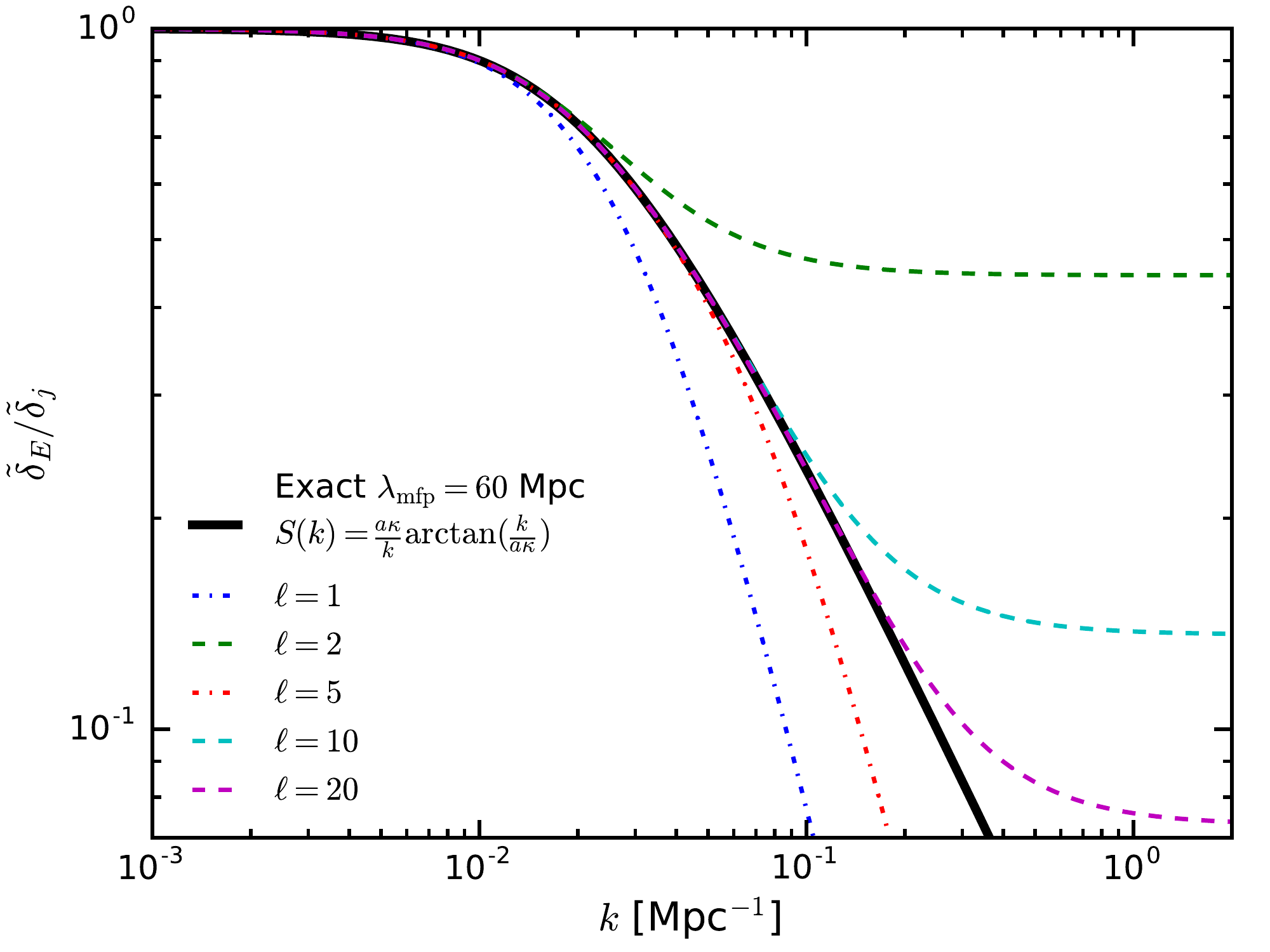}
\hfill
\caption{\label{fig:higher_order_solution} ``Transfer functions'' $\fdE/\fdj$ when closing the moment equations at angular multiple orders, for the uniform mean free path problem with $\lambda_{\rm mfp}=60$~Mpc. The black solid line shows the exact solution $S(k)$. The blue, green, red, cyan, and magenta lines represent the solution when closing at order $\ell=1, 2, 5, 10, 20$, respectively.}
\end{figure}

To close the above moment equations at order $\ell$, we set the next higher Legendre moment to zero (i.e. $I_{\ell+1}=0$). The matrix equations can be solved analytically\footnote{The inverse of a tridiagonal matrix can be calculated using a recurrence relation \url{https://en.wikipedia.org/wiki/Tridiagonal_matrix}.}. Focusing on the uniform mean free path case with $\lambda_{\rm mfp}=60$~Mpc, Figure~\ref{fig:higher_order_solution} shows the ``transfer functions'' $\fdE/\fdj$, when closing the moment equations at order $\ell=1,2,5,10,20$, illustrated by the blue, green, red, cyan, and magenta lines respectively. The $\ell=1$ order solution corresponds to the M1/isotropic solution. The black solid line shows the exact solution $S(k)$. 
A rough estimate for the wavenumber at which the $\ell$-th order solution starts deviating from the exact solution is $k \sim \ell/\lambda_{\rm mfp}$, since information on angular scales $\lesssim \pi / \ell$ is not captured when closing the equations at order $\ell$. The convergence to the exact solution is thus very slow, and capturing the post-reionization ionizing background fluctuations accurately requires following the radiation field on very small angular scales.

\paragraph{Note added.} %This is also a good position for notes added
% after the paper has been written.

% The bibliography will probably be heavily edited during typesetting.
% We'll parse it and, using the arxiv number or the journal data, will
% query inspire, trying to verify the data (this will probalby spot
% eventual typos) and retrive the document DOI and eventual errata.
% We however suggest to always provide author, title and journal data:
% in short all the informations that clearly identify a document.

% \begin{thebibliography}{99}

% \bibitem{a}
% Author, \emph{Title}, \emph{J. Abbrev.} {\bf vol} (year) pg.

% \bibitem{b}
% Author, \emph{Title},
% arxiv:1234.5678.

% \bibitem{c}
% Author, \emph{Title},
% Publisher (year).

% % Please avoid comments such as "For a review'', "For some examples",
% % "and references therein" or move them in the text. In general,
% % please leave only references in the bibliography and move all
% % accessory text in footnotes.

% % Also, please have only one work for each \bibitem.

% \end{thebibliography}

\bibliography{References}

\providecommand{\href}[2]{#2}\begingroup\raggedright\begin{thebibliography}{100}

\bibitem{trac11}
H.~Y. {Trac} and N.~Y. {Gnedin}, \emph{{Computer Simulations of Cosmic
  Reionization}}, \href{https://doi.org/10.1166/asl.2011.1214}{\emph{Advanced
  Science Letters} {\bfseries 4} (2011) 228}
  [\href{https://arxiv.org/abs/0906.4348}{{\ttfamily 0906.4348}}].

\bibitem{mcquinn16}
M.~{McQuinn}, \emph{{The Evolution of the Intergalactic Medium}},
  \href{https://doi.org/10.1146/annurev-astro-082214-122355}{\emph{\araa}
  {\bfseries 54} (2016) 313}
  [\href{https://arxiv.org/abs/1512.00086}{{\ttfamily 1512.00086}}].

\bibitem{gnedin14}
N.~Y. {Gnedin}, \emph{{Cosmic Reionization on Computers. I. Design and
  Calibration of Simulations}},
  \href{https://doi.org/10.1088/0004-637X/793/1/29}{\emph{\apj} {\bfseries 793}
  (2014) 29} [\href{https://arxiv.org/abs/1403.4245}{{\ttfamily 1403.4245}}].

\bibitem{chardin15}
J.~{Chardin}, M.~G. {Haehnelt}, D.~{Aubert} and E.~{Puchwein},
  \emph{{Calibrating cosmological radiative transfer simulations with Ly
  {\ensuremath{\alpha}} forest data: evidence for large spatial UV background
  fluctuations at z {\ensuremath{\sim}} 5.6-5.8 due to rare bright sources}},
  \href{https://doi.org/10.1093/mnras/stv1786}{\emph{\mnras} {\bfseries 453}
  (2015) 2943} [\href{https://arxiv.org/abs/1505.01853}{{\ttfamily
  1505.01853}}].

\bibitem{chardin17}
J.~{Chardin}, E.~{Puchwein} and M.~G. {Haehnelt}, \emph{{Large-scale opacity
  fluctuations in the Ly{\ensuremath{\alpha}} forest: evidence for QSOs
  dominating the ionizing UV background at z {\ensuremath{\sim}} 5.5-6?}},
  \href{https://doi.org/10.1093/mnras/stw2943}{\emph{\mnras} {\bfseries 465}
  (2017) 3429} [\href{https://arxiv.org/abs/1606.08231}{{\ttfamily
  1606.08231}}].

\bibitem{chardin18a}
J.~{Chardin}, G.~{Kulkarni} and M.~G. {Haehnelt}, \emph{{Self-shielding of
  hydrogen in the IGM during the epoch of reionization}},
  \href{https://doi.org/10.1093/mnras/sty992}{\emph{\mnras} {\bfseries 478}
  (2018) 1065} [\href{https://arxiv.org/abs/1707.06993}{{\ttfamily
  1707.06993}}].

\bibitem{chardin18b}
J.~{Chardin}, M.~G. {Haehnelt}, S.~E.~I. {Bosman} and E.~{Puchwein}, \emph{{A
  tale of seven narrow spikes and a long trough: constraining the timing of the
  percolation of H II bubbles at the tail end of reionization with ULAS
  J1120+0641}}, \href{https://doi.org/10.1093/mnras/stx2362}{\emph{\mnras}
  {\bfseries 473} (2018) 765}
  [\href{https://arxiv.org/abs/1707.03841}{{\ttfamily 1707.03841}}].

\bibitem{keating18}
L.~C. {Keating}, E.~{Puchwein} and M.~G. {Haehnelt}, \emph{{Spatial
  fluctuations of the intergalactic temperature-density relation after hydrogen
  reionization}}, \href{https://doi.org/10.1093/mnras/sty968}{\emph{\mnras}
  {\bfseries 477} (2018) 5501}
  [\href{https://arxiv.org/abs/1709.05351}{{\ttfamily 1709.05351}}].

\bibitem{keating19}
L.~C. {Keating}, G.~{Kulkarni}, M.~G. {Haehnelt}, J.~{Chardin} and D.~{Aubert},
  \emph{{Constraining the second half of reionization with the Lyman-$\beta$
  forest}}, {\emph{arXiv e-prints} (2019) arXiv:1912.05582}
  [\href{https://arxiv.org/abs/1912.05582}{{\ttfamily 1912.05582}}].

\bibitem{keating20}
L.~C. {Keating}, L.~H. {Weinberger}, G.~{Kulkarni}, M.~G. {Haehnelt},
  J.~{Chardin} and D.~{Aubert}, \emph{{Long troughs in the
  Lyman-{\ensuremath{\alpha}} forest below redshift 6 due to islands of neutral
  hydrogen}}, \href{https://doi.org/10.1093/mnras/stz3083}{\emph{\mnras}
  {\bfseries 491} (2020) 1736}
  [\href{https://arxiv.org/abs/1905.12640}{{\ttfamily 1905.12640}}].

\bibitem{kulkarni19}
G.~{Kulkarni}, L.~C. {Keating}, M.~G. {Haehnelt}, S.~E.~I. {Bosman},
  E.~{Puchwein}, J.~{Chardin} et~al., \emph{{Large Ly {\ensuremath{\alpha}}
  opacity fluctuations and low CMB {\ensuremath{\tau}} in models of late
  reionization with large islands of neutral hydrogen extending to z \&lt;
  5.5}}, \href{https://doi.org/10.1093/mnrasl/slz025}{\emph{\mnras} {\bfseries
  485} (2019) L24} [\href{https://arxiv.org/abs/1809.06374}{{\ttfamily
  1809.06374}}].

\bibitem{roy20}
A.~{Roy}, G.~{Kulkarni}, P.~D. {Meerburg}, A.~{Challinor}, C.~{Baccigalupi},
  A.~{Lapi} et~al., \emph{{Revised estimates of CMB $B$-mode polarization
  induced by patchy reionization}}, {\emph{arXiv e-prints} (2020)
  arXiv:2004.02927} [\href{https://arxiv.org/abs/2004.02927}{{\ttfamily
  2004.02927}}].

\bibitem{wu19a}
X.~{Wu}, R.~{Kannan}, F.~{Marinacci}, M.~{Vogelsberger} and L.~{Hernquist},
  \emph{{Simulating the effect of photoheating feedback during reionization}},
  \href{https://doi.org/10.1093/mnras/stz1726}{\emph{\mnras} {\bfseries 488}
  (2019) 419} [\href{https://arxiv.org/abs/1903.06167}{{\ttfamily
  1903.06167}}].

\bibitem{wu19b}
X.~{Wu}, M.~{McQuinn}, R.~{Kannan}, A.~{D'Aloisio}, S.~{Bird}, F.~{Marinacci}
  et~al., \emph{{Imprints of temperature fluctuations on the z ̃ 5
  Lyman-{\ensuremath{\alpha}} forest: a view from radiation-hydrodynamic
  simulations of reionization}},
  \href{https://doi.org/10.1093/mnras/stz2807}{\emph{\mnras} {\bfseries 490}
  (2019) 3177} [\href{https://arxiv.org/abs/1907.04860}{{\ttfamily
  1907.04860}}].

\bibitem{aubert10}
D.~{Aubert} and R.~{Teyssier}, \emph{{Reionization Simulations Powered by
  Graphics Processing Units. I. On the Structure of the Ultraviolet Radiation
  Field}}, \href{https://doi.org/10.1088/0004-637X/724/1/244}{\emph{\apj}
  {\bfseries 724} (2010) 244}
  [\href{https://arxiv.org/abs/1004.2503}{{\ttfamily 1004.2503}}].

\bibitem{ocvirk16}
P.~{Ocvirk}, N.~{Gillet}, P.~R. {Shapiro}, D.~{Aubert}, I.~T. {Iliev},
  R.~{Teyssier} et~al., \emph{{Cosmic Dawn (CoDa): the First
  Radiation-Hydrodynamics Simulation of Reionization and Galaxy Formation in
  the Local Universe}},
  \href{https://doi.org/10.1093/mnras/stw2036}{\emph{\mnras} {\bfseries 463}
  (2016) 1462} [\href{https://arxiv.org/abs/1511.00011}{{\ttfamily
  1511.00011}}].

\bibitem{ocvirk18}
P.~{Ocvirk}, D.~{Aubert}, J.~G. {Sorce}, P.~R. {Shapiro}, N.~{Deparis},
  T.~{Dawoodbhoy} et~al., \emph{{Cosmic Dawn II (CoDa II): a new
  radiation-hydrodynamics simulation of the self-consistent coupling of galaxy
  formation and reionization}}, {\emph{arXiv e-prints} (2018) arXiv:1811.11192}
  [\href{https://arxiv.org/abs/1811.11192}{{\ttfamily 1811.11192}}].

\bibitem{rosdahl18}
J.~{Rosdahl}, H.~{Katz}, J.~{Blaizot}, T.~{Kimm}, L.~{Michel-Dansac},
  T.~{Garel} et~al., \emph{{The SPHINX cosmological simulations of the first
  billion years: the impact of binary stars on reionization}},
  \href{https://doi.org/10.1093/mnras/sty1655}{\emph{\mnras} {\bfseries 479}
  (2018) 994} [\href{https://arxiv.org/abs/1801.07259}{{\ttfamily
  1801.07259}}].

\bibitem{katz18}
H.~{Katz}, T.~{Kimm}, M.~{Haehnelt}, D.~{Sijacki}, J.~{Rosdahl} and
  J.~{Blaizot}, \emph{{A Census of the LyC photons that form the UV background
  during reionization}},
  \href{https://doi.org/10.1093/mnras/sty1225}{\emph{\mnras} {\bfseries 478}
  (2018) 4986} [\href{https://arxiv.org/abs/1802.01586}{{\ttfamily
  1802.01586}}].

\bibitem{so14}
G.~C. {So}, M.~L. {Norman}, D.~R. {Reynolds} and J.~H. {Wise}, \emph{{Fully
  Coupled Simulation of Cosmic Reionization. II. Recombinations, Clumping
  Factors, and the Photon Budget for Reionization}},
  \href{https://doi.org/10.1088/0004-637X/789/2/149}{\emph{\apj} {\bfseries
  789} (2014) 149} [\href{https://arxiv.org/abs/1311.2152}{{\ttfamily
  1311.2152}}].

\bibitem{norman15}
M.~L. {Norman}, D.~R. {Reynolds}, G.~C. {So}, R.~P. {Harkness} and J.~H.
  {Wise}, \emph{{Fully Coupled Simulation of Cosmic Reionization. I. Numerical
  Methods and Tests}},
  \href{https://doi.org/10.1088/0067-0049/216/1/16}{\emph{\apjs} {\bfseries
  216} (2015) 16} [\href{https://arxiv.org/abs/1306.0645}{{\ttfamily
  1306.0645}}].

\bibitem{norman18}
M.~L. {Norman}, P.~{Chen}, J.~H. {Wise} and H.~{Xu}, \emph{{Fully Coupled
  Simulation of Cosmic Reionization. III. Stochastic Early Reionization by the
  Smallest Galaxies}},
  \href{https://doi.org/10.3847/1538-4357/aae30b}{\emph{\apj} {\bfseries 867}
  (2018) 27} [\href{https://arxiv.org/abs/1705.00026}{{\ttfamily 1705.00026}}].

\bibitem{petkova11}
M.~{Petkova} and V.~{Springel}, \emph{{Simulations of galaxy formation with
  radiative transfer: hydrogen reionization and radiative feedback}},
  \href{https://doi.org/10.1111/j.1365-2966.2010.17955.x}{\emph{\mnras}
  {\bfseries 412} (2011) 935}
  [\href{https://arxiv.org/abs/1008.4459}{{\ttfamily 1008.4459}}].

\bibitem{bauer15}
A.~{Bauer}, V.~{Springel}, M.~{Vogelsberger}, S.~{Genel}, P.~{Torrey},
  D.~{Sijacki} et~al., \emph{{Hydrogen reionization in the Illustris
  universe}}, \href{https://doi.org/10.1093/mnras/stv1893}{\emph{\mnras}
  {\bfseries 453} (2015) 3593}
  [\href{https://arxiv.org/abs/1503.00734}{{\ttfamily 1503.00734}}].

\bibitem{finlator09b}
K.~{Finlator}, F.~{{\"O}zel}, R.~{Dav{\'e}} and B.~D. {Oppenheimer}, \emph{{The
  late reionization of filaments}},
  \href{https://doi.org/10.1111/j.1365-2966.2009.15521.x}{\emph{\mnras}
  {\bfseries 400} (2009) 1049}
  [\href{https://arxiv.org/abs/0908.1576}{{\ttfamily 0908.1576}}].

\bibitem{finlator11}
K.~{Finlator}, R.~{Dav{\'e}} and F.~{{\"O}zel}, \emph{{Galactic Outflows and
  Photoionization Heating in the Reionization Epoch}},
  \href{https://doi.org/10.1088/0004-637X/743/2/169}{\emph{\apj} {\bfseries
  743} (2011) 169} [\href{https://arxiv.org/abs/1106.4321}{{\ttfamily
  1106.4321}}].

\bibitem{finlator12}
K.~{Finlator}, S.~P. {Oh}, F.~{{\"O}zel} and R.~{Dav{\'e}}, \emph{{Gas clumping
  in self-consistent reionization models}},
  \href{https://doi.org/10.1111/j.1365-2966.2012.22114.x}{\emph{\mnras}
  {\bfseries 427} (2012) 2464}
  [\href{https://arxiv.org/abs/1209.2489}{{\ttfamily 1209.2489}}].

\bibitem{finlator18}
K.~{Finlator}, L.~{Keating}, B.~D. {Oppenheimer}, R.~{Dav{\'e}} and
  E.~{Zackrisson}, \emph{{Reionization in Technicolor}},
  \href{https://doi.org/10.1093/mnras/sty1949}{\emph{\mnras} {\bfseries 480}
  (2018) 2628} [\href{https://arxiv.org/abs/1805.00099}{{\ttfamily
  1805.00099}}].

\bibitem{gnedin01}
N.~Y. {Gnedin} and T.~{Abel}, \emph{{Multi-dimensional cosmological radiative
  transfer with a Variable Eddington Tensor formalism}},
  \href{https://doi.org/10.1016/S1384-1076(01)00068-9}{\emph{\na} {\bfseries 6}
  (2001) 437} [\href{https://arxiv.org/abs/astro-ph/0106278}{{\ttfamily
  astro-ph/0106278}}].

\bibitem{levermore84}
C.~D. {Levermore}, \emph{{Relating Eddington factors to flux limiters.}},
  \href{https://doi.org/10.1016/0022-4073(84)90112-2}{\emph{\jqsrt} {\bfseries
  31} (1984) 149}.

\bibitem{fan06}
X.~{Fan}, M.~A. {Strauss}, R.~H. {Becker}, R.~L. {White}, J.~E. {Gunn}, G.~R.
  {Knapp} et~al., \emph{{Constraining the Evolution of the Ionizing Background
  and the Epoch of Reionization with z\raisebox{-0.5ex}\textasciitilde6
  Quasars. II. A Sample of 19 Quasars}},
  \href{https://doi.org/10.1086/504836}{\emph{\aj} {\bfseries 132} (2006) 117}
  [\href{https://arxiv.org/abs/astro-ph/0512082}{{\ttfamily
  astro-ph/0512082}}].

\bibitem{mcgreer15}
I.~D. {McGreer}, A.~{Mesinger} and V.~{D'Odorico}, \emph{{Model-independent
  evidence in favour of an end to reionization by z {\ensuremath{\approx}} 6}},
  \href{https://doi.org/10.1093/mnras/stu2449}{\emph{\mnras} {\bfseries 447}
  (2015) 499} [\href{https://arxiv.org/abs/1411.5375}{{\ttfamily 1411.5375}}].

\bibitem{bosman18}
S.~E.~I. {Bosman}, X.~{Fan}, L.~{Jiang}, S.~{Reed}, Y.~{Matsuoka}, G.~{Becker}
  et~al., \emph{{New constraints on Lyman-{\ensuremath{\alpha}} opacity with a
  sample of 62 quasarsat z \&gt; 5.7}},
  \href{https://doi.org/10.1093/mnras/sty1344}{\emph{\mnras} {\bfseries 479}
  (2018) 1055} [\href{https://arxiv.org/abs/1802.08177}{{\ttfamily
  1802.08177}}].

\bibitem{becker15}
G.~D. {Becker}, J.~S. {Bolton}, P.~{Madau}, M.~{Pettini}, E.~V. {Ryan-Weber}
  and B.~P. {Venemans}, \emph{{Evidence of patchy hydrogen reionization from an
  extreme Ly{\ensuremath{\alpha}} trough below redshift six}},
  \href{https://doi.org/10.1093/mnras/stu2646}{\emph{\mnras} {\bfseries 447}
  (2015) 3402} [\href{https://arxiv.org/abs/1407.4850}{{\ttfamily 1407.4850}}].

\bibitem{becker18}
G.~D. {Becker}, F.~B. {Davies}, S.~R. {Furlanetto}, M.~A. {Malkan}, E.~{Boera}
  and C.~{Douglass}, \emph{{Evidence for Large-scale Fluctuations in the
  Metagalactic Ionizing Background Near Redshift Six}},
  \href{https://doi.org/10.3847/1538-4357/aacc73}{\emph{\apj} {\bfseries 863}
  (2018) 92} [\href{https://arxiv.org/abs/1803.08932}{{\ttfamily 1803.08932}}].

\bibitem{meiksin20}
A.~{Meiksin}, \emph{{The influence of metagalactic ultraviolet background
  fluctuations on the high-redshift Ly{\ensuremath{\alpha}} forest}},
  \href{https://doi.org/10.1093/mnras/stz3395}{\emph{\mnras} {\bfseries 491}
  (2020) 4884} [\href{https://arxiv.org/abs/1912.00828}{{\ttfamily
  1912.00828}}].

\bibitem{gnedin17}
N.~Y. {Gnedin}, G.~D. {Becker} and X.~{Fan}, \emph{{Cosmic Reionization on
  Computers: Properties of the Post-reionization IGM}},
  \href{https://doi.org/10.3847/1538-4357/aa6c24}{\emph{\apj} {\bfseries 841}
  (2017) 26} [\href{https://arxiv.org/abs/1605.03183}{{\ttfamily 1605.03183}}].

\bibitem{zahn07}
O.~{Zahn}, A.~{Lidz}, M.~{McQuinn}, S.~{Dutta}, L.~{Hernquist},
  M.~{Zaldarriaga} et~al., \emph{{Simulations and Analytic Calculations of
  Bubble Growth during Hydrogen Reionization}},
  \href{https://doi.org/10.1086/509597}{\emph{\apj} {\bfseries 654} (2007) 12}
  [\href{https://arxiv.org/abs/astro-ph/0604177}{{\ttfamily
  astro-ph/0604177}}].

\bibitem{zahn11}
O.~{Zahn}, A.~{Mesinger}, M.~{McQuinn}, H.~{Trac}, R.~{Cen} and L.~E.
  {Hernquist}, \emph{{Comparison of reionization models: radiative transfer
  simulations and approximate, seminumeric models}},
  \href{https://doi.org/10.1111/j.1365-2966.2011.18439.x}{\emph{\mnras}
  {\bfseries 414} (2011) 727}
  [\href{https://arxiv.org/abs/1003.3455}{{\ttfamily 1003.3455}}].

\bibitem{furlanetto05}
S.~R. {Furlanetto} and S.~P. {Oh}, \emph{{Taxing the rich: recombinations and
  bubble growth during reionization}},
  \href{https://doi.org/10.1111/j.1365-2966.2005.09505.x}{\emph{\mnras}
  {\bfseries 363} (2005) 1031}
  [\href{https://arxiv.org/abs/astro-ph/0505065}{{\ttfamily
  astro-ph/0505065}}].

\bibitem{sobacchi14}
E.~{Sobacchi} and A.~{Mesinger}, \emph{{Inhomogeneous recombinations during
  cosmic reionization}},
  \href{https://doi.org/10.1093/mnras/stu377}{\emph{\mnras} {\bfseries 440}
  (2014) 1662} [\href{https://arxiv.org/abs/1402.2298}{{\ttfamily 1402.2298}}].

\bibitem{shukla16}
H.~{Shukla}, G.~{Mellema}, I.~T. {Iliev} and P.~R. {Shapiro}, \emph{{The
  effects of Lyman-limit systems on the evolution and observability of the
  epoch of reionization}},
  \href{https://doi.org/10.1093/mnras/stw249}{\emph{\mnras} {\bfseries 458}
  (2016) 135} [\href{https://arxiv.org/abs/1602.01144}{{\ttfamily
  1602.01144}}].

\bibitem{iliev06}
I.~T. {Iliev}, B.~{Ciardi}, M.~A. {Alvarez}, A.~{Maselli}, A.~{Ferrara}, N.~Y.
  {Gnedin} et~al., \emph{{Cosmological radiative transfer codes comparison
  project - I. The static density field tests}},
  \href{https://doi.org/10.1111/j.1365-2966.2006.10775.x}{\emph{\mnras}
  {\bfseries 371} (2006) 1057}
  [\href{https://arxiv.org/abs/astro-ph/0603199}{{\ttfamily
  astro-ph/0603199}}].

\bibitem{iliev09}
I.~T. {Iliev}, D.~{Whalen}, G.~{Mellema}, K.~{Ahn}, S.~{Baek}, N.~Y. {Gnedin}
  et~al., \emph{{Cosmological radiative transfer comparison project - II. The
  radiation-hydrodynamic tests}},
  \href{https://doi.org/10.1111/j.1365-2966.2009.15558.x}{\emph{\mnras}
  {\bfseries 400} (2009) 1283}
  [\href{https://arxiv.org/abs/0905.2920}{{\ttfamily 0905.2920}}].

\bibitem{petkova09}
M.~{Petkova} and V.~{Springel}, \emph{{An implementation of radiative transfer
  in the cosmological simulation code GADGET}},
  \href{https://doi.org/10.1111/j.1365-2966.2009.14843.x}{\emph{\mnras}
  {\bfseries 396} (2009) 1383}
  [\href{https://arxiv.org/abs/0812.1801}{{\ttfamily 0812.1801}}].

\bibitem{worseck14}
G.~{Worseck}, J.~X. {Prochaska}, J.~M. {O'Meara}, G.~D. {Becker}, S.~L.
  {Ellison}, S.~{Lopez} et~al., \emph{{The Giant Gemini GMOS survey of z$_{em}$
  \&gt; 4.4 quasars - I. Measuring the mean free path across cosmic time}},
  \href{https://doi.org/10.1093/mnras/stu1827}{\emph{\mnras} {\bfseries 445}
  (2014) 1745} [\href{https://arxiv.org/abs/1402.4154}{{\ttfamily 1402.4154}}].

\bibitem{meiksin18}
A.~{Meiksin} and M.~{McQuinn}, \emph{{Time-dependent fluctuations in the
  metagalactic photoionization background}},
  \href{https://doi.org/10.1093/mnras/sty2907}{\emph{\mnras} {\bfseries 482}
  (2019) 4777} [\href{https://arxiv.org/abs/1809.08645}{{\ttfamily
  1809.08645}}].

\bibitem{finlator09}
K.~{Finlator}, F.~{{\"O}zel} and R.~{Dav{\'e}}, \emph{{A new moment method for
  continuum radiative transfer in cosmological re-ionization}},
  \href{https://doi.org/10.1111/j.1365-2966.2008.14190.x}{\emph{\mnras}
  {\bfseries 393} (2009) 1090}
  [\href{https://arxiv.org/abs/0808.3578}{{\ttfamily 0808.3578}}].

\bibitem{levermore81}
C.~D. {Levermore}, \emph{{Relating Eddington factors to flux limiters.}},
  \href{https://doi.org/10.1016/0022-4073(84)90112-2}{\emph{\jqsrt} {\bfseries
  31} (1984) 149}.

\bibitem{aubert08}
D.~{Aubert} and R.~{Teyssier}, \emph{{A radiative transfer scheme for
  cosmological reionization based on a local Eddington tensor}},
  \href{https://doi.org/10.1111/j.1365-2966.2008.13223.x}{\emph{\mnras}
  {\bfseries 387} (2008) 295}
  [\href{https://arxiv.org/abs/0709.1544}{{\ttfamily 0709.1544}}].

\bibitem{rosdahl13}
J.~{Rosdahl}, J.~{Blaizot}, D.~{Aubert}, T.~{Stranex} and R.~{Teyssier},
  \emph{{RAMSES-RT: radiation hydrodynamics in the cosmological context}},
  \href{https://doi.org/10.1093/mnras/stt1722}{\emph{\mnras} {\bfseries 436}
  (2013) 2188} [\href{https://arxiv.org/abs/1304.7126}{{\ttfamily 1304.7126}}].

\bibitem{rosdahl15}
J.~{Rosdahl} and R.~{Teyssier}, \emph{{A scheme for radiation pressure and
  photon diffusion with the M1 closure in RAMSES-RT}},
  \href{https://doi.org/10.1093/mnras/stv567}{\emph{\mnras} {\bfseries 449}
  (2015) 4380} [\href{https://arxiv.org/abs/1411.6440}{{\ttfamily 1411.6440}}].

\bibitem{aubert15}
D.~{Aubert}, N.~{Deparis} and P.~{Ocvirk}, \emph{{EMMA: an adaptive mesh
  refinement cosmological simulation code with radiative transfer}},
  \href{https://doi.org/10.1093/mnras/stv1896}{\emph{\mnras} {\bfseries 454}
  (2015) 1012} [\href{https://arxiv.org/abs/1508.07888}{{\ttfamily
  1508.07888}}].

\bibitem{kannan19}
R.~{Kannan}, M.~{Vogelsberger}, F.~{Marinacci}, R.~{McKinnon}, R.~{Pakmor} and
  V.~{Springel}, \emph{{AREPO-RT: radiation hydrodynamics on a moving mesh}},
  \href{https://doi.org/10.1093/mnras/stz287}{\emph{\mnras} {\bfseries 485}
  (2019) 117} [\href{https://arxiv.org/abs/1804.01987}{{\ttfamily
  1804.01987}}].

\bibitem{abel02}
T.~{Abel} and B.~D. {Wandelt}, \emph{{Adaptive ray tracing for radiative
  transfer around point sources}},
  \href{https://doi.org/10.1046/j.1365-8711.2002.05206.x}{\emph{\mnras}
  {\bfseries 330} (2002) L53}
  [\href{https://arxiv.org/abs/astro-ph/0111033}{{\ttfamily
  astro-ph/0111033}}].

\bibitem{mellema06}
G.~{Mellema}, I.~T. {Iliev}, M.~A. {Alvarez} and P.~R. {Shapiro}, \emph{{C
  $^{2}$-ray: A new method for photon-conserving transport of ionizing
  radiation}}, \href{https://doi.org/10.1016/j.newast.2005.09.004}{\emph{\na}
  {\bfseries 11} (2006) 374}
  [\href{https://arxiv.org/abs/astro-ph/0508416}{{\ttfamily
  astro-ph/0508416}}].

\bibitem{mcquinn07}
M.~{McQuinn}, A.~{Lidz}, O.~{Zahn}, S.~{Dutta}, L.~{Hernquist} and
  M.~{Zaldarriaga}, \emph{{The morphology of HII regions during reionization}},
  \href{https://doi.org/10.1111/j.1365-2966.2007.11489.x}{\emph{\mnras}
  {\bfseries 377} (2007) 1043}
  [\href{https://arxiv.org/abs/astro-ph/0610094}{{\ttfamily
  astro-ph/0610094}}].

\bibitem{trac07}
H.~{Trac} and R.~{Cen}, \emph{{Radiative Transfer Simulations of Cosmic
  Reionization. I. Methodology and Initial Results}},
  \href{https://doi.org/10.1086/522566}{\emph{\apj} {\bfseries 671} (2007) 1}
  [\href{https://arxiv.org/abs/astro-ph/0612406}{{\ttfamily
  astro-ph/0612406}}].

\bibitem{wise11}
J.~H. {Wise} and T.~{Abel}, \emph{{ENZO+MORAY: radiation hydrodynamics adaptive
  mesh refinement simulations with adaptive ray tracing}},
  \href{https://doi.org/10.1111/j.1365-2966.2011.18646.x}{\emph{\mnras}
  {\bfseries 414} (2011) 3458}
  [\href{https://arxiv.org/abs/1012.2865}{{\ttfamily 1012.2865}}].

\bibitem{pawlik08}
A.~H. {Pawlik} and J.~{Schaye}, \emph{{TRAPHIC - radiative transfer for
  smoothed particle hydrodynamics simulations}},
  \href{https://doi.org/10.1111/j.1365-2966.2008.13601.x}{\emph{\mnras}
  {\bfseries 389} (2008) 651}
  [\href{https://arxiv.org/abs/0802.1715}{{\ttfamily 0802.1715}}].

\bibitem{pawlik11}
A.~H. {Pawlik} and J.~{Schaye}, \emph{{Multifrequency, thermally coupled
  radiative transfer with TRAPHIC: method and tests}},
  \href{https://doi.org/10.1111/j.1365-2966.2010.18032.x}{\emph{\mnras}
  {\bfseries 412} (2011) 1943}
  [\href{https://arxiv.org/abs/1008.1071}{{\ttfamily 1008.1071}}].

\bibitem{petkova11code}
M.~{Petkova} and V.~{Springel}, \emph{{A novel approach for accurate radiative
  transfer in cosmological hydrodynamic simulations}},
  \href{https://doi.org/10.1111/j.1365-2966.2011.18986.x}{\emph{\mnras}
  {\bfseries 415} (2011) 3731}
  [\href{https://arxiv.org/abs/1012.1017}{{\ttfamily 1012.1017}}].

\bibitem{iliev06sim}
I.~T. {Iliev}, G.~{Mellema}, U.~L. {Pen}, H.~{Merz}, P.~R. {Shapiro} and M.~A.
  {Alvarez}, \emph{{Simulating cosmic reionization at large scales - I. The
  geometry of reionization}},
  \href{https://doi.org/10.1111/j.1365-2966.2006.10502.x}{\emph{\mnras}
  {\bfseries 369} (2006) 1625}
  [\href{https://arxiv.org/abs/astro-ph/0512187}{{\ttfamily
  astro-ph/0512187}}].

\bibitem{mellama06sim}
G.~{Mellema}, I.~T. {Iliev}, U.-L. {Pen} and P.~R. {Shapiro}, \emph{{Simulating
  cosmic reionization at large scales - II. The 21-cm emission features and
  statistical signals}},
  \href{https://doi.org/10.1111/j.1365-2966.2006.10919.x}{\emph{\mnras}
  {\bfseries 372} (2006) 679}
  [\href{https://arxiv.org/abs/astro-ph/0603518}{{\ttfamily
  astro-ph/0603518}}].

\bibitem{jensen13}
H.~{Jensen}, K.~K. {Datta}, G.~{Mellema}, E.~{Chapman}, F.~B. {Abdalla}, I.~T.
  {Iliev} et~al., \emph{{Probing reionization with LOFAR using 21-cm redshift
  space distortions}},
  \href{https://doi.org/10.1093/mnras/stt1341}{\emph{\mnras} {\bfseries 435}
  (2013) 460} [\href{https://arxiv.org/abs/1303.5627}{{\ttfamily 1303.5627}}].

\bibitem{giri19}
S.~K. {Giri}, G.~{Mellema}, T.~{Aldheimer}, K.~L. {Dixon} and I.~T. {Iliev},
  \emph{{Neutral island statistics during reionization from 21-cm tomography}},
  \href{https://doi.org/10.1093/mnras/stz2224}{\emph{\mnras} {\bfseries 489}
  (2019) 1590} [\href{https://arxiv.org/abs/1903.01294}{{\ttfamily
  1903.01294}}].

\bibitem{trac08}
H.~{Trac}, R.~{Cen} and A.~{Loeb}, \emph{{Imprint of Inhomogeneous Hydrogen
  Reionization on the Temperature Distribution of the Intergalactic Medium}},
  \href{https://doi.org/10.1086/595678}{\emph{\apjl} {\bfseries 689} (2008)
  L81} [\href{https://arxiv.org/abs/0807.4530}{{\ttfamily 0807.4530}}].

\bibitem{battaglia13}
N.~{Battaglia}, A.~{Natarajan}, H.~{Trac}, R.~{Cen} and A.~{Loeb},
  \emph{{Reionization on Large Scales. III. Predictions for Low-l Cosmic
  Microwave Background Polarization and High-l Kinetic Sunyaev-Zel'dovich
  Observables}}, \href{https://doi.org/10.1088/0004-637X/776/2/83}{\emph{\apj}
  {\bfseries 776} (2013) 83} [\href{https://arxiv.org/abs/1211.2832}{{\ttfamily
  1211.2832}}].

\bibitem{molaro19}
M.~{Molaro}, R.~{Dav{\'e}}, S.~{Hassan}, M.~G. {Santos} and K.~{Finlator},
  \emph{{Artist: fast radiative transfer for large-scale simulations of the
  epoch of reionization}},
  \href{https://doi.org/10.1093/mnras/stz2171}{\emph{\mnras} {\bfseries 489}
  (2019) 5594} [\href{https://arxiv.org/abs/1901.03340}{{\ttfamily
  1901.03340}}].

\bibitem{daloisio20}
A.~{D'Aloisio}, M.~{McQuinn}, H.~{Trac}, C.~{Cain} and A.~{Mesinger},
  \emph{{Hydrodynamic Response of the Intergalactic Medium to Reionization}},
  {\emph{arXiv e-prints} (2020) arXiv:2002.02467}
  [\href{https://arxiv.org/abs/2002.02467}{{\ttfamily 2002.02467}}].

\bibitem{mao20}
Y.~{Mao}, J.~{Koda}, P.~R. {Shapiro}, I.~T. {Iliev}, G.~{Mellema}, H.~{Park}
  et~al., \emph{{The impact of inhomogeneous subgrid clumping on cosmic
  reionization}}, \href{https://doi.org/10.1093/mnras/stz2986}{\emph{\mnras}
  {\bfseries 491} (2020) 1600}
  [\href{https://arxiv.org/abs/1906.02476}{{\ttfamily 1906.02476}}].

\bibitem{alvarez12}
M.~A. {Alvarez} and T.~{Abel}, \emph{{The Effect of Absorption Systems on
  Cosmic Reionization}},
  \href{https://doi.org/10.1088/0004-637X/747/2/126}{\emph{\apj} {\bfseries
  747} (2012) 126} [\href{https://arxiv.org/abs/1003.6132}{{\ttfamily
  1003.6132}}].

\bibitem{mhr00}
J.~{Miralda-Escud{\'e}}, M.~{Haehnelt} and M.~J. {Rees}, \emph{{Reionization of
  the Inhomogeneous Universe}},
  \href{https://doi.org/10.1086/308330}{\emph{\apj} {\bfseries 530} (2000) 1}
  [\href{https://arxiv.org/abs/astro-ph/9812306}{{\ttfamily
  astro-ph/9812306}}].

\bibitem{miraldaescude03}
J.~{Miralda-Escud{\'e}}, \emph{{On the Evolution of the Ionizing Emissivity of
  Galaxies and Quasars Required by the Hydrogen Reionization}},
  \href{https://doi.org/10.1086/378286}{\emph{\apj} {\bfseries 597} (2003) 66}
  [\href{https://arxiv.org/abs/astro-ph/0211071}{{\ttfamily
  astro-ph/0211071}}].

\bibitem{rahmati14}
A.~{Rahmati} and J.~{Schaye}, \emph{{Predictions for the relation between
  strong HI absorbers and galaxies at redshift 3}},
  \href{https://doi.org/10.1093/mnras/stt2235}{\emph{\mnras} {\bfseries 438}
  (2014) 529} [\href{https://arxiv.org/abs/1310.3317}{{\ttfamily 1310.3317}}].

\bibitem{mcquinn11}
M.~{McQuinn}, S.~P. {Oh} and C.-A. {Faucher-Gigu{\`e}re}, \emph{{On Lyman-limit
  Systems and the Evolution of the Intergalactic Ionizing Background}},
  \href{https://doi.org/10.1088/0004-637X/743/1/82}{\emph{\apj} {\bfseries 743}
  (2011) 82} [\href{https://arxiv.org/abs/1101.1964}{{\ttfamily 1101.1964}}].

\bibitem{munoz16}
J.~A. {Mu{\~n}oz}, S.~P. {Oh}, F.~B. {Davies} and S.~R. {Furlanetto},
  \emph{{The flatness and sudden evolution of the intergalactic ionizing
  background}}, \href{https://doi.org/10.1093/mnras/stv2355}{\emph{\mnras}
  {\bfseries 455} (2016) 1385}
  [\href{https://arxiv.org/abs/1410.2249}{{\ttfamily 1410.2249}}].

\bibitem{zheng02}
Z.~{Zheng} and J.~{Miralda-Escud{\'e}}, \emph{{Self-shielding Effects on the
  Column Density Distribution of Damped Ly{\ensuremath{\alpha}} Systems}},
  \href{https://doi.org/10.1086/340330}{\emph{\apjl} {\bfseries 568} (2002)
  L71} [\href{https://arxiv.org/abs/astro-ph/0201275}{{\ttfamily
  astro-ph/0201275}}].

\bibitem{rahmati13}
A.~{Rahmati}, A.~H. {Pawlik}, M.~{Rai{\v{c}}evi{\'c}} and J.~{Schaye},
  \emph{{On the evolution of the H I column density distribution in
  cosmological simulations}},
  \href{https://doi.org/10.1093/mnras/stt066}{\emph{\mnras} {\bfseries 430}
  (2013) 2427} [\href{https://arxiv.org/abs/1210.7808}{{\ttfamily 1210.7808}}].

\bibitem{becker13}
G.~D. {Becker} and J.~S. {Bolton}, \emph{{New measurements of the ionizing
  ultraviolet background over 2 \&lt; z \&lt; 5 and implications for hydrogen
  reionization}}, \href{https://doi.org/10.1093/mnras/stt1610}{\emph{\mnras}
  {\bfseries 436} (2013) 1023}
  [\href{https://arxiv.org/abs/1307.2259}{{\ttfamily 1307.2259}}].

\bibitem{wyithe11}
J.~S.~B. {Wyithe} and J.~S. {Bolton}, \emph{{Near-zone sizes and the rest-frame
  extreme ultraviolet spectral index of the highest redshift quasars}},
  \href{https://doi.org/10.1111/j.1365-2966.2010.18030.x}{\emph{\mnras}
  {\bfseries 412} (2011) 1926}
  [\href{https://arxiv.org/abs/1008.1107}{{\ttfamily 1008.1107}}].

\bibitem{calverley11}
A.~P. {Calverley}, G.~D. {Becker}, M.~G. {Haehnelt} and J.~S. {Bolton},
  \emph{{Measurements of the ultraviolet background at 4.6 \&lt; z \&lt; 6.4
  using the quasar proximity effect}},
  \href{https://doi.org/10.1111/j.1365-2966.2010.18072.x}{\emph{\mnras}
  {\bfseries 412} (2011) 2543}
  [\href{https://arxiv.org/abs/1011.5850}{{\ttfamily 1011.5850}}].

\bibitem{daloisio18}
A.~{D'Aloisio}, M.~{McQuinn}, F.~B. {Davies} and S.~R. {Furlanetto},
  \emph{{Large fluctuations in the high-redshift metagalactic ionizing
  background}}, \href{https://doi.org/10.1093/mnras/stx2341}{\emph{\mnras}
  {\bfseries 473} (2018) 560}
  [\href{https://arxiv.org/abs/1611.02711}{{\ttfamily 1611.02711}}].

\bibitem{prochaska09}
J.~X. {Prochaska} and A.~M. {Wolfe}, \emph{{On the (Non)Evolution of H I Gas in
  Galaxies Over Cosmic Time}},
  \href{https://doi.org/10.1088/0004-637X/696/2/1543}{\emph{\apj} {\bfseries
  696} (2009) 1543} [\href{https://arxiv.org/abs/0811.2003}{{\ttfamily
  0811.2003}}].

\bibitem{crighton15}
N.~H.~M. {Crighton}, M.~T. {Murphy}, J.~X. {Prochaska}, G.~{Worseck},
  M.~{Rafelski}, G.~D. {Becker} et~al., \emph{{The neutral hydrogen
  cosmological mass density at z = 5}},
  \href{https://doi.org/10.1093/mnras/stv1182}{\emph{\mnras} {\bfseries 452}
  (2015) 217} [\href{https://arxiv.org/abs/1506.02037}{{\ttfamily
  1506.02037}}].

\bibitem{bird17}
S.~{Bird}, R.~{Garnett} and S.~{Ho}, \emph{{Statistical properties of damped
  Lyman-alpha systems from Sloan Digital Sky Survey DR12}},
  \href{https://doi.org/10.1093/mnras/stw3246}{\emph{\mnras} {\bfseries 466}
  (2017) 2111} [\href{https://arxiv.org/abs/1610.01165}{{\ttfamily
  1610.01165}}].

\bibitem{padmanabhan16}
H.~{Padmanabhan}, T.~R. {Choudhury} and A.~{Refregier}, \emph{{Modelling the
  cosmic neutral hydrogen from DLAs and 21-cm observations}},
  \href{https://doi.org/10.1093/mnras/stw353}{\emph{\mnras} {\bfseries 458}
  (2016) 781} [\href{https://arxiv.org/abs/1505.00008}{{\ttfamily
  1505.00008}}].

\bibitem{davies16}
F.~B. {Davies} and S.~R. {Furlanetto}, \emph{{Large fluctuations in the
  hydrogen-ionizing background and mean free path following the epoch of
  reionization}}, \href{https://doi.org/10.1093/mnras/stw931}{\emph{\mnras}
  {\bfseries 460} (2016) 1328}
  [\href{https://arxiv.org/abs/1509.07131}{{\ttfamily 1509.07131}}].

\bibitem{davies18}
F.~B. {Davies}, G.~D. {Becker} and S.~R. {Furlanetto}, \emph{{Determining the
  Nature of Late Gunn-Peterson Troughs with Galaxy Surveys}},
  \href{https://doi.org/10.3847/1538-4357/aac2d6}{\emph{\apj} {\bfseries 860}
  (2018) 155} [\href{https://arxiv.org/abs/1708.08927}{{\ttfamily
  1708.08927}}].

\bibitem{nasir20}
F.~{Nasir} and A.~{D'Aloisio}, \emph{{Observing the tail of reionization:
  neutral islands in the z = 5.5 Lyman-{\ensuremath{\alpha}} forest}},
  \href{https://doi.org/10.1093/mnras/staa894}{\emph{\mnras} {\bfseries 494}
  (2020) 3080} [\href{https://arxiv.org/abs/1910.03570}{{\ttfamily
  1910.03570}}].

\bibitem{onorbe19}
J.~{O{\~n}orbe}, F.~B. {Davies}, {Luki{\'c}}, Z.~{}, J.~F. {Hennawi} and
  D.~{Sorini}, \emph{{Inhomogeneous reionization models in cosmological
  hydrodynamical simulations}},
  \href{https://doi.org/10.1093/mnras/stz984}{\emph{\mnras} {\bfseries 486}
  (2019) 4075} [\href{https://arxiv.org/abs/1810.11683}{{\ttfamily
  1810.11683}}].

\bibitem{croft99}
R.~A.~C. {Croft}, D.~H. {Weinberg}, M.~{Pettini}, L.~{Hernquist} and N.~{Katz},
  \emph{{The Power Spectrum of Mass Fluctuations Measured from the
  Ly{\ensuremath{\alpha}} Forest at Redshift z = 2.5}},
  \href{https://doi.org/10.1086/307438}{\emph{\apj} {\bfseries 520} (1999) 1}
  [\href{https://arxiv.org/abs/astro-ph/9809401}{{\ttfamily
  astro-ph/9809401}}].

\bibitem{croft04}
R.~A.~C. {Croft}, \emph{{Ionizing Radiation Fluctuations and Large-Scale
  Structure in the Ly{\ensuremath{\alpha}} Forest}},
  \href{https://doi.org/10.1086/421839}{\emph{\apj} {\bfseries 610} (2004) 642}
  [\href{https://arxiv.org/abs/astro-ph/0310890}{{\ttfamily
  astro-ph/0310890}}].

\bibitem{meiksin04}
A.~{Meiksin} and M.~{White}, \emph{{The effects of ultraviolet background
  correlations on Ly{\ensuremath{\alpha}} forest flux statistics}},
  \href{https://doi.org/10.1111/j.1365-2966.2004.07724.x}{\emph{\mnras}
  {\bfseries 350} (2004) 1107}
  [\href{https://arxiv.org/abs/astro-ph/0307289}{{\ttfamily
  astro-ph/0307289}}].

\bibitem{mcdonald05}
P.~{McDonald}, U.~{Seljak}, R.~{Cen}, P.~{Bode} and J.~P. {Ostriker},
  \emph{{Physical effects on the Ly{\ensuremath{\alpha}} forest flux power
  spectrum: damping wings, ionizing radiation fluctuations and galactic
  winds}},
  \href{https://doi.org/10.1111/j.1365-2966.2005.09141.x}{\emph{\mnras}
  {\bfseries 360} (2005) 1471}
  [\href{https://arxiv.org/abs/astro-ph/0407378}{{\ttfamily
  astro-ph/0407378}}].

\bibitem{slosar09}
A.~{Slosar}, S.~{Ho}, M.~{White} and T.~{Louis}, \emph{{The acoustic peak in
  the Lyman alpha forest}},
  \href{https://doi.org/10.1088/1475-7516/2009/10/019}{\emph{\jcap} {\bfseries
  2009} (2009) 019} [\href{https://arxiv.org/abs/0906.2414}{{\ttfamily
  0906.2414}}].

\bibitem{maselli05}
A.~{Maselli} and A.~{Ferrara}, \emph{{Radiative transfer effects on the
  Ly{\ensuremath{\alpha}} forest}},
  \href{https://doi.org/10.1111/j.1365-2966.2005.09682.x}{\emph{\mnras}
  {\bfseries 364} (2005) 1429}
  [\href{https://arxiv.org/abs/astro-ph/0510258}{{\ttfamily
  astro-ph/0510258}}].

\bibitem{mesinger09}
A.~{Mesinger} and S.~{Furlanetto}, \emph{{The inhomogeneous ionizing background
  following reionization}},
  \href{https://doi.org/10.1111/j.1365-2966.2009.15547.x}{\emph{\mnras}
  {\bfseries 400} (2009) 1461}
  [\href{https://arxiv.org/abs/0906.3020}{{\ttfamily 0906.3020}}].

\bibitem{pontzen14}
A.~{Pontzen}, \emph{{Scale-dependent bias in the
  baryonic-acoustic-oscillation-scale intergalactic neutral hydrogen}},
  \href{https://doi.org/10.1103/PhysRevD.89.083010}{\emph{\prd} {\bfseries 89}
  (2014) 083010} [\href{https://arxiv.org/abs/1402.0506}{{\ttfamily
  1402.0506}}].

\bibitem{gontcho14}
S.~{Gontcho A Gontcho}, J.~{Miralda-Escud{\'e}} and N.~G. {Busca}, \emph{{On
  the effect of the ionizing background on the Ly{\ensuremath{\alpha}} forest
  autocorrelation function}},
  \href{https://doi.org/10.1093/mnras/stu860}{\emph{\mnras} {\bfseries 442}
  (2014) 187} [\href{https://arxiv.org/abs/1404.7425}{{\ttfamily 1404.7425}}].

\bibitem{cooray02}
A.~{Cooray} and R.~{Sheth}, \emph{{Halo models of large scale structure}},
  \href{https://doi.org/10.1016/S0370-1573(02)00276-4}{\emph{\physrep}
  {\bfseries 372} (2002) 1}
  [\href{https://arxiv.org/abs/astro-ph/0206508}{{\ttfamily
  astro-ph/0206508}}].

\bibitem{mcquinn14}
M.~{McQuinn} and G.~{Worseck}, \emph{{The case against large intensity
  fluctuations in the z ̃ 2.5 He II Ly{\ensuremath{\alpha}} forest}},
  \href{https://doi.org/10.1093/mnras/stu242}{\emph{\mnras} {\bfseries 440}
  (2014) 2406} [\href{https://arxiv.org/abs/1306.4985}{{\ttfamily 1306.4985}}].

\bibitem{mcquinn18}
M.~{McQuinn} and A.~{D'Aloisio}, \emph{{The observable 21cm signal from
  reionization may be perturbative}},
  \href{https://doi.org/10.1088/1475-7516/2018/10/016}{\emph{\jcap} {\bfseries
  2018} (2018) 016} [\href{https://arxiv.org/abs/1806.08372}{{\ttfamily
  1806.08372}}].

\bibitem{eilers18}
A.-C. {Eilers}, F.~B. {Davies} and J.~F. {Hennawi}, \emph{{The Opacity of the
  Intergalactic Medium Measured along Quasar Sightlines at z
  {\ensuremath{\sim}} 6}},
  \href{https://doi.org/10.3847/1538-4357/aad4fd}{\emph{\apj} {\bfseries 864}
  (2018) 53} [\href{https://arxiv.org/abs/1807.04229}{{\ttfamily 1807.04229}}].

\bibitem{deparis19}
N.~{Deparis}, D.~{Aubert}, P.~{Ocvirk}, J.~{Chardin} and J.~{Lewis},
  \emph{{Impact of the reduced speed of light approximation on ionization front
  velocities in cosmological simulations of the epoch of reionization}},
  \href{https://doi.org/10.1051/0004-6361/201832889}{\emph{\aap} {\bfseries
  622} (2019) A142} [\href{https://arxiv.org/abs/1803.01634}{{\ttfamily
  1803.01634}}].

\bibitem{ocvirk19}
P.~{Ocvirk}, D.~{Aubert}, J.~{Chardin}, N.~{Deparis} and J.~{Lewis},
  \emph{{Impact of the reduced speed of light approximation on the post-overlap
  neutral hydrogen fraction in numerical simulations of the epoch of
  reionization}},
  \href{https://doi.org/10.1051/0004-6361/201832923}{\emph{\aap} {\bfseries
  626} (2019) A77} [\href{https://arxiv.org/abs/1803.02434}{{\ttfamily
  1803.02434}}].

\bibitem{garaldi19}
E.~{Garaldi}, N.~Y. {Gnedin} and P.~{Madau}, \emph{{Constraining the Tail End
  of Reionization Using Ly{\ensuremath{\alpha}} Transmission Spikes}},
  \href{https://doi.org/10.3847/1538-4357/ab12dc}{\emph{\apj} {\bfseries 876}
  (2019) 31} [\href{https://arxiv.org/abs/1902.07713}{{\ttfamily 1902.07713}}].

\bibitem{furlanetto09}
S.~R. {Furlanetto} and A.~{Mesinger}, \emph{{The ionizing background at the end
  of reionization}},
  \href{https://doi.org/10.1111/j.1365-2966.2009.14449.x}{\emph{\mnras}
  {\bfseries 394} (2009) 1667}
  [\href{https://arxiv.org/abs/0809.4493}{{\ttfamily 0809.4493}}].

\bibitem{gnedin16}
N.~Y. {Gnedin}, \emph{{On the Proper Use of the Reduced Speed of Light
  Approximation}},
  \href{https://doi.org/10.3847/1538-4357/833/1/66}{\emph{\apj} {\bfseries 833}
  (2016) 66} [\href{https://arxiv.org/abs/1607.07869}{{\ttfamily 1607.07869}}].

\end{thebibliography}\endgroup
\bibliographystyle{JHEP}

\end{document}